\documentclass[a4paper,fleqn,usenatbib]{mnras}

\usepackage[T1]{fontenc}
\usepackage{ae,aecompl}
\usepackage{graphicx}	
\usepackage{amsmath}	
\usepackage{amssymb}	
\usepackage{pdflscape}
\usepackage{pifont}
\usepackage{float}
\usepackage{caption}
\usepackage{booktabs}

\usepackage{color}
\definecolor{myblue}{RGB}{30,144,255}

\definecolor{myred}{RGB}{239, 86, 48}

\newcommand{\ebmv}{$E(B-V)$}
\newcommand{\mbh}{$M_{BH}$}
\newcommand{\Mdot}{$\dot{M}$}
\newcommand{\mdot}{$\dot{m}$}
\newcommand{\msun}{$M_{\odot}$}
\newcommand{\hbeta}{H$\beta$}

\newcommand{\Ledd}{$L_{\text{AGN}}/L_{\text{Edd}}$}

\newcommand{\mAA}{\textup{\AA}}

\title[super-Eddington AGN SED]{Super- and sub-Eddington accreting massive black holes: A comparison
of slim and thin accretion discs through study of the spectral energy distribution.}

\author[Castell\'o-Mor et al.]{
N. Castell\'o-Mor$^{1}$\thanks{E-mail: nuria@wise.tau.ac.il (TAU)},
H. Netzer$^{1}$
and S. Kaspi$^{1,2}$
\\
$^{1}$School of Physics and Astronomy, Tel Aviv University, Tel Aviv 69978, Israel\\
$^{2}$Wise Observatory, School of Physics and Astronomy, Tel Aviv University, Tel Aviv 69978, Israel
}

\date{Accepted XXX. Received YYY; in original form ZZZ}

\pubyear{2015}

\begin{document}
\label{firstpage}
\pagerange{\pageref{firstpage}--\pageref{lastpage}}
\maketitle

\begin{abstract}
We employ optical and UV observations to present SEDs for two reverberation-mapped
samples of super-Eddington and sub-Eddington AGN with similar luminosity
distributions. The samples are fitted with accretion disc models in order to look
for SED differences that depend on the Eddington ratio. The fitting takes into
account measured BH mass and accretion rates, BH spin and intrinsic reddening of
the sources. All objects in both groups can be fitted by thin AD models over the
range 0.2-1$\,\mu$m with reddening as a free parameter. The intrinsic reddening
required to fit the data are relatively small, $E(B-V)\leq0.2$~mag, except for one
source. Super-Eddington AGN seem to require more reddening. The distribution of
$E(B-V)$ is similar to what is observed in larger AGN samples. The best fit disc
models recover very well the BH mass and accretion for the two groups. However, 
the SEDs are very different, with super-Eddington sources requiring much more luminous 
far-UV continuum. The exact amount depends on the possible
saturation of the UV radiation in slim discs. In particular, we derive for the
super-Eddington sources a typical bolometric correction at 5100\AA{} of 60--150
compared with a median of $\sim$20 for the sub-Eddington AGN. The measured torus
luminosity relative to $\lambda L_{\lambda}(5100\mAA{}$) are similar in both
groups. The $\alpha_{OX}$ distribution is similar too. However, we find extremely
small torus covering factors for super-Eddington sources, an order of magnitude
smaller than those of sub-Eddington AGN. The small differences between the groups
regarding the spectral range 0.2-22$\,\mu$m, and the significant differences related
to the part of the SED that we cannot observe may be consistent with some slim disc
models. An alternative explanation is that present day slim-disc models over-estimate
the far UV luminosity of such objects by a large amount.
\end{abstract}


\section{Introduction}
\label{Introduction}

Optically thick accretion flows in the vicinity of the central black hole (BH) are believed to be the main
power-house of active galactic nuclei (AGN). The emitted radiation from such systems is determined by
the BH mass ($M_{BH}$),
BH spin ($a_{\star}$), and the mass accretion rate, \Mdot. The accretion
rate can be expressed in normalized units, or Eddington ratio, $\dot{m}=$\Ledd, where $L_{\text{AGN}}$ is the bolometric
luminosity of the system. Optically thick geometrically thin accretion discs (ADs) have been proposed to explain
the observed spectral energy distribution (SED) of many AGN with low Eddington ratios
\citep[\Ledd$\le 0.3$,][]{Koratkar1999,Blaes2001,Shang2005,Davis2007,Davis2011,Laor2011,Jin2012,Slone2012,Netzer2014,Capellupo2015}.
At higher accretion rates, the disc becomes ``thick'' or ``slim'' and
the nature of the accretion changes dramatically with processes like photon trapping and advection becoming important
\cite[][see review by \citealt{Wang2014a}]{Abramowicz1988,Sadowski2014}.  Slim ADs are thought to have  SEDs that are
different from thin ADs, with an energy cut-off that extends to higher energies and a strong anisotropy of the emitted
radiation. We follow \citet{Wang2014b} and coin such objects ``Super-Eddington Accreting Massive Black Holes''
(SEAMBHs).

The radiation efficiency of thin ADs around BHs, $\eta$, defined by  $L_{\text{AGN}}=\eta\dot{M}_{BH}c^2$, is
obtained from  the standard AD theory \citep[e.g.][]{Shakura1973,Thorne1974}. $\eta$ depends on the location of the innermost
stable circular orbit (ISCO) which in turn depends on the BH spin. For maximally rotating BHs, with spin parameter
$a_{\star}=0.998$, $\eta=0.32$ and for retrograde discs, with $a_{\star}=-1$, $\eta=0.038$.  This is not the case for slim ADs that
are not well understood. In such cases, there is an
ill-defined radiation efficiency that may not depend on the BH spin and is suggested to be considerably smaller than the
corresponding thin AD efficiency due to the so-called ``photon trapping''. The issue is particularly important for very high accretion rates where the
theoretical models suggest that the emitted radiation follows an expression of the type \citep{Mineshige2000}
\begin{equation}
    L_{\text{AGN}}\approx 3 \times 10^{38}  \left[1+\ln{\dot{\mathcal{M}}/\dot{\mathcal{M}}_{crit}}\right]\, M_{BH}
\label{eq.saturation}
\end{equation}
where $\dot{\mathcal{M}}$ is the dimensionless accretion rate defined as $\dot{m}=\eta \dot{\mathcal{M}}$ and $\dot{\mathcal{M}}_{crit}\approx20$ \citep{DuPu2015}.
Such theoretical approximations are yet to be confirmed
by numerical simulations. For example, the recent numerical simulations of such objects by
\citet{Sadowski2015} suggest a relatively
small drop in efficiency up to extremely high accretion rates.
Finally, all AGN are known to be powerful X-ray sources with $L_{2-10\,\text{keV}}$ luminosity that, in most observed
cases, is considerably below the integrated optical-UV luminosity \citep[e.g.][]{Marconi2004,Steffen2006,Vasudevan2007,
Vasudevan2009,Grupe2010,Marchese2012}.
The dimension and variability of the X-ray source suggest that it is intimately related to the accretion process,
thus the emitted X-ray radiation is most probably drawn from the accretion process itself. This must be taken into account
when comparing the radiation efficiencies of thin and slim ADs.

The observed optical-UV SEDs of some AGN resemble the prediction of the thin AD model. In particular, they show a prominent
bump in the optical-ultraviolet, the \emph{Big Blue Bump} (BBB), which peaks at a BH-mass dependent frequency
and declines at higher energies.
Unfortunately, the comparison with theoretical AD SEDs has been hampered by the lack of simultaneous observations of these highly variable
sources, and the limited wavelength range of most observations. Another source of uncertainty is intrinsic reddening
due to dust in the host galaxy of the AGN. Because of this, many earlier studies failed to reach a  conclusion regarding the
origin of the observed SED with only marginal indications for a disc-like spectrum \citep[see][]{Koratkar1999,Davis2007}.
Moreover, an empirical SED made of a broken power-law with slopes $-0.5<\alpha<1.5$, where $F_{\nu} \propto \nu^{-\alpha}$,
have been shown to give better fits to many observed, limited wavelength spectra \citep{Zheng1997}.

The recent work of \citet{Capellupo2015} shed new light on this issue. The work is based on VLT/Xshooter observations of
39 type-I AGN with $z\approx 1.55$ and a large range of $L_{\text{AGN}}$ and \mdot. They provide simultaneous information on the
rest-frame wavelength range of 1100-9200\AA{} which is large enough to test the AD model predictions.
More than 90\% of the sources in this sample are well fitted with a thin AD model. The fitting requires moderate intrinsic reddening in
$\sim$30\% of the sources with extinction in the range $0.1<A_{V}<0.5\,$mag. This indicates that many earlier attempts to
discover the unique spectral signature of thin ADs failed as a result of AGN variability and non-simultaneous data that cover
only a limited wavelength range. All the sources in the \citet{Capellupo2015} sample are relatively low accretion
rate systems with 90\% of the sources in the range $10^{-2}<\dot{m}<0.3$ and not a single source
with $\dot{m}>1$.

The most accurate information about \mdot{} in type-I AGN is obtained for sources with directly measured  BH mass through
reverberation mapping (RM). Such information is now available for about 40 objects with $\dot{m}<0.1$ \citep[see][]{Bentz2013} and
for 15 objects with $\dot{m}>0.1$ \citep[][the objects we refer to as SEAMBHs]{DuPu2015}. The comparison of the two groups suggests
that for a given optical
luminosity, the emissivity weighted radius of the broad line region (BLR) is considerably smaller in SEAMBHs with differences that
can amount
to a factor of $\approx 3$. As explained in \cite{Wang2014b} and \cite{DuPu2015}, this can be interpreted as a change in the nature
of the power-house
where the lower \mdot{} systems are powered by thin ADs and the higher \mdot{} objects by slim ADs.

There are several other well known difference between high and low \mdot{} systems.
The first  is the nature of their X-ray continuum manifested by higher variability amplitude, steeper
soft X-ray slope and larger $\alpha_{OX}$ (the slope connecting the flux at 2500\AA{} and 2 keV) for
the higher \mdot{} objects. These differences have been studied, extensively, in numerous
papers \citep{Pogge2000,Steffen2006,Gallo2006}, especially among lower luminosity
AGN. Clear spectroscopic differences are also found at optical-UV wavelengths where, 
for a given monochromatic luminosity, the higher \mdot{} sources show narrower broad emission lines
and much stronger Fe{\sc ii} lines. Such objects have been named ``narrow line Seyfert 1 galaxies'' (NLS1s).
Their spectral differences are well characterized in the ``eigen-vector 1'' scheme \citep{Sulentic2000}.

The goal of the present work is to make a detailed comparison of the optical-UV SEDs
of sub-Eddington and super-Eddington AGN. The number of accurately measured BH mass in SEAMBHs
has reached a stage where such a comparison is feasible and can be used to test various models
and scenarios related to the AGN power-house and the accretion process itself. In particular,
we want to test whether thin and slim AD models reliably explain the observed
optical-UV SED of such objects and the properties of their dusty tori, and test the
differences, if any, between the two groups. An important aim of the present work is to
improve the X-ray-to-optical SED measurements and hence the
estimates of the bolometric luminosity of AGN in sources powered by thin or slim ADs.
The pioneering study of \citet{Elvis1994} presented radio-to-X-ray SEDs for quasar, is restricted to
bright X-ray sources and cover both the intrinsic AGN continuum and the processed torus emission.
Unfortunately, the processed radiation is added to the intrinsic emission which results in double
counting and too large bolometric correction factors. In later works, double counting was avoided,
but intrinsic reddening and host galaxy contribution was not taken into account
\citep[see e.g.][]{Marconi2004,Hopkins2004,Richards2006,Vasudevan2007}. This affects the bolometric
luminosity measurements, underestimating its value. More recent works
\citep[][\citealt{Vasudevan2009b} hereafter V09, \citealt{Jin2012}, among others]{Brocksopp2006,Vasudevan2009},
presented optical-to-X-ray SED emphasizing interesting trends between SED signature and Eddington ratio.
However, most samples were restricted to low accretion rate AGN. In particular, there was a lack
of super-Eddington sources with reliable, RM-based BH mass estimates since such objects were not known
at the time.

The sample of AGN presented here is restricted to those AGN which are bright
enough in the optical/UV and all have direct mass measurements. There are 29 
objects selected by their spectroscopic properties with 16 of those showing 
extremely high accretion rates (referred to as super-Eddington AGN). This makes 
our approach different than all previous studies.
From all previous studies, the work of V09 is the one most similar to ours.
It takes into account reddening and host galaxy corrections in the optical/UV,
but focus on low accretion rate sources and does not discuss the torus properties. Below we show
the similarity between V09 and our study regarding sub-Eddington AGN, and the missing information
about the torus, and make detailed comparison with the new group of super-Eddington sources.
The most recent works of \citet{Marchese2012} and \citet{Fanali2013} are other attempts
to correlate accretion disc and X-ray properties of luminous AGN. There are several fundamental
differences between these papers and the present work. First, the BH mass estimate is based on single-epoch
(virial) method. Second, the accretion disc model is not a fit to individual SEDs but rather the same
model for all sources with an assumed stationary BH but an efficiency $\eta$ of 0.1 which is more
appropriate to rotating BH. Third, and most important, the three bands used for the estimate bolometric
luminosity (X-ray XMM-Newton, UV GALEX and optical SDSS) are not obtained simultaneously which,
as we show below, introduces an uncertainty which is impossible to estimate. Because of these differences
we do not attempt to compare the results obtained in these papers to the one presented below.
Recently, \citet{Jin2016} presented a new result for a super-Eddington source
RX~J1140.1+0307, where upper limit on the BH mass was obtained through RM (the upper limit
is on the $H\beta$ time lag). \citet{Jin2016} present several SED fits based on
accretion disc models using non-simultaneous data. There are also difficulties in host
galaxy subtraction and reddening correction. \citet{Jin2016} found that
the best-fit SED overestimate the mass of the black hole by an order of magnitude
relative to the RM results. Because of these and the fact that the source is in the
lowest black hole mass regime we do not attempt to compare their results to the ones
presented here.

The structure of the paper is as follows. In Section~\ref{SampleSelection} we describe the sample selection and the observational data.
In section~\ref{intrinsicSEDs} we describe the intrinsic disc SED that we use and the fitting procedure is presented in
section~\ref{SEDmodelling}. The principal results of the thin AD modelling and some additional spectral properties are
presented in Section~\ref{Results}. Finally, in Section~\ref{Conclusions}, we summarize our main conclusions from this work.
Throughout this paper we assume a $\Lambda$CDM cosmology with $\Omega_{\Lambda}=0.7$, $\Omega_{m}=0.3$, and
$H_0=70$ km/s/Mpc.

\section{Sample Selection}
\label{SampleSelection}

The goal of this work is to compare various properties of high Eddington ratio AGN
with RM-measured BH mass with a similar group of low Eddington ratio AGN.
Our high Eddington ratio sample consists of 16 radio-quiet AGN
that are basically all the potential SEAMBH candidates listed in the \cite{DuPu2015} sample. The BLR sizes of 13
of the sources were measured in a two-year RM-campaign on the Lijiang 2.4m telescope by the ``SEAMBH Collaboration''.
The objects were selected by their spectroscopic properties. They are all NLS1s with suspected high
Eddington ratio, $\dot{m}$. The selection assumes a conservative method to estimate the
normalized accretion rate $\dot{\mathcal{M}}\geq3$.
The value was chosen to make sure that $\dot{m}$ for the lowest efficiency discs (those with spin -1 and $\eta=0.038$) exceeds 0.1
\citep{DuPu2014,Wang2014b,DuPu2015}. Three additional SEAMBHs candidates, PG~2130+099, PG~0844+349 and Mrk~110, with similar
properties, were selected from the general catalogue of $\sim 50$ RM-AGN. As of early 2015, these are the best candidates
for being super-Eddington accretors. We coin this sample the ``super-Eddington group''.

\begin{figure*}
    \centering
    \includegraphics[width=0.33\linewidth]{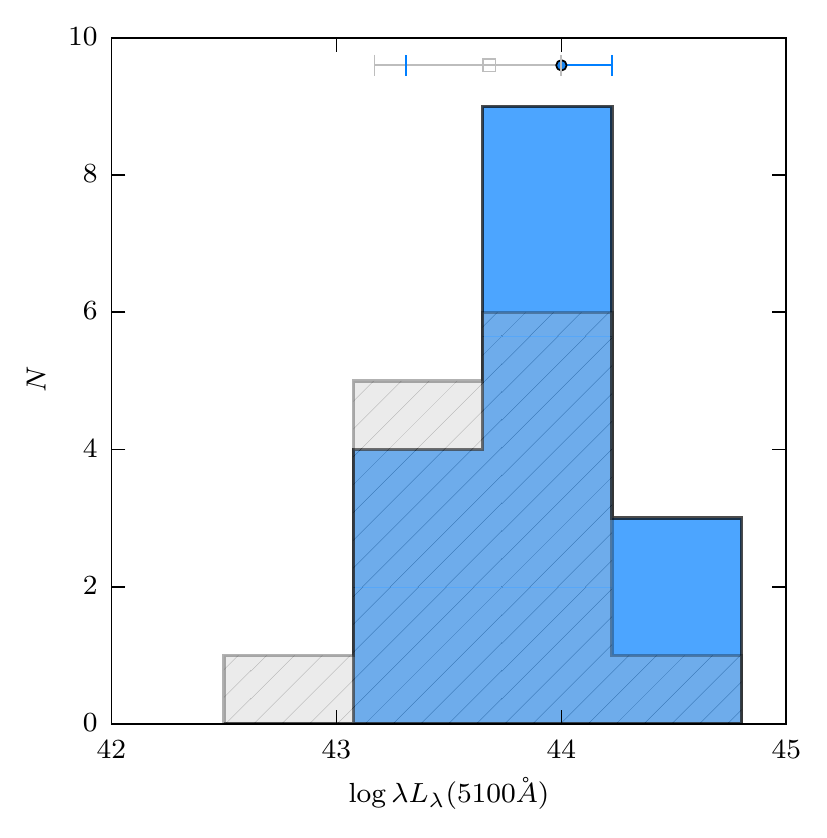}
    \includegraphics[width=0.33\linewidth]{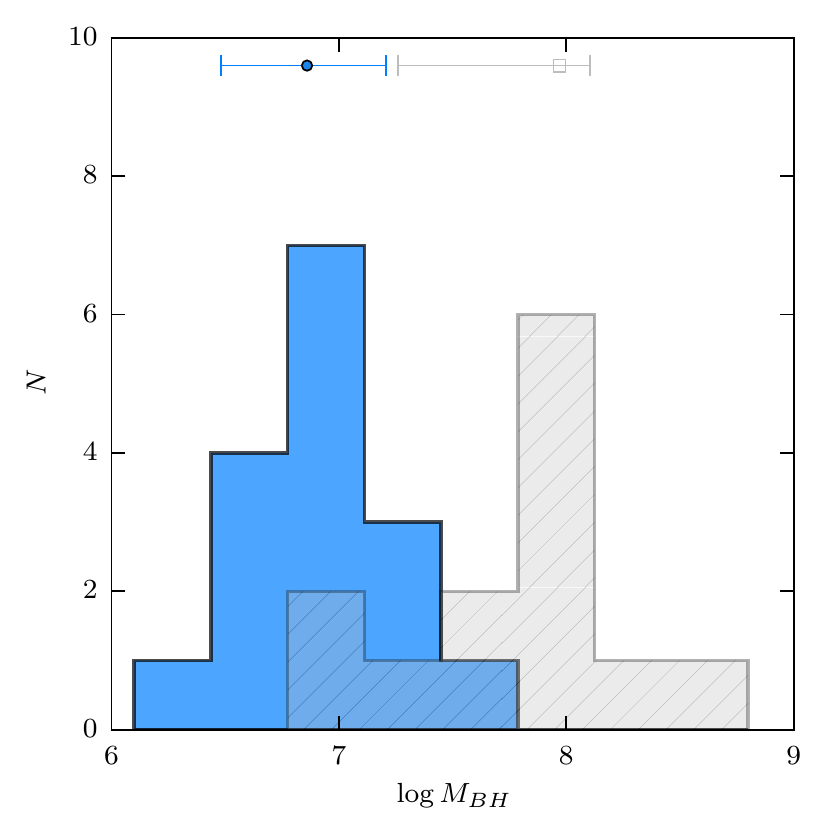}
    \includegraphics[width=0.33\linewidth]{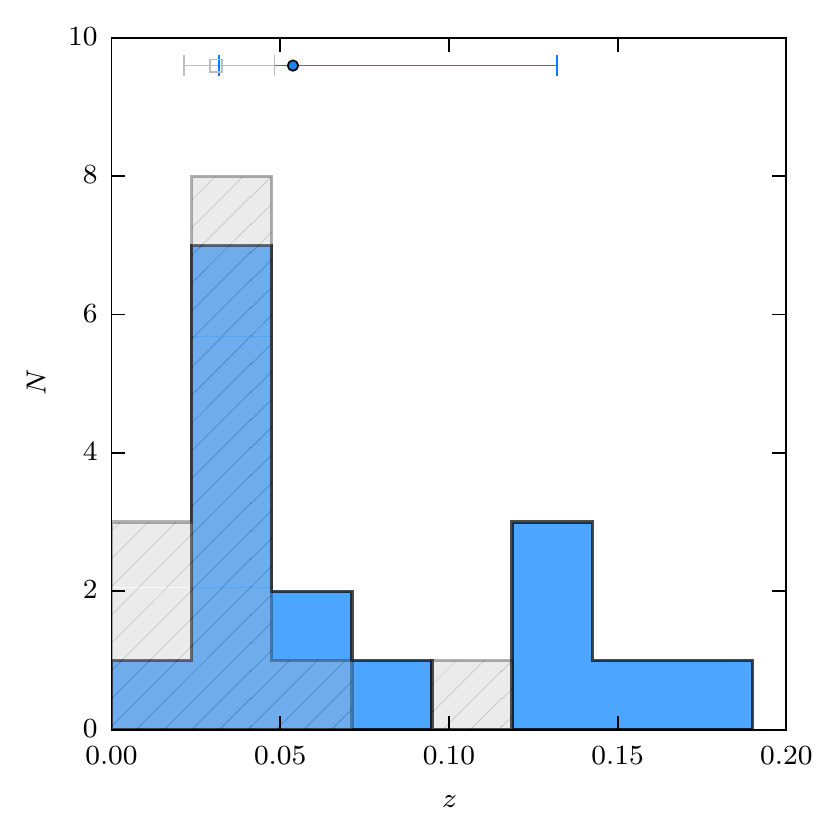}
    \caption{\emph{From left to right}: Distributions of $\log \lambda L_{\lambda}(5100\mAA{})$,
    virial black hole mass, and redshift for both samples. Filled (blue) histogram represents super-Eddington
    AGN and hatched (grey) histogram sub-Eddington AGN. The point with error bars on the top of each panel is
    the median and the 16$^{th}$ and 84$^{th}$ percentiles for each group.}\label{figure.L5100A}
\end{figure*}

We have also selected a control sample of 13 type-I AGN from the general group of $\sim$50 RM objects with a much
smaller Eddington ratio (defined in the same way) and $\dot{\mathcal{M}}<3$. This group, named as super-Eddington,
was selected to overlap in optical luminosity, defined as $\lambda L_{\lambda}$ at the rest-frame wavelength 5100\AA,
the luminosity of the super-Eddington group. The luminosity distributions of the two groups
are compared in Figure~\ref{figure.L5100A}. Although the Kolmogorov-Smirnov test can
not reject the hypothesis that both luminosity distributions are drawn from the same distribution,
there is a factor two on its average rest-frame 5100\AA{} luminosity which can not be improved due to the
limited RM-AGN sample ($10^{44}$ erg/s and 4.8$\times10^{44}$ erg/s for super-Eddington and
sub-Eddington groups, respectively). The redshift distribution of the sub-Eddington group is
quite similar to the range covered by our super-Eddington objects. The median redshifts are
0.054 and 0.031 for the super-Eddington and the sub-Eddington samples, respectively.
Figure~\ref{figure.L5100A} also shows the distribution of the virial BH mass in the two
samples. Due to the selection method which is based on source luminosity, the typical BH mass in
the control sample is much larger.

\subsection{Photometric and Spectroscopic Data}
\label{ObservationalData}

We have collected  optical-UV photometric and spectroscopic data from various surveys and data bases. The data are used
to define and separate stellar from non-stellar sources of continuum radiation and to constrain the properties of the SEDs
of all sources. The optical spectrum of 17 RM objects were taken from the Sloan Digital Sky Survey (SDSS) DR10.
For the remaining 12 sources, we used a mean optical spectrum from previous reverberation mapping campaigns. From
the AGN Watch database\footnote{The AGN Watch is a consortium of astronomers who have studied the inner structure of
AGN through continuum and emission-line variability. See http://www.astronomy.ohio-state.edu/$\sim$agnwatch/} we selected
the mean spectrum of Mrk~279, Mrk~509, NGC~7469 and Fairall 9. The mean optical spectrum of Mrk~79 and  Mrk~817 were
given by Peterson M.B. (private communication, \citealt{Peterson1998}). The mean optical spectrum of the Palomar Green (PG)
quasars (PG~2130+099, PG~0844+349, and PG~1617+175) were taken from \citet{Kaspi2000}. Finally, the mean optical spectrum of Mrk~486,
Mrk~335 and Mrk~1044 are those published by the
``SEAMBH Collaboration'' \citep{DuPu2014}. All objects in our sample, except for Mrk~79 and PG~1617+175, have high quality
optical spectra with broad optical wavelength range.

All objects in this work, except for Mrk~486, PG~0844+349 and SDSS~J080101.14+184840.7, have NUV and FUV
photometry (2306\AA{} and 1551\AA{},
respectively) obtained by the NASA Galaxy Evolution Explorer (GALEX). The GALEX fluxes were collected from the online
database and all image were inspected visually. We found  23 objects with observations in the all-sky imaging survey (AIS).
In addition, three objects (Mrk~1044, IRAS~04416+1215, Mrk~590) are also included in the medium imaging survey (MIS). No discrepancy
was found between both observations (a difference between AIS and MIS photometry less than 0.15 magnitudes in both GALEX bands).
Three out of these 23 objects have been detected in different surveys: PG~2130+099 was found in the MIS, while NGC~5548
and PG~1229+204 were selected from the Guest Investigator (GI) program. For 17 objects we also assembled optical/UV
photometry from the {\it XMM-Newton} Optical Monitor (OM): V~5430\AA{}, B~4500\AA{}, U~3440\AA{}, UVW1~2910\AA{},
UVM2~2310\AA{} and UVW2~2120\AA{}. From visually inspection of the available Hubble Space Telescope (HST) data
we cannot extend our dataset into the extreme UV because of variability, i.e. there is at least a factor 2 on flux between
the HST spectrum and the selected optical/UV dataset.

Summarizing, of our 29 sources we have optical spectra in conjunction with GALEX photometry for 26 sources and 16 sources
which, in addition to these data, also have XMM-OM photometry, and in 8 cases simultaneous optical/UV photometry.
A detailed list of all the observations is given in Table~\ref{table.Observations}.

\begin{table*}
  {\renewcommand{\arraystretch}{1.19}
  \caption{Optical and UV observations used to model the SED of super-Eddington (upper block) and
  sub-Eddington AGN (lower block).}\label{table.Observations}
  \hspace*{-0.7cm}
  \resizebox{1.05\linewidth}{!}{
\begin{tabular}{lccccccc}\toprule
  & & \multicolumn{2}{c}{Optical Spectrum} & \multicolumn{2}{c}{GALEX} & \multicolumn{2}{c}{{\it XMM-Newton} OM} \\\cmidrule(r{.75em}l){3-4}\cmidrule(r{.75em}l){5-6}\cmidrule(r{0.75em}l){7-8}
    Object & $E(B-V)_{\text{Gal}}$ & survey & date & survey & date & XMMOM & date \\ \midrule
    Mrk~142  &  0.016  &  SDSS DR10 & 2003-03-09   & AIS  &  2004-01-22 \\
    Mrk~335  &  0.035  &  SEAMBH Collaboration & 2012 Oct -- 2013 Feb & AIS  &  2007-04-02 & J000619.5+201211 & 2007-07-10 \\
    Mrk~382  &  0.048  &  SDSS DR10 & 2001-10-19 & AIS  &  2007-01-21 & J075525.3+391110 & 2011-11-02 \\
    Mrk~486  &  0.015  &  SEAMBH Collaboration & 2013 Mar -- 2013 Jul & &  \\
    Mrk~493  &  0.025  &  SDSS DR10 & 2004-05-16 & AIS  &  2007-05-26 & &  \\
    Mrk~1044 &  0.033  &  SEAMBH Collaboration & 2012 Oct -- 2013 Feb & MIS & 2008-10-11 \\
    IRAS~04416+1215  & 0.436  &  SDSS DR10 & 2006-12-17 & MIS  &  2008-12-17 & &  \\
    IRAS~F12397+3333 & 0.019  &  SDSS DR10 & 2005-03-02 & AIS  &  2004-04-15 & J124210.6+331702 & 2005-06-20 \\
    SDSS~J075101.42+291419.1 & 0.042  &  SDSS DR10 & 2002-12-28 & AIS  &  2007-01-18 & &  \\
    SDSS~J080101.41+184840.7 & 0.032  &  SDSS DR10 & 2004-11-10 & &   & &  \\
    SDSS~J081441.91+212918.5 & 0.039  &  SDSS DR10 & 2004-11-18 & AIS  &  2006-01-27 & &  \\
    SDSS~J081456.10+532533.5 & 0.032  &  SDSS DR10 & 2004-10-19 & AIS  &  2004-01-13 & &  \\
    SDSS~J093922.89+370943.9 & 0.014  &  SDSS DR10 & 2003-12-23 & AIS  &  2005-08-12 & J093922.9+370945 & 2006-11-01 \\
    PG2130+099  &  0.044  &  \citet{Kaspi2000} & 1991 Aug -- 1997 Oct & MIS  &  2009-08-22 & J213227.8+100819 & 2003-05-16 \\
    PG0844+349  &  0.037  &  \citet{Kaspi2000} & 1991 Aug -- 1997 Oct & &  & J084742.4+344504 & 2009-05-03 \\
    Mrk110  &  0.012  &  SDSS DR10 & 2001-12-09 & AIS  &  2007-01-23 & J092512.9+521711 & 2004-11-15  \\
    \addlinespace\\
    Mrk~79   &  0.071  &  \citet{Peterson1998} &  1983 -- 1985  & AIS  &  2007-01-20 & J074232.8+494835 & 2008-04-26 \\
    Mrk~279  &  0.016  &  AGN Watch & 1996 Dec -- 1997 Jan & AIS  &  2004-01-25 & J135303.5+691830 & 2005-11-19 \\
    Mrk~290  &  0.014  &  SDSS DR10 & 2002-03-14 & AIS  & 2005-10-22  & J153552.3+575409 & 2006-05-04 \\
    Mrk~509  &  0.057  &  AGN Watch & 1988 Sep -- 1993 Dec & AIS  &  2007-02-09 & J204409.7-104324 & 2009-11-14 \\
    Mrk~590  &  0.037  &  SDSS DR10 & 2003-01-08 & MIS  &  2008-10-23 & & \\
    Mrk~817  &  0.007  &  \citet{Peterson1998} & 1983 -- 1985 & AIS &  2007-02-21 & & \\
    Mrk~1511 &  0.041  &  SDSS DR10 & 2007-04-17 & AIS  &  2007-04-17 & J153118.1+072729 & 2012-02-23 \\
    NGC~5548 &  0.020  &  SDSS DR10 & 2006-05-04 & GII  &  2006-10-30 & J141759.5+250813 & 2013-07-29 \\
    NGC~7469 &  0.068  &  AGN Watch & 1996 Jun -- 1996 Jul &  AIS  &  2007-03-22 & J230315.6+085226 & 2004-11-30 \\
    PG~1229+204 &  0.027  & SDSS DR10 & 2008-01-16 & GII  &  2007-05-22 & J123203.7+200929 & 2005-07-09 \\
    PG~1617+175  &  0.042  &  \citet{Kaspi2000} & 1991 Aug -- 1997 Oct & AIS  &  2006-06-14 & & \\
    Fairall9 &  0.025  &  AGN Watch & 1994 May -- 1995 Jan & AIS  &  2007-08-17 & J012345.7-584820 & 2013-12-19 \\
    MCG+06-26-012 &  0.019  &  SDSS DR10 & 2005-03-31 & AIS  &  2007-03-22 & & \\
\bottomrule
\end{tabular}
}}
\end{table*}

\section{Observed and Intrinsic SEDs}
\label{intrinsicSEDs}

A central goal of the present work is to compare the multi-wavelength SEDs of super-Eddington
and sub-Eddington AGN. For this we need to correct for non-simultaneous observations,
to subtract the stellar and emission line contributions, and to consider the possibility of
intrinsic reddening of the sources. As for  Galactic interstellar reddening, this is done
assuming the \citet{Cardelli1989} extinction law using the Galactic extinction colour excess
\ebmv{} obtained from the NASA/IPAC Infrared Science Archive\footnote{http://irsa.ipac.caltech.edu/applications/DUST/}.

\subsection{Long term optical-UV variability}
\label{OC.Variability}

The presence of variability in the optical/UV continuum of all the sources in our sample is well documented.
Whilst we can expect variability of order  a few per cent in the optical-UV continuum on short-time of days to a week,
much larger factors are expected on time scales of months and years. For example, \citet{SantosLleo1997}
found a factor of $\sim$2 for Fairall 9 in the optical continuum flux over a three month period and even larger factors
have been found for NGC~5548 \citep[a factor of $\sim$7 over a seven years period, ][]{Peterson1999}. Therefore
variability is the main challenge for the work presented here.

The information that is available for most sources is a combination of optical spectroscopy and GALEX photometry. Most
of these data are not simultaneous. In order to avoid as much as possible the non-simultaneity, and to complete the UV
GALEX information, we used data obtained by the {\it XMM-Newton} Optical Monitor (OM). OM photometry
was available for 17 out of 29 objects and contemporaneous optical/UV observations (at least 5 simultaneous OM points) for
only 6 sources (see Table~\ref{table.Observations}). These contemporaneous optical/UV observations provide reliable
constraints on the source SEDs and help to reduce the fitting uncertainty (see Appendix~\ref{IndividualNotes}).

The heterogeneity of our optical/UV datasets dictate a somewhat different procedure for each of the fitted sources.
Three different procedures were used: {\sc i)} a simultaneous and non-simultaneous SED were fitted in order to quantify
the non-contemporaneous SED shape; {\sc ii)} when simultaneous dataset was not possible, both GALEX and OM photometry
in conjunction with the optical spectrum were fitted as long as there is no signs of variability, i.e. all segments of the
fitted continuum join smoothly; {\sc iii)} for those sources without {\it XMM-Newton} OM observations we fitted
the non-simultaneous optical-GALEX SED assuming  they were observed at a similar flux epoch. Comments on individual sources
are given in Appendix~\ref{IndividualNotes}. This general procedure works well in all the objects discussed in the present
paper although non-simultaneity is still the main limiting factor.

\subsection{Host Galaxy Contribution}
\label{OC.HG}
The optical/UV emission may be contaminated by starlight from the host galaxy. The relative contribution depends on
the aperture size, AGN luminosity, and the stellar population in the host. Thus careful galaxy subtraction
is necessary if we are to determine the SED shape. Such corrections are often made by subtracting galaxy
bulge template spectra. Star formation in the central part must also be considered since the host of many AGN
are star formation galaxies \citep[e.g. ][]{Sani2010}.

There are three methods that can be used to correct the observed spectra for host-galaxy contamination. The first
is a direct subtraction of the stellar light measured from HST images. Such data are available for 25 of the sources
\citep[][]{Peterson1998,Bentz2009,DuPu2014,Wang2014b}. Published galaxy light profiles were used
in these cases and normalized to the aperture in question. This correction is normally small (20-70\%) for the
3\arcsec{} SDSS fiber and considerably larger for the 6\arcsec{} aperture of the optical OM filter. The second method
makes use of empirical expressions derived by \citet{Shen2011} and \citet{Elvis2012}. The uncertainty here is much
larger because of the considerable range in host properties. Finally, one can use the fact that the broad \hbeta{} lines
show no Baldwin effect \citep{Dietrich2002}. This means that the observed equivalent width of the line, EW(\hbeta{}),
can be used to derive the fraction of the non-AGN light entering the aperture at 4861\AA{}. \citet{DuPu2015}
show the distributions of EW(\hbeta{}) in low Eddington ratio AGN and in SEAMBHs. The two differ by a considerable amount
with medians that are $(122\pm44)\mAA$ and $(89\pm31)\mAA$, respectively.
For those sources without HST observations,
we prefer the use of this method over the expressions given by \citet{Shen2011}. The main concern is how to treat
the five objects without HST observations.

The subtraction procedure starts with an estimate of the non-AGN flux at 4861\AA{} followed by a subtraction of a single
simple stellar model spectrum which is scaled to this flux. A nominal 20\% uncertainty on this flux was adopted due to a combination
of the host-galaxy modelling and the uncertainties on the emission line fluxes. The SDSS fibers include only the innermost few
kpc of the host, hence we chose a quiescent galaxy model from the evolutionary spectral library of \citet{Charlot1991}. For 15
out of 29 objects, the instantaneous-burst model with an age of 11 Gyr and solar metallicity (Z=0.02) provided sufficiently good fit
to the stellar spectrum. For the remaining 14 objects, this template gives a ``flux excess'' at the longest observed wavelengths.
A template with 11 Gyr and Z=0.05 provides better fit to such spectra and was adopted in these cases. We also experimented with
adding younger stellar population components but did not find significant improvements over the old population templates.

The difference on aperture between the SDSS fibres (3\arcsec) and the OM broadband filters (12\arcsec{} and 35\arcsec{}
diameter for the optical and UV filter, respectively) prevent us from using the estimated host-galaxy contribution to the optical
spectrum to subtract the star-light from the OM photometric data. For the OM photometric dataset, the host galaxy emission was
modeled by adopting the published galaxy-light radial profiles \citep[][]{Bentz2009} which were integrated over the specific OM
aperture to estimate the host-galaxy contribution to the total observed flux.

\subsection{Emission line Contributions}
\label{OC.PhotometricCorrections}
The wavelength range for the photometry are wide enough that the underlying nuclear continuum might be contaminated
by broad and narrow emission lines, i.e. Balmer lines, C$\,${\sc iv}, Mg$\,${\sc ii}, Fe$\,${\sc ii},
as well as the Balmer continuum.

In order to exclude the underlying continuum, we used the composite quasar spectrum of \citet{VandenBerk2001}
to estimate the emission line fraction for each photometric waveband. We assumed only the emitted flux in the spectral
window over which the effective transmission is greater than 10\% of the peak effective transmission: FUV~1343--1786\AA{},
NUV~1771--2831\AA{}, UVW2~1805--2454\AA{}, UVM2~1970--2675\AA{}, UVW1~2410--3565\AA{}, U~3030--3890\AA{},
B~3815--4910\AA{} and V~5020--5870\AA{}. For comparison, at redshift $z=0.06$, which is the mean of our
sample, the contributions from the emission line regions at each photometric window are 5\%, 11.9\%, 9.4\%, 13.0\%, 23.4\%, 24.6\%,
15.9\% and 24.9\% for FUV, NUV, UVW2, UVM2, UVW1, U, B  and V, respectively. Since line variations follow continuum
variations (albeit with somewhat different lags), we do not expect large fluctuations in these fractions.

\subsection{Intrinsic reddening}
\label{OC.IR}

Intrinsic extinction in AGN can be significant which can make it an important factor when determining the SED of the
optical/UV continuum. Previous studies \citep[][Mejia-Restrepo et al. 2015 (submitted), and references
therein]{Lusso2013,Capellupo2015,Collinson2015} found a range of properties with most AGN showing little
dust attenuation, i.e. \ebmv<0.1, independent of the reddening law. Such reddening does not affect much the optical fluxes, but its
effect on the UV part of the spectrum is significantly larger. Not accounting for this effect will result in fitting the wrong accretion
disc models and the miss-calculation of source luminosity and accretion rate.

Extinction curves which are commonly used in AGN studies include: Milky Way-like extinction with its known broad
bump at $\sim$2175\AA{}, Small Magellanic Cloud-type (SMC) curves, a simple power-law extinction, or a combination of a
power law with a flatter curve in the far-UV \citep{Gaskell2004}. Several earlier
works \citep{Hopkins2004,Glikman2012} claimed that the typical bump at 2175\AA{} of the MW-like extinction curve is not
observed in AGN spectra. Recent, higher quality spectra and more carefully fitted SEDs by \citet{Capellupo2015} clearly show
this bump in some of the spectra.

Since the wavelength dependence of the extinction is unknown, a priori, we experimented with three possibilities: {\sc i)}
the \citet{Cardelli1989} as a MW-like curve with $R_V=3.1$, {\sc ii)} a simple power-law $A(\lambda)=A_{0} \lambda^{-1}$,
and {\sc iii)} the SMC extinction curve as given in \citet{Gordon2003}. According to the wavelength dependence of the extinction
curve, the intrinsic reddening will be larger for the SMC curve than for the MW. The most consistent approach is to add this extinction
as an additional parameter in the SED modelling analysis. Clearly, if extinction is important, the intrinsic underlying continuum
shape will depend significantly on the form of the reddening curve.
As a practical point we note that the extremely broad GALEX
bands do not allow us to take into account spectral features like the 2175\AA{} absorption and weaker ISM lines and we only treat
the total observed flux in these bands.

\section{SED model fitting}
\label{SEDmodelling}

\subsection{Accretion disc models}

In this work, we use the numerical code described in \citet{Slone2012} to calculate thin AD spectra. The calculations assume a
\citet{Shakura1973} disc with a variable viscosity parameter (chosen in this paper to be $\alpha = 0.1$). As in all models of this type
(see Sec.~\ref{Introduction}), the spin-dependent ISCO determines the mass-to-energy conversion efficiency, $\eta$.
The focus of the \citet{Slone2012} work is the effect of disc winds on the emitted SEDs. Here we do not consider
disc winds because of the lack of far UV spectroscopy required to deduce their presence and because we focus on
slim disc where there are additional large uncertainties (see below). As in \citet{Slone2012}, our calculations
include Comptonization of the emitted radiation at every point in the disc atmosphere and,
for BH spin values of $a_{\star}>0$, full General
Relativistic corrections. For retrograde discs with $a_{\star}<0$, the general relativity effects are not
included, which is a fair approximation given the large size of the ISCO ($>6r_{g}$, where $r_{g}=GM_{BH}/c^2$ is the gravitational
radius of the black hole).
Apart from Comptonization, the calculations do not include any other radiative transfer
in the disc atmosphere that can make significant changes to the far UV SED \citep[e.g.][Fig. 2]{Davis2011}.

Standard, \citet{Shakura1973} thin accretion disc models are limited to $\dot{m} \le 0.3$. Beyond this accretion rate, the disc geometry
becomes thick and many of the approximation used in the  model no-longer hold \citep[see][and references therein]{Laor1989}.
As explained in Section~\ref{Introduction}, the main differences between thin and slim discs are the thicker geometry,
due to the much larger radiation pressure in the disc, and the way the radiation escapes the system. This includes radial advection
and perhaps saturation of the emitted luminosity. Unfortunately, detailed SED calculations of slim discs are highly simplified and
hardly available. A recent detailed model of this type is discussed in \citet{Wang2014b}. According to this and earlier models, the
long wavelength part of the SED originates outside of the thick part of the disc and is, therefore, very similar in its shape to the
thin AD SED. For small BH mass systems, significant differences between the two appear only at very short wavelengths, beyond the
Lyman limit. The special geometry dictates strong anisotropy in such systems, much beyond the standard AD anisotropy due to inclination.
This, again, is most noticeable at short wavelengths \citep[see Fig. 4 in][]{Wang2014a}.
The observations discussed here do not include the wavelength range below $\lambda$=1000\AA{} and hence we use the thin
AD model for the fitting of the SED over this range. Later on, when we discuss the Lyman continuum emission, we
consider the various possibilities regarding slim discs.

As explained earlier, theoretical slim disc models are still highly simplistic and are not in very good agreement with state-of-the-art
numerical calculations like those of \citet{Sadowski2015}. Given this fundamental uncertainty, we treat all derived quantities that
depend on the short wavelength part of the model, e.g. the bolometric luminosity and Eddington ratio, as the most uncertain parameters
for the sub-sample of 16 SEAMBHs described in this work.

\subsection{Fitting procedure}

We used the \citet{Slone2012} code described earlier to calculate a large number of thin disc spectra that include the entire range
of BH mass, accretion rate and spin expected in our sample. The input for the fit include, for each source,  the black hole mass
(\mbh{} in units of $M_{\odot}$), and the mass accretion rate (\Mdot{} in units of $M_{\odot}$/yr). We used the RM-based masses
and the measured (from the 5100\AA{} continuum) accretion rates listed in \citet[see Table 7]{DuPu2015}. The uncertainty on the mass
is estimated to be a factor of $\sim$3 due to the the uncertainties on the measured time lags, the measured FWHM(H$\beta$), and the
uncertainties in the conversion of observed broad line profiles to a ``mean gas velocity'' (the $f_{_{BLR}}$ term in
$M_{BH}=f_{_{BLR}}\,c\,\tau\,FWHM(H\beta)^{2}/G$).
The accretion rate is obtained from the luminosity at rest-wavelength 5100\AA{} using the method described in \citet{Davis2011} as
detailed in \citet[][see Eqn. 1 there]{Netzer2014}. We assumed a factor three uncertainty on the accretion rate derived in this way which
is a combination of flux uncertainty (mostly stellar light subtraction), the unknown inclination, and the fact that the chosen wavelength
(5100\AA{}) is not on the $L_{\nu} \propto \nu^{1/3}$ part of the SED.

All sources fitted in this work are type-I AGN which are assumed to be observed close to face-on. The range of inclination
is roughly 0--60 degree and the range in $\cos{(\theta)}$ 0.5--1. We assumed the anisotropy function proposed by
\citet{Netzer2014},
\begin{equation}
    f(\theta) = \frac{f_0F_{\nu}}{F_\nu{}\text{(face-on)}} = f_{0}\;\frac{\cos{\theta}\left(1+b(\nu)\cos{\theta}\right)}{1+b(\nu)}
\end{equation}
with $b(\nu)=2$, $f_0 = 1.2\times10^{30}$~erg/s/Hz, and $\cos{\theta}=0.75$ for all sources. The remaining parameters are the BH spin
and intrinsic reddening. We experimented with the full range of spin parameters, from -1 to 0.998. Because of the fitted range of wavelengths,
which is far from the frequency of maximum disc temperature, there is very little difference between different spin value. We therefore show, for
each source only two values, $a_{\star}=-1$ and $a_{\star}=0.998$.

The fitting procedure includes the comparison of the observed SED with various combinations of disc SEDs covering the range of mass,
accretion rate, and the two chosen spins and assumed reddening. The reddening is taken into account by changing $E(B-V)$ in steps
of 0.004 mag, calculating, for each value, a new mass accretion rate and its range of uncertainty. A simple $\chi^{2}$ procedure was used
to find the best-fit combination of reddening and thin AD models. We use at least three line-free windows covering the optical
spectroscopic data, and all the available photometric data. The line-free windows are centred on 4205\AA{}, 5100\AA{} and 6855\AA{},
with widths
ranging from 10 to 30\AA{}. For those objects with no photometric data, three additional line-free continuum windows were used: 5620\AA{}, 6205\AA{}
and 6860\AA{}. For the error on each continuum point, we combine the standard error from the Poison noise, an assumed 5\% error on the flux calibration,
and the relative error of 20\%  on the combination of the uncertainties on the host-galaxy contribution and the unknown stellar population.
To allow for the large elapsed time between the GALEX and the optical observations, which increases the uncertainty due to source variability,
we added an uncertainty of 20\% to the GALEX fluxes. Note that this is not meant to take into account the real variations between epoch since this is
dealt with, in our special method described in Section~\ref{OC.Variability} where we provide more information about the way we used
the best SED that avoids, as much as possible, the variability issue. We refer the reader to Appendix~\ref{IndividualNotes} for a detailed
description of the fitting results for individual sources.

\section{Results}
\label{Results}

The samples described here were selected from the RM AGN sample with high accretion rate,
$\mathcal{\dot{M}}\geq3$ (super-Eddington) and with sub-Eddington accretion rate
$\mathcal{\dot{M}}<3$. In the following we use the normalized Eddington ratio,
$\dot{m}=\eta\dot{\mathcal{M}}$, which allow an easy comparison with previous works.
While not all data are of the same quality, we were able to secure photometric and spectroscopic
data that cover the 0.2-20$\,\mu$m range for all sources. For 8 sources
(4 sub-Eddington and 4 super-Eddington) we have both simultaneous and non-simultaneous
optical/UV photometry. In the following discussions we do not distinguish between simultaneous
and non-simultaneous SEDs.

Our best fitted disc SEDs are shown in Figure~\ref{figure.bfSED} and all the model parameters,
including the virialized \mbh{} and \Mdot{}, are listed in Table~\ref{table.bfSED}.
Two models are listed per source, corresponding to the minimum and maximum
spin parameter. The error bars on the intrinsic reddening enclose the 68\% confidence range.
Finally, the average spectral properties for both groups and for the entire sample are summarized
in Table~\ref{table.statisitcs}. In the following plots, for every derived thin AD model
parameter we display the mean between the higher and lower values corresponding to the minimum
and maximum spin, and the uncertainly which represents the range of possible values.

\begin{table*}
{\renewcommand{\arraystretch}{1.3}
\caption{Median Parameters for super-Eddington (13), sub-Eddington (16) AGN and the entire sample (29).
 The uncertainties reflect the 16$^{th}$ and 84$^{th}$ percentiles.
 When saturation (Eqn.~\ref{eq.saturation}) is taken into account the parameter is tagged with the term $_{sat}$.
 Two numbers are listed for each thin AD model corresponding to the minimum and the
 maximum spin. The unabsorbed luminosity at 2$\,$keV is given by the best-fit
 power law over the hard X-ray band (2-10$\,$keV).\label{table.statisitcs}}
\begin{tabular}{lrrrrrr}\toprule
        {\sc parameter} & \multicolumn{2}{c}{super-Eddington} &  \multicolumn{2}{c}{sub-Eddington} & \multicolumn{2}{c}{all} \\
        & $a_{\star}=0.998$ & $a_{\star}=-1$ & $a_{\star}=0.998$ & $a_{\star}=-1$ & $a_{\star}=0.998$ & $a_{\star}=-1$ \\
    \midrule
    \multicolumn{7}{l}{reverberation-mapped resutls}\\
    $\log \lambda L_{\lambda}(5100\mAA{})$  & $44.00^{+0.23}_{-0.69}$  & $44.00^{+0.23}_{-0.69}$  & $43.68^{+0.32}_{-0.51}$  & $43.68^{+0.32}_{-0.51}$  & $43.71^{+0.50}_{-0.55}$  & $43.71^{+0.50}_{-0.55}$ \\
    $z$  & $0.054^{+0.078}_{-0.022}$ & $0.054^{+0.078}_{-0.022}$ & $0.031^{+0.017}_{-0.009}$ & $0.031^{+0.017}_{-0.009}$ & $0.035^{+0.081}_{-0.009}$ & $0.035^{+0.081}_{-0.009}$\\
    $\log M_{BH}$  & $6.86^{+0.35}_{-0.38}$  & $6.86^{+0.35}_{-0.38}$  & $7.97^{+0.13}_{-0.71}$  & $7.97^{+0.13}_{-0.71}$  & $7.16^{+0.85}_{-0.51}$  & $7.16^{+0.85}_{-0.51}$ \\
    $\log\dot{\mathcal{M}}$  & $1.61^{+0.98}_{-0.77}$  & $1.60^{+0.98}_{-0.77}$  & $-0.70^{+0.35}_{-0.38}$  & $-0.72^{+0.37}_{-0.36}$  & $0.84^{+1.24}_{-1.68}$  & $0.83^{+1.23}_{-1.69}$ \\
    \addlinespace[0.7em]
    \multicolumn{7}{l}{power-law modelling}\\
    $\beta_{UV}$  & $2.13^{+0.37}_{-0.36}$  & $2.13^{+0.37}_{-0.36}$  & $2.12^{+0.36}_{-0.30}$  & $2.12^{+0.36}_{-0.30}$  & $2.12^{+0.39}_{-0.34}$  & $2.12^{+0.39}_{-0.34}$ \\
    $E(B-V)$ /PL/  & $0.00^{+0.06}_{-0.00}$  & $0.00^{+0.06}_{-0.00}$  & $0.00^{+0.00}_{-0.00}$  & $0.00^{+0.00}_{-0.00}$  & $0.00^{+0.00}_{-0.00}$  & $0.00^{+0.00}_{-0.00}$ \\
    \addlinespace[0.7em]
    \multicolumn{7}{l}{thin AD modelling}\\
    $E(B-V)$  & $0.07^{+0.06}_{-0.03}$  & $0.06^{+0.06}_{-0.02}$  & $0.01^{+0.03}_{-0.01}$  & $0.00^{+0.00}_{-0.00}$  & $0.05^{+0.04}_{-0.05}$  & $0.04^{+0.06}_{-0.04}$ \\
    $\log M_{BH}$  & $7.25^{+0.28}_{-0.46}$  & $7.26^{+0.26}_{-0.46}$  & $7.88^{+0.13}_{-0.60}$  & $7.80^{+0.44}_{-0.52}$  & $7.32^{+0.65}_{-0.52}$  & $7.36^{+0.59}_{-0.56}$ \\
    $\log\dot{\mathcal{M}}$  & $1.81^{+0.46}_{-1.43}$  & $1.80^{+0.58}_{-1.22}$  & $-0.62^{+0.98}_{-0.44}$  & $-0.32^{+0.66}_{-0.58}$  & $0.36^{+1.72}_{-1.07}$  & $0.50^{+1.62}_{-1.03}$ \\
    $\log L_{\text{Lyman}}$  & $46.55^{+0.46}_{-0.92}$  & $45.57^{+0.62}_{-0.96}$  & $44.96^{+0.45}_{-0.50}$  & $43.70^{+0.32}_{-0.96}$  & $45.59^{+1.29}_{-0.81}$  & $44.48^{+1.40}_{-1.11}$ \\
    $\log L_{\text{Lyman,sat}}$  & $45.84^{+0.19}_{-0.39}$  & $45.78^{+0.29}_{-1.17}$  & $44.96^{+0.45}_{-0.50}$  & $43.70^{+0.32}_{-0.96}$  & $45.52^{+0.37}_{-0.74}$  & $44.48^{+1.43}_{-1.11}$ \\
    $\langle h\nu \rangle$  & $11.8^{+2.7}_{-7.0}$  & $3.2^{+1.0}_{-1.5}$  & $2.4^{+2.7}_{-0.3}$  & $1.3^{+0.5}_{-0.1}$  & $5.3^{+8.5}_{-3.0}$  & $1.8^{+2.3}_{-0.5}$ \\
    $\kappa_{5100\mAA{}}$  & $208^{+475}_{-171}$  & $61^{+71}_{-53}$  & $17^{+45}_{-5}$  & $3^{+7}_{-1}$  & $62^{+373}_{-46}$  & $10^{+90}_{-7}$ \\
    $\kappa_{5100\mAA{},sat}$  & $53^{+51}_{-32}$  & $65^{+61}_{-57}$  & $18^{+46}_{-5}$  & $3^{+7}_{-1}$  & $32^{+51}_{-19}$  & $10^{+95}_{-7}$ \\
    $\log \lambda L_{\lambda}(5100\mAA{})$ (erg/s)  & $44.23^{+0.37}_{-0.52}$  & $43.93^{+0.44}_{-0.47}$  & $43.64^{+0.63}_{-0.33}$  & $43.63^{+0.64}_{-0.41}$  & $43.90^{+0.59}_{-0.51}$  & $43.83^{+0.51}_{-0.55}$ \\
    $\log \lambda L_{\lambda}(2500\mAA{})$  (erg/s)  & $44.51^{+0.24}_{-0.25}$  & $44.48^{+0.28}_{-0.25}$  & $43.95^{+0.24}_{-0.25}$  & $43.89^{+0.29}_{-0.27}$  & $44.15^{+0.46}_{-0.43}$  & $44.12^{+0.44}_{-0.49}$ \\
    $L_{AGN}$ (erg/s)  & $46.56^{+0.46}_{-0.90}$  & $45.65^{+0.57}_{-0.79}$  & $45.07^{+0.49}_{-0.44}$  & $44.28^{+0.35}_{-0.44}$  & $45.65^{+1.23}_{-0.78}$  & $44.84^{+1.10}_{-0.72}$ \\
    $L_{AGN,sat}$ (erg/s)  & $45.89^{+0.19}_{-0.42}$  & $45.82^{+0.29}_{-0.95}$  & $45.07^{+0.49}_{-0.44}$  & $44.28^{+0.35}_{-0.44}$  & $45.56^{+0.38}_{-0.69}$  & $44.84^{+1.11}_{-0.72}$ \\
    ${\Delta\log M_{BH}}^{\dagger}$  & $0.31^{+0.19}_{-0.24}$  & $0.31^{+0.19}_{-0.22}$  & $-0.11^{+0.37}_{-0.19}$  & $-0.04^{+0.29}_{-0.14}$  & $0.15^{+0.34}_{-0.36}$  & $0.19^{+0.28}_{-0.31}$ \\
    ${\Delta\log\dot{\mathcal{M}}}^{\ddagger}$  & $-0.24^{+0.58}_{-0.33}$  & $-0.13^{+0.47}_{-0.32}$  & $0.59^{+0.23}_{-1.06}$  & $0.54^{+0.15}_{-0.62}$  & $-0.01^{+0.68}_{-0.55}$  & $0.13^{+0.49}_{-0.46}$ \\
    \addlinespace[0.7em]
    \multicolumn{7}{l}{Luminosities}\\
    $L_{\text{trous}}$ (erg/s)  & $44.52^{+0.50}_{-0.36}$  & $44.52^{+0.50}_{-0.36}$  & $44.53^{+0.37}_{-0.47}$  & $44.53^{+0.37}_{-0.47}$  & $44.53^{+0.48}_{-0.44}$  & $44.53^{+0.48}_{-0.44}$ \\
    $\log L_{\nu}(5\mu m)$ (erg/s)  & $43.99^{+0.50}_{-0.38}$  & $43.99^{+0.50}_{-0.38}$  & $43.86^{+0.52}_{-0.39}$  & $43.86^{+0.52}_{-0.39}$  & $43.86^{+0.62}_{-0.35}$  & $43.86^{+0.62}_{-0.35}$ \\
    $\log L_{\nu}(2keV)$ (erg/s)  & $43.23^{+0.10}_{-0.40}$  & $43.23^{+0.10}_{-0.40}$  & $42.85^{+0.57}_{-0.40}$  & $42.85^{+0.57}_{-0.40}$  & $43.19^{+0.22}_{-0.64}$  & $43.19^{+0.22}_{-0.64}$ \\
\bottomrule
\addlinespace[0.8em]
\multicolumn{7}{l}{$^{\dagger}$ Defined as  $\Delta \log{M_{BH}}\equiv (\log{M_{BH}})^{\text{fit}}-(\log{M_{BH}})^{\text{virial}} $}\\
\multicolumn{7}{l}{$^{\ddagger}$ Defined as  $\Delta \log{\dot{\mathcal{M}}}\equiv (\log{\dot{\mathcal{M}}})^{\text{fit}}-(\log{\dot{\mathcal{M}}})^{\text{virial}}$}
\end{tabular}}
\end{table*}

\subsection{Accretion Disc SED}
\label{section.thinAD}
A major goal of this project is to determine what fraction of sub-Eddington and super-Eddington
accretors in our sample can be fit by the simple optically thick, geometrically thin AD model
based on their long wavelength SEDs. By allowing intrinsic reddening as a free parameter of the
thin AD model, we can fit all the 29 sources of our sample. For 23 sources, our modelling requires
some reddening and for 6 sources, which are all sub-Eddington AGN, the amount of reddening
is consistent with zero. We found that 14\% of the RM-selected AGN are consistent with
\ebmv$>0.1$ in good agreement with the work of \citet{Krawczyk2015} where the SED was modelled
by a single power law. Indeed, the reddening distribution of our entire sample is also consistent
with that presented by \citet[][Figure~\ref{figure.Av}]{Lusso2013}.
Several of the declared good fits still include small deviations of the model from the local
continuum at some wavelengths. This is not surprising given the uncertainties on AD models,
especially the radiative transfer in the disc atmosphere that was not treated here,
as well as on the choice of the host galaxy template. In general, the simple $\chi^2$-fitting
procedure cannot distinguish between $a_{\star}=-1$ and $a_{\star}=0.998$, given the assumed
uncertainties on BH mass, BH accretion rate, reddening and the long wavelengths used for the fitting.
Observations at shorter wavelengths (i.e. extreme-UV $\lambda<2000\mAA{}$) are clearly
required to make such a distinction.

\begin{figure}
    \centering
    \includegraphics[width=0.9\linewidth]{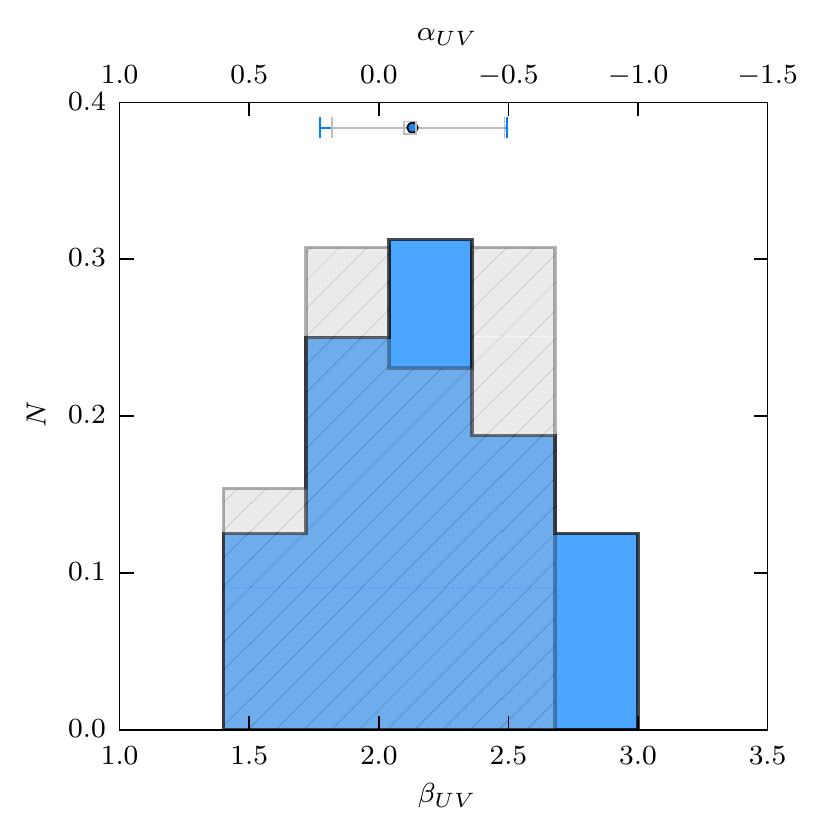}
    \caption{The distribution of optical to UV spectral indices in the
    super-Eddington and sub-Eddington objects. The spectral
	indices are defined such that $L_{\lambda}\propto \lambda^{-\beta_{UV}}$
	and $\beta_{UV}=2-\alpha_{UV}$. Colour code as Figure~\ref{figure.L5100A}.}
	\label{fig.PLresults}
\end{figure}

\begin{figure*}
    \centering
    \includegraphics[width=0.45\linewidth]{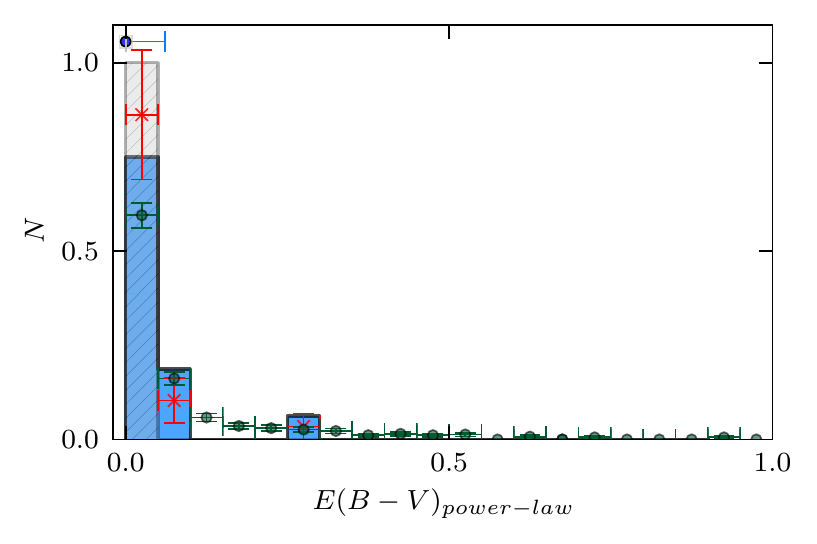}
    \includegraphics[width=0.45\linewidth]{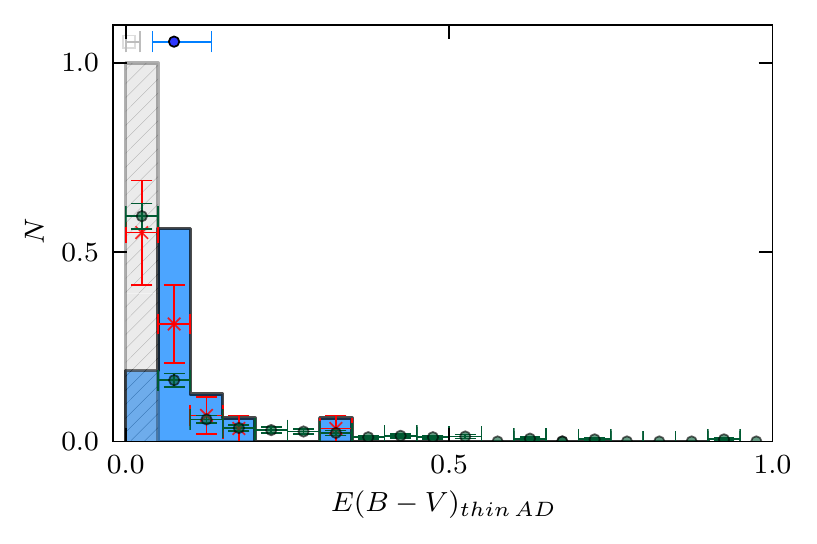}\\
    \caption{Intrinsic reddening E(B-V) distribution modelled by a classical SMC-like extinction curve with $R_{V}=2.74$. The
	amount of reddening was derived from model fittings assuming a single power law (\emph{left}) and a
	thin AD model (\emph{right}). The red crosses with error bars represent the reddening distribution for
	the entire sample (29 objects) and the green circles with error bars show the distribution of \ebmv{} presented by \citet{Lusso2013}.
	Histogram colour code as in Fig.~\ref{figure.L5100A}.}
    \label{figure.Av}
\end{figure*}

For consistency with previous works, the 0.2-1$\,\mu$m SED was also fitted using a single reddened
power-law model ($L_{\nu}\propto \nu^{-\alpha_{UV}}$).
A satisfactory fit ($\chi^2_{\nu}<2$) with $A_{V}=0$ is found for 86\% (25/29) of the sources.
The remaining four objects (which are super-Eddington) were successfully fitted with $A_{V}>0$.
The reddening distribution of the entire sample is quite consistent, as in the case of the thin
AD modelling, with that presented in various earlier studies, e.g. \citet{Lusso2013}. The range
of slopes is large, from $\alpha_{UV}=0.55$ to $\alpha_{UV}=-0.98$.
The Kolmogorov-Smirnov test shows that the optical to UV spectral index corrected for intrinsic
reddening for SEAMBHs and sub-Eddington
accretors are fully consistent (see Figure~\ref{fig.PLresults}), with means $\langle\alpha_{UV}\rangle=-0.14$ and $-0.10$ and a standard deviations of
$\sigma=0.40$ and $0.33$, respectively. A comparison with earlier works on much larger samples
\citep{VandenBerk2001,Grupe2010} shows that our sample has bluer optical/UV continua.

\subsection{Properties of the obscuring dust}
\label{SED.IR}

If the adopted thin AD model does indeed explain the emitted SED of AGN in our sample,
then dust in the host galaxies of approximately 50\% of the AGN in our sample is
contributing to the reddening of the optical-UV spectrum. It is therefore interesting
to test the nature of the dust which is causing the extinction. Several earlier studies
\citep{Hopkins2004,Glikman2012} claimed that an SMC-type extinction curve best accounts
for the reddening, in preference to MW-type or \citet{Gaskell2004} reddening,
\citet{Capellupo2015} find that in their sample of 39 z=1.5 AGN, there were no cases
where the SMC-type curve allowed for a better fit than either the MW or simple power-law
extinction curve. Our observations indicate that for those AGN that require intrinsic
reddening, there is no preference for one particular extinction curve. This is probably
because the UV observations are based on photometry instead of spectroscopy and may also
be the result of the limited wavelength range which is essential to differentiate between
the various extinction laws.

As noted above, the amount of extinction seems to be more significant in super-Eddington
AGN. Figure~\ref{figure.Av} compares the reddening distribution of the super-Eddington
and the sub-Eddington groups assuming a SMC extinction curve. The median reddening for
the super-Eddington sources is $\langle E(B-V)\rangle=0.07$~mag with a scatter of
$\sigma=0.08$, while the sub-Eddington AGN sample shows $\langle E(B-V)\rangle
\approx0.01$~mag with $\sigma=0.01$. A Kolmogorov-Smirnov test rules out the hypothesis
that both groups are drawn from the same distribution of \ebmv{} with a probability
of $>99.9$\%. Experimenting with other extinction curves give basically the same
distribution shown in Figure~\ref{figure.Av}.

\subsection{Uncertainties in black hole mass}

\begin{figure}
    \centering
    \includegraphics[width=0.8\linewidth]{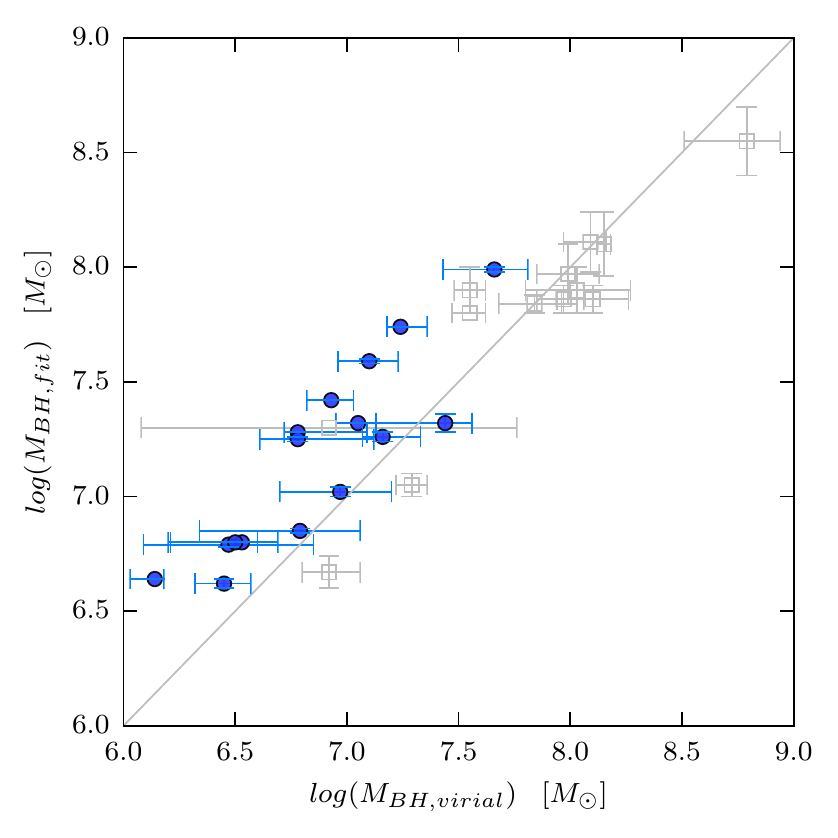}\\
    \includegraphics[width=0.8\linewidth]{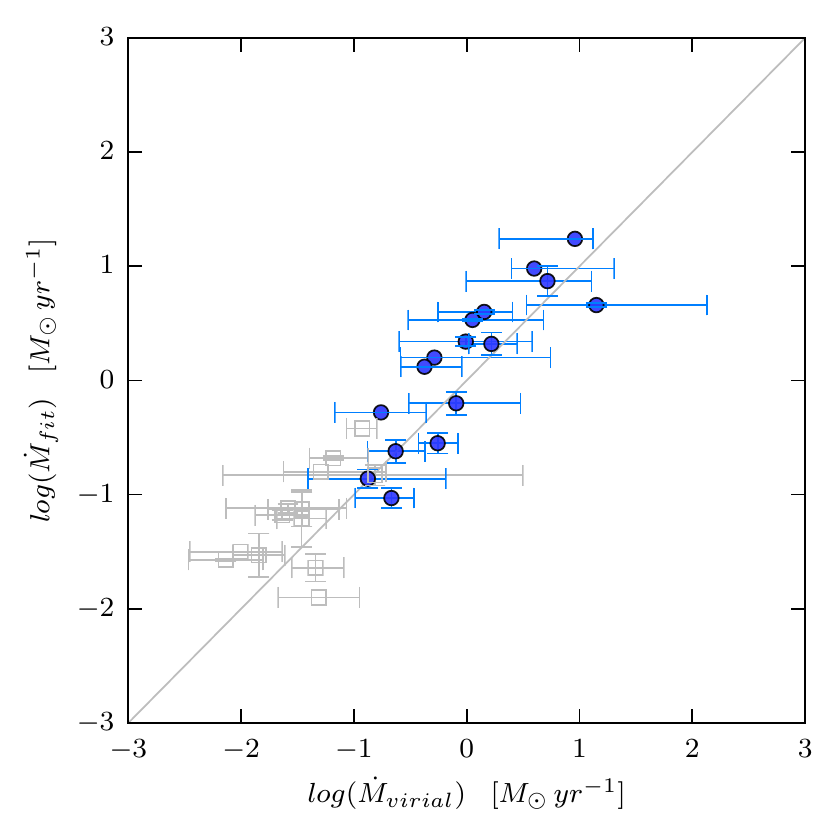}\\
    \caption{Comparison of the virialized black hole mass (\emph{top}) and the dimensionless mass
        accretion rate (\emph{bottom}) against those given by our best-fit thin AD model.
        The straight lines indicate the 1:1 ratios. Gray-blue circles represent super-Eddington
        AGN and gray-empty squares sub-Eddington AGN.}\label{figure.Mbh}
\end{figure}

In order to evaluate the goodness of our fit, we have analysed the black hole properties recovered by the thin AD fit.
We find that the measured ``virialized'' mass black hole and mass accretion rate, within the allowed uncertainties,
lead to a suitable disc model for all the 29 AGN. The comparisons are shown in Figure~\ref{figure.Mbh}. 
The median deviation of the black hole mass ($\Delta \log{M_{BH}}\equiv \log{M_{BH,\text{fit}}}-\log{M_{BH,\text{virial}}}$) 
for the super-Eddington sources is 0.305$\,$dex with a 16$^{th}$ and 84$^{th}$ percentile of 0.07 and 0.5$\,$dex, 
while the median for the sub-Eddington sources is -0.07$\,$dex with its percentiles at -0.23 and 0.25$\,$dex. There is no systematic 
shift on the mass accretion rate which is uniformly distributed. This is because the accretion rate in the disc model is 
directly related to the observed luminosity and the best mass estimate. Note also that the assumed inclination of 
$\cos(\theta)=0.75$ only introduce an additional uncertainty on the fitting results and can not explain the shift on the 
black hole mass observed in the super-Eddington group.\\
Finally, it is interesting to note that our thin AD fit recover almost perfectly both the black 
hole mass and the mass accretion rate. This is in contrast with what has been found in the  
work of \citet{Calderone2013}, which claims that the best-fit black hole mass is on average a 
factor of $\sim$6 greater than the corresponding virialized mass estimation.

\subsection{Ionizing Continuum}

An additional consistency check of the fitted SEDs can be obtained by studying the disc ionizing
continuum which is directly related to the observed emission lines. The relative intensity of the lines, as well
as their equivalent widths are related to the mean energy of the ionizing photons, the ionization parameter and
the covering factor by gas near the BH. The hydrogen Lyman continuum as the part of the SED responsible for most
of the heating and ionization of the broad emission-line gas can be derived from our disc SEDs. In
particular, we can calculate both Lyman continuum, $L_{Lyman}$, and the mean energy of an ionizing photon,
$\langle h\nu\rangle$, and compare the distribution of these numbers in the two groups.

\begin{figure}
    \centering
    \includegraphics{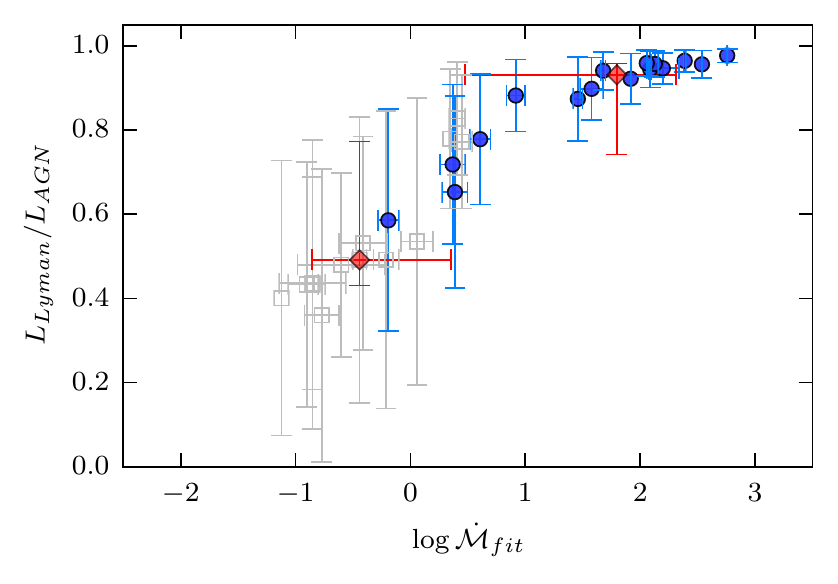}\\
    \includegraphics{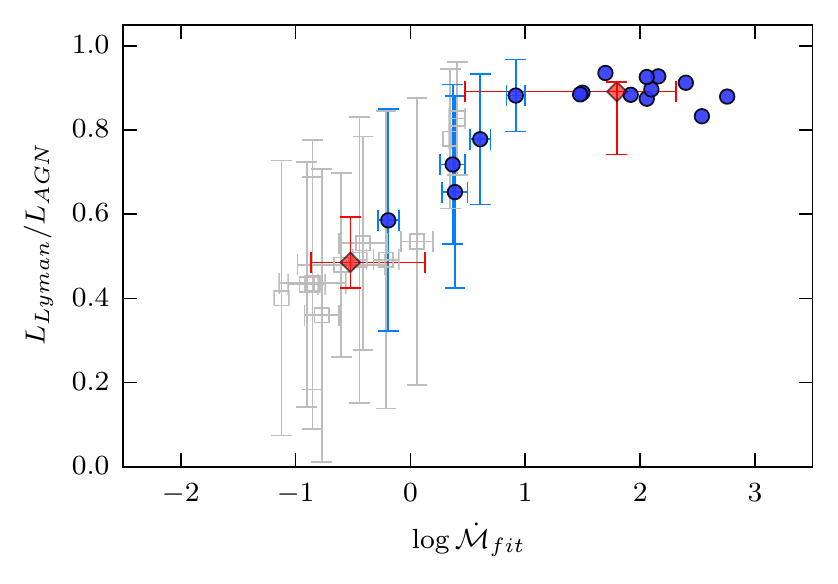}\\
    \includegraphics{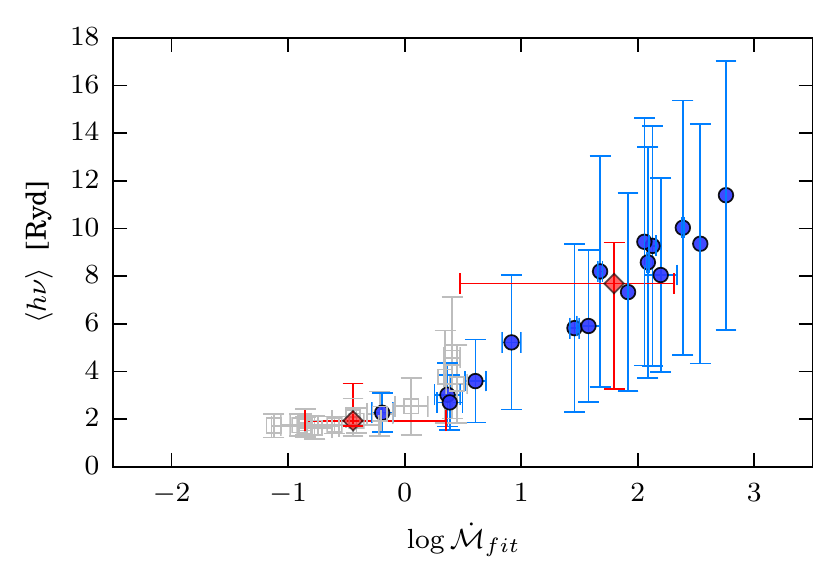}
    \caption{Lyman continuum luminosity fraction for cases of no saturation (\emph{top}) and saturation according to Eqn.~\ref{eq.saturation} (\emph{midle}). The mean energy of an ionizing photon
    as a function of the normalized Eddington ratio for the cases of the no saturation is shown at the bottom. The red diamond represents
    the median parameter for each group and the error bars correspond to the 16$^{th}$ and 84$^{th}$ percentile.
    Gray-blue circles represent super-Eddington AGN and grey-empty squares sub-Eddington AGN.
    }\label{figure.Llyman}
\end{figure}

A comparison of the Lyman continuum properties of the two groups is shown in
Figure~\ref{figure.Llyman} where both parameters are plotted against the normalized Eddington ratio.
For comparison, the mean energy of an ionizing photon with an optical/UV power-law SED of slope
$\alpha_{\nu}=1.5$ is 2.31 Ryd.
As expected, the super-Eddington group, with its higher accretion rate and smaller BH mass,
are predicted to have larger $L_{\text{Lyman}}$/$L_{\text{AGN}}$ and higher $\langle h\nu\rangle$.
The diagrams show a very strong trends with the mass accretion rate. This is in line with the results of
V09, although their sample is in the lower-Eddington ratio regime. The average ionization fraction
is $\sim$0.5 for the lowest Eddington ratio $\dot{\mathcal{M}}\leq3$ (similar with V09) and rise to $\sim$0.9
for the highest $\dot{\mathcal{M}}>3$.

Here, and in all other results related to the Lyman continuum, the soft X-ray emission, and the bolometric correction factors and torus covering factors discussed below, we must
take into account two important factors related to the highly simplified assumption of the high energy
part of the SED in super-Eddington sources. First, the entire disc structure can change at high accretion rate,
leading to two-component disc like the one proposed by \citet{Done2012}. This possibility is not investigated
here, partly because of the large uncertainties on such models and mostly because of the lack of short wavelength observations.
Second, saturation, if exists, affects the far-UV SED much more than the near-UV and optical
continuum \citep[][and references therein]{DuPu2015}. In this scenario the bolometric luminosity for sources with
$\dot{\mathcal{M}}>20$ has only a weak logarithmic dependency on the mass accretion rate (Eqn.~\ref{eq.saturation}).
Such values of $\dot{\mathcal{M}}$ are found in most of the super-Eddington sources in our sample (11/16, see Table~\ref{table.bfSED}).
The case of saturation is compared with the thin AD predictions in Figure~\ref{figure.Llyman}, and comparison with the other relevant quantities are shown in Figs.~\ref{figure.b5100A} and \ref{figure.Ltorus}. Indeed it results in lower Lyman continuum bolometric
correction factors and torus covering factors (see below). However,
all these are still considerably higher than the values estimated for
the sub-Eddington sources. Note that we do not have a simple way of
estimating the changes in the mean energy of the ionizing photons in
cases of saturation and hence do no comment of this additional effects.

In principle, the higher expected $L_{\text{Lyman}}/L_{\text{AGN}}$ and $\langle h\nu\rangle$ for super-Eddington sources could be
investigated by comparing line intensities and EWs between the two groups. One prediction is that, given similar BLR covering factors, super-Eddington sources will show larger line EWs and stronger lines of highly ionized species. While this
general issue is beyond the scope of the present work, we note that such differences in EWs have never been reported.
In fact, several studies show tends in the opposite directions \citep[see e.g. ][]{DuPu2014}.\\

\subsection{Bolometric Correction factors}
Our disc models can also be used to determine the bolometric correction factors for the objects in question. Such factors are used
in many AGN studies \citep{Marconi2004,Netzer2014} as a way to deduce the bolometric luminosity, $L_{\text{AGN}}$, in those
cases (basically
all cases) where much of the SED is not directly observed. Bolometric corrections are frequently given as
\begin{equation}
    \kappa_{\nu} = \frac{L_{\text{AGN}}}{\nu L_{\nu}}
\end{equation}
where $L_{\text{AGN}}$ is the intrinsic bolometric luminosity obtained from the best-fit thin AD model. Here we report
$\kappa_{\nu}$ relative to
5100\AA{}, 3000\AA{} and 1400\AA{} (see Table~\ref{table.bfSED}). Unfortunately, our limited wavelength coverage does not allow us to
distinguish between the spins which is the main uncertainty on the bolometric correction factors. The difference in efficiency between the
two extreme spin values amount to about a factor 8.4.

\begin{figure}
    \centering
    \includegraphics[width=0.95\linewidth]{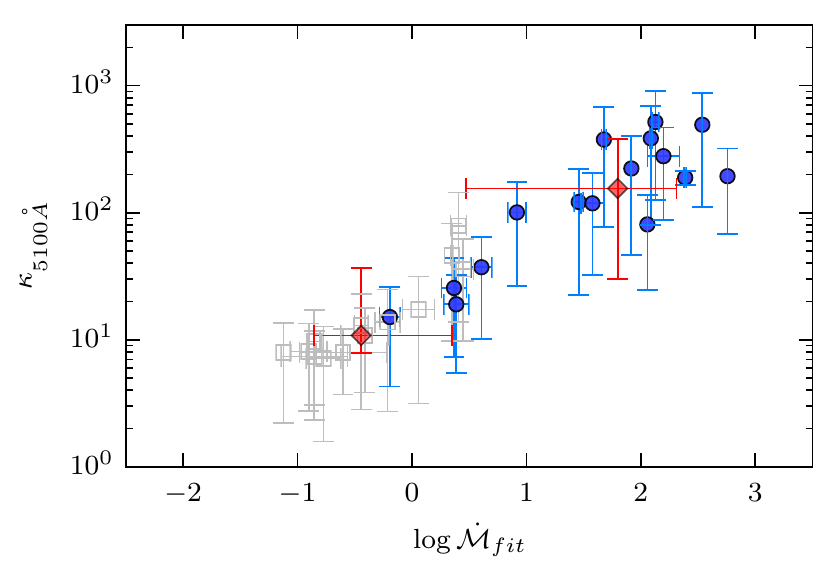}\\
    \includegraphics[width=0.95\linewidth]{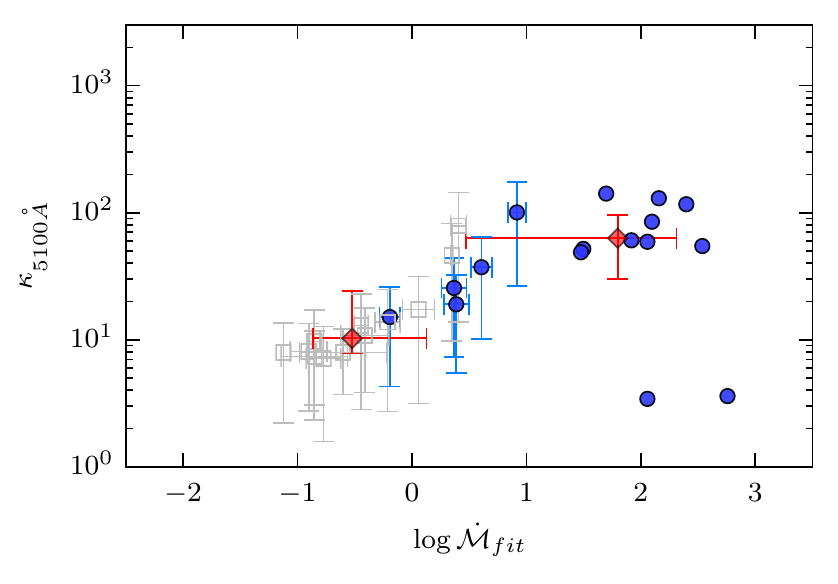}
    \caption{Bolometric correction at 5100\AA{} as a function of the Eddington ratio for the best-fit thin AD models on the top panel and the case of saturation at the bottom. Colour code as in Fig.~\ref{figure.Llyman}.}
    \label{figure.b5100A}
\end{figure}

Figure~\ref{figure.b5100A} shows the distribution of the bolometric correction factor as a function of mass
accretion rate. In general, $\kappa_{5100\AA{}}$ has a very wide distribution where, for the sub-Eddington group, the values
are distributed around the ``standard'' value of about 10 used in many earlier studies like \citet{Kaspi2000,Marconi2004,Trakhtenbrot2012}.
There is a clear tendency for a larger $\kappa_{5100\AA{}}$ in the group of super-Eddington sources due to the significantly higher accretion
rate for a given BH mass. This results in a shift of the SED to shorter wavelength and a considerable increase in $\kappa_{5100\AA{}}$.
For this group, a more typical value is about 150. As explained, there are large uncertainties in calculated disc SEDs for such sources
since their large $L_{\text{AGN}}/L_{Edd}$
suggests they contain slim rather than thin ADs.
In particular, we must consider the possibility of saturation for $\dot{\mathcal{M}}>20$.
In this case the bolometric correction factor at 5100\AA{} is considerably smaller, about 60 instead of 150 (see bottom panel of
Fig.~\ref{figure.b5100A}).


\subsection{Torus emission and SED}
\label{section.torus.SED}

We also tested the possible differences between sub-Eddington and super-Eddington sources by investigating their torus emission.
This is done by using the most recent available WISE All-Sky Data Release that covers $>$99\% of the sky. The torus SED
was constructed from the four WISE bands, at 3.6, 4.5, 12 and 22$\,\mu$m, using only data with a signal-to-noise ratio SNR>20. The observed
IR SED were modelled by the torus template of \citet{Mor2012} which is made of three components: a clumpy torus, a dusty narrow line region,
and a hot pure-graphite dust component which represent the innermost part of the torus. In general, this SED is similar to other
torus SEDs, e.g. those used by \citet{Mullaney2011} and \citet{Lira2013} except that none of these models include the
hot graphite dust and hence cannot provide
a proper fit to the short wavelength part at around 1-3$\,\mu$m. Given the lack of adequate spectral resolution, there is no way to take into account
the contribution of star-formation emission in the host galaxy. In addition, we do not have good enough
spatial information to isolate the nuclear AGN emission from star formation in the host galaxies. These points were
discussed in detail in \citet{Sani2011} and \citet{Mor2012} where a much larger number of
low and high Eddington ratio (NLS1s) AGN are considered.

The fitting procedure is based on the use of three templates, the mean torus SED and its 25$^{th}$ and
75$^{th}$ percentiles. In this procedure, the normalization is the only additional free parameter and its value
is determined by a $\chi^{2}$ analysis. This procedure has been recently discussed in great detail
by Nezer et al. (2015). Earlier attempts are described in
\citet{Mullaney2011,Lusso2013,Roseboom2013,Tsai2015} and others.

We find that in all 29 cases, one of the three templates gave a satisfactory fit to the observed
near- and mid-IR. Figure~\ref{figure.medianSED} shows the median torus AGN for both groups which was
constructed from the best-fit templates. In general, the observed SEDs are very similar but the mid-IR
spectrum of SEAMBHs is somewhat bluer.

\begin{figure}
    \centering
    \includegraphics[width=0.95\linewidth]{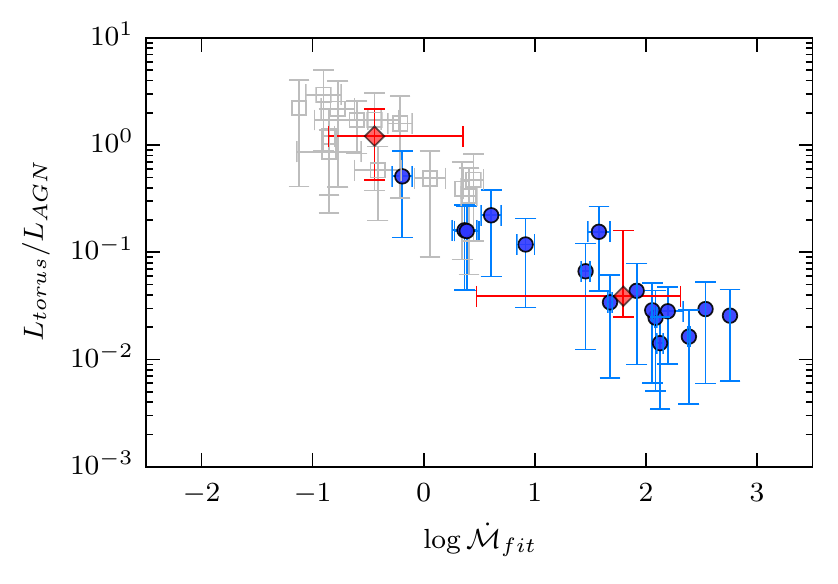}\\
    \includegraphics[width=0.95\linewidth]{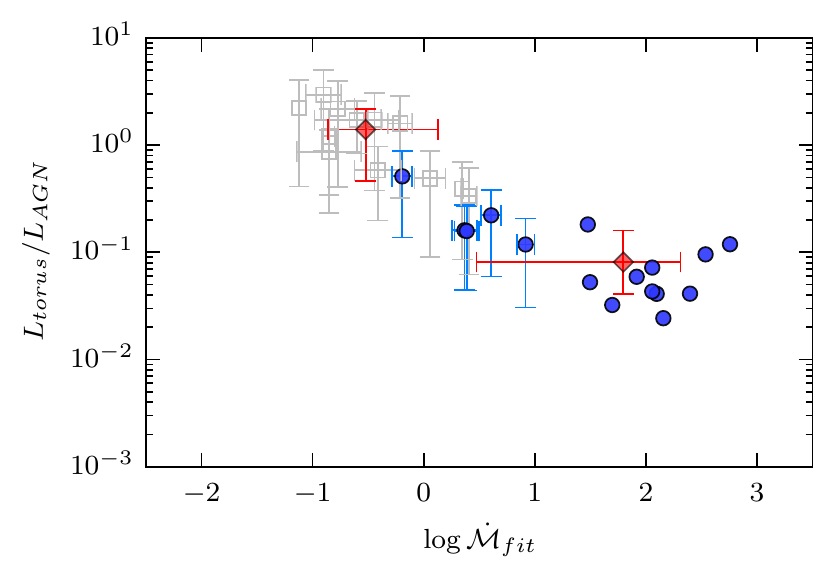}\\
    \caption{The torus luminosity fraction, defined as $L_{torus}/L_{\text{AGN}}$ and assuming
    complete isotropy for the dust emission, as a function of the normalized Eddington ratio (top no saturation, bottom saturation according to Eqn.~\ref{eq.saturation}).
    Colour code as in Fig.~\ref{figure.Llyman}.}
    \label{figure.Ltorus}
\end{figure}

\begin{figure}
   \includegraphics[width=\linewidth]{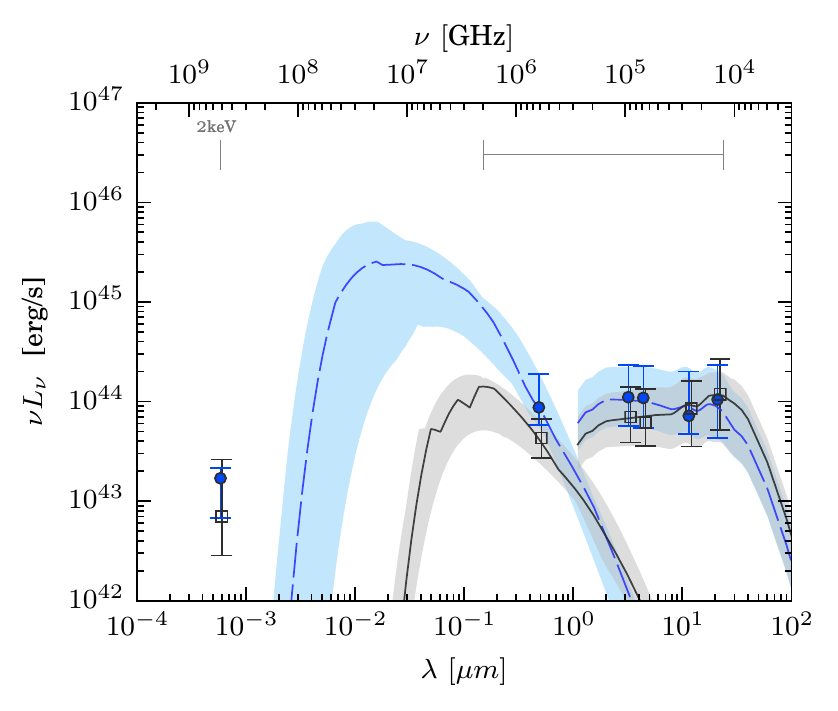}
   \caption{Median AGN SEDs for the two groups: blue, dashed line represents
   super-Eddington and black, solid line sub-Eddington. The shaded areas show
   the 25$^{th}$ and 75$^{th}$ percentiles. The circle (super-Eddington) and square (sub-Eddington)
   points represent the median luminosity at 2$\,$keV, 5100\AA{}, 3.6$\,\mu$m, 4.5$\,\mu$m, 12$\,\mu$m and 22$\,\mu$m.
   The X-ray luminosity at 2$\,$keV was computed assuming a single power-law between 2-10$\,$keV
   corrected for galactic absorption. The horizontal line displays the spectral window used to
   fit both thin AD model and the torus templates. The displayed median disc SED
   is assuming the minimum spin ($a_{\star}=-1$).}\label{figure.medianSED}
\end{figure}

Given the torus SED we can compare the torus covering factor of super-Eddington and sub-Eddington sources.
In general, this depends on the emission pattern of the torus and hence on deviations from isotropy
\citep[][and references therein]{Netzer2015}. For the purpose of the present paper we assume complete
isotropy and compare the integrated 1-200$\,\mu$m emission of the torus, $L_{\text{torus}}$, with
$\lambda L_{\lambda}(5100\mAA{})$ and $L_{\text{AGN}}$ of each source. For the torus model used here
$L_{torus}/L_{5\mu m}=3.58$ for the median composite, 3.18 for the bluer composite and 4.27 for the redder
composite. The comparison of $L_{torus}/\lambda L_{\lambda}(5100\mAA{})$ in the two groups suggest
very similar distribution (see Table~\ref{table.statisitcs}). On the other hand
$L_{\text{torus}}/L_{\text{AGN}}$, which is related to the torus covering factor, show lower fractions
for super-Eddington AGN (see Figure~\ref{figure.Ltorus}) because of their larger $\kappa_{5100\AA{}}$.
As discussed below, this difference suggest either saturation (Eqn.~\ref{eq.saturation}) or
that the far-UV luminosity of the super-Eddington is
over-estimated by a large factor.

\subsection{Simultaneous Optical/UV and X-ray Observations}
The AGN paradigm includes a thermal accretion disc emitting in the optical-UV and a corona which
up-scatters the emitted disc photons to X-ray energies resulting in a power-law X-ray emission.
Studies of large AGN samples show a clear dependence of $\alpha_{OX}$ (the spectral index connecting
the 2500\AA{} and 2$\,$keV emission) on the UV luminosity such that higher $L_{\nu}(2500\AA{})$ corresponds to weaker
2~keV emission \citep[see][and references therein]{Steffen2006,Grupe2010}.
In this paper we treat
the 2-10~keV emission in a simplistic way and focus on $\alpha_{OX}$. In a forthcoming publication
we provide more information about the X-ray spectral shape including new XMM-Newton observations for
some of the sources.

\begin{figure}
    \centering
    \includegraphics[width=0.8\linewidth]{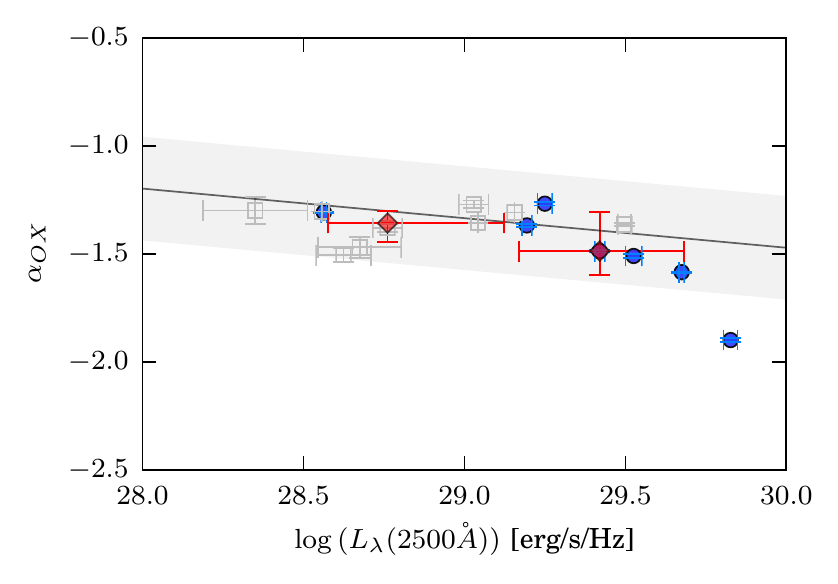}\\
    \includegraphics[width=0.8\linewidth]{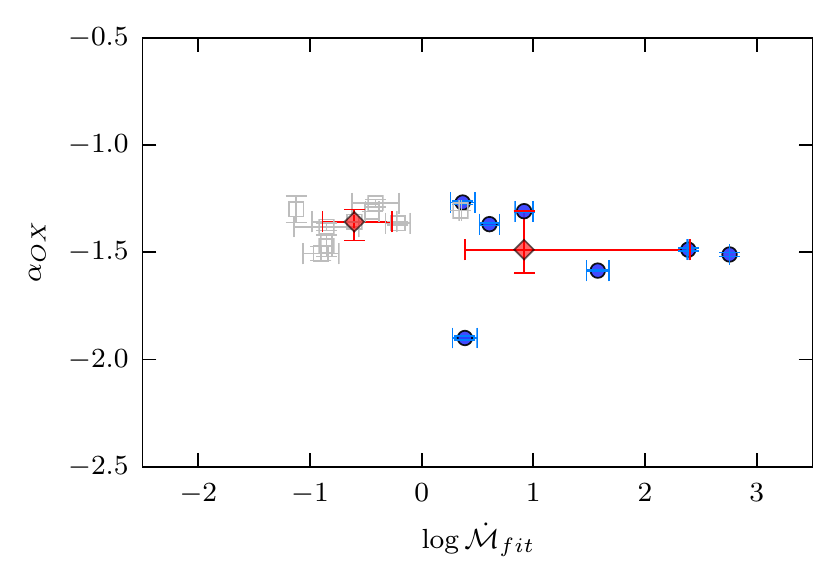}\\
    \includegraphics[width=0.8\linewidth]{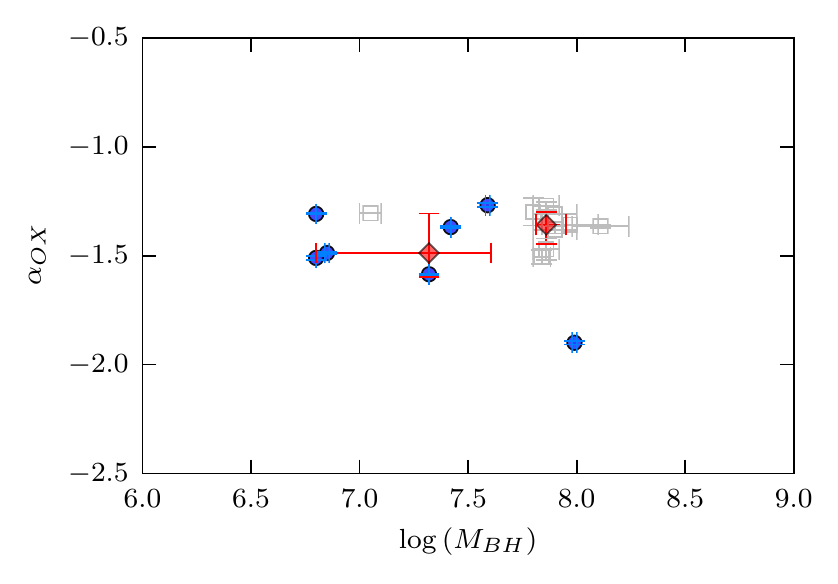}
    \caption{$\alpha_{OX}$ as a function of the rest-frame luminosity at 2500\AA{}
    (\emph{top}), the normalized Eddington ratio (\emph{middle}) and the black hole mass
    (\emph{bottom}). The solid line and shaded area in the top panel show the best fit
    and spread obtained by \citet{Steffen2006} for a sample of optically selected AGN.
    Colour code as in Fig.~\ref{figure.L5100A}. The two outliers with the smallest
    $\alpha_{OX}$ are NGC~7469 and PG~0844.}
    \label{figure.alphaOX}
\end{figure}

We calculate the optical-to-X-ray spectral slope $\alpha_{OX}$ for 17 sources that have simultaneous
X-ray spectrum and OM photometry and show them in Figure~\ref{figure.alphaOX} against the rest-frame
luminosity at 2500\AA{}, the normalized Eddington ratio and the black hole mass. The X-ray luminosities
were derived from the absorption-corrected rest-frame 2-10~keV fluxes which was modelled with a single
power law accounting for Galactic absorption. Except for two sources (NGC~7469, and PG~0844+349),
the values of $\alpha_{OX}$ corrected for intrinsic reddening lie within the scatter expected from
\citet{Steffen2006} which reinforce the importance of accounting for dust extinction
\citep[V09,][]{Grupe2010}. There are reasons to suspect the reliability of $\alpha_{OX}$ for
the two outliers. The X-ray spectrum of NGC~7469 shows sign of high absorption which could
explain their lower optical-to-X-ray spectral slope. For PG~0844 there is an inconsistency between
the optical spectrum and the OM photometry that suggest a large uncertainty due to
SED variations. We point the reader to Appendix~\ref{appendix.Notes} for a more detailed explanation.
Interestingly, the distribution of $\alpha_{OX}$ shows no large differences between sub-Eddington
and super-Eddington sources with an average $\alpha_{OX}$ of -1.36 and -1.48, respectively.
In contrast with V09, we do not find any correlation between $\alpha_{OX}$ and the black hole mass.

\section{Discussion and Conclusions}
\label{Conclusions}

We have presented SEDs for a sample of AGN selected for their accretion rate from a variety
of reverberation mapping campaigns. We define a group of super-Eddington AGN thought to be powered
by slim accretion discs because of their extremely high Eddington ratio, and compared them to
sub-Eddington AGN thought to be powered by thin accretion discs. The unique aspect of the work
is the detailed comparison with high accretion rate AGN with BH mass measured from direct RM. We have 16 objects
of this group which triples the number of such objects known in previous studies.
The 5~keV to 10000\AA{} spectral energy distribution modelled here is, in our opinion, the best way to compare
thin and slim disc models and to test the various suggestions for the extreme properties of slim
discs at the far-UV. A very detailed study of the intrinsic SED of very high accretion rate AGN, like those studied here,
has never been done before. We are able to improve the SED modelling by employing the most
accurate measurements of mass accretion rate and black hole mass, as well as simultaneous
optical-to-UV photometry in conjunction with high resolution optical spectra. These are useful
for determining the degree of host-galaxy contamination and for constrain better the intrinsic
reddening of the sources.

In general we find that the host galaxy contribution spans a fairly wide range of values, but the
optical/UV emission lines are unlikely to contaminate the underlying disc continuum. The
simultaneous and non-simultaneous optical-to-UV (0.2-1$\,\mu$m)
data, allowed us to fit thin AD models, including both Galactic and intrinsic extinction, to all
29 RM-selected AGN: 13 sub-Eddington sources and 16 super-Eddington sources.

We find that in the lower-Eddington ratio regime ($\dot{m}<0.3$) the estimated intrinsic dust
extinction is consistent with zero in approximately 46\% of the sources. The fractional
number, and the median $E(B-V)$ are similar to those obtained in earlier and larger samples of AGN
that used different types of SEDs. For the high accretion rate AGN ($\dot{m}\ge0.3$) we find that
at least 75\% (12/16) of these require significant optical-UV dust reddening. On average, the
level of dust extinction in the super-Eddington group is higher than that of the slow accreting AGN.

As in earlier studies that combine accretion disc and corona models, we confirm that those AGN with higher
accretion rate and smaller BH mass present larger reddening-corrected ionizing fractions,
$L_{\text{Lyman}}/L_{\text{AGN}}$. The actual number depends on the importance of saturation in the super-Eddington sources.
Another related conclusion is that super-Eddington sources show higher
mean energy of the ionizing photons and larger bolometric correction factors. In this fast accreting
AGN sample, the 5100\AA{} bolometric corrections is
an order of magnitude larger than the factors found for sub-Eddington sources, depending on the importance of saturation. This correlation
factor is considerably larger than the ``standard'' value used in many earlier studies based on low Eddington ratio sample. Note that our approach does not take into account
the contribution of the X-ray emission and extrapolate the thin AD models to far-UV energies.
A useful extension of this work would be to acquire simultaneous HST observations
at shorter wavelengths, potentially improving the extreme-UV SED shape and reveal clues about the nature
of slim accretion disc.

Our study shows that for 15/17 objects with simultaneous UV-X-ray observations, there is no correlation
between $\alpha_{OX}$ with accretion rate or black hole mass. Part of this may be related to the reddening
correction used in this work mostly for large $\dot{m}$ AGN. In particular, our study does not confirm the
correlation of $\alpha_{OX}$ with black hole mass suggested in V09.

We find no statistical evidence that the torus SED in super-Eddington AGN
is different from the one in sub-Eddington AGN, although the median torus SED is marginally bluer
(see Figure~\ref{figure.medianSED}). Moreover, the medians $L_{5\mu m}/L_{5100\mAA}$ of the two groups
are very similar. However, there is a very significant difference in the derived tours covering factors,
with super-Eddington sources showing much smaller $L_{\text{torus}}/L_{\text{AGN}}$ by as much as an
order of magnitude even when considering saturation. This is basically
the same result obtained for the $L_{5100\mAA}$ bolometric correction factor and is directly related
to the predicted, but so far unobserved far-UV emission of super-Eddington objects.

There are two major conclusions to the present study. First, thin AD models can be used to obtain
satisfactory fits to the 0.2-1$\,\mu$m SEDs of all AGN, whether sub-Eddington or super-Eddington accretors.
This is generally consistent with the few available, albeit not very accurate calculated slim accretion
disc 0.2-1$\,\mu$m SEDs that show them to be generally similar to SEDs of thin ADs
\citep[][and references therein]{Wang2014a}. Second, several
of the important results obtained here, that are related to the high energy (Lyman continuum) part of the
SED, including $L_{\text{Lyman}}/L_{\text{AGN}}$, the mean energy of the ionizing photons, and the
bolometric correction factors, all suggest large differences between the two groups. However, these are not
direct observations but rather inferences from model assumptions about the part of the continuum, below
about 0.2-1$\,\mu$m, which is not accessible to us. It is quite likely that
the total emitted high frequency radiation in super-Eddington AGN is smaller than assumed here, and its angular dependence
may be different too.

We find two indications that the above differences between the groups found here may be related
to wrong or uncertain model assumptions. The emission line intensities, or line EWs of super-Eddington
sources are not significantly different from those of sub-Eddington sources. This point was not
addressed in detail in the present work, but is well known in earlier studies of large samples of NLS1,
like \citet{Grupe2010}. In fact, various studies \citep[e.g.][]{DuPu2015} show that the EW(H$\beta$)
in super-Eddington sources is somewhat smaller than in sub-Eddington AGN. Strong angular dependence of
the SED in a system with a flat BLR whose covering factor is small is one possible explanation.
The second is the finding that $L_{\text{torus}}/L_{\text{AGN}}$ (or, equivalently the torus covering
factor) is smaller in super-Eddington sources. However,
$L_{\text{torus}}/\lambda L_{\lambda}(5100\mAA{})$ of the two groups is very similar. We find it hard
to believe that the increase in $L_{\text{AGN}}$ is exactly compensated for by a decrease in the
(geometrical) covering factor of the torus. A more likely explanation is that the emitted
$L_{\text{Lyman}}$ radiation in SEAMBHs is either much weaker than
assumed here, due to saturation, or strongly angle dependent -two possibilities that have been raised in earlier studies
\citep[e.g.][]{Wang2014a}. While we prefer the latter explanation, the very good agreement of
$L_{\text{torus}}/\lambda L_{\lambda}(5100\mAA{})$ between the two groups suggest fine tuning that is hard to accept.

\section*{Acknowledgements}
We thank the anonymous referee for providing a positive and useful report which
has clarified our discussion and improved this paper.
Funding for this work has been provided by a joint ISF-NSFC grant number 83/13. We are grateful to Du Pu for help with
the Lijiang data and to Brad Peterson for help with the AGN Watch data.




\bibliographystyle{mnras}
\bibliography{references} 

\appendix

\section{Thin AD Results}
In this Appendix, we present the results of our fitting thin AD models to all sources.
Two numbers are listed per source and parameter, corresponding
to minimum ($a_{\star}=-1.0$) and maximum ($a_{\star}=0.998$) spin parameters.
The difference in efficiency between the two amount to about a factor $8.4$.
For each source, the best-fit SED is also shown in Figure~\ref{figure.bfSED}.

\begin{landscape}
\begin{table}
{\renewcommand{\arraystretch}{1.55}
\caption{{\bf super-Eddington group:} Measured and deduced physical parameters modelling the intrinsic reddening \ebmv{}
    with a classical SMC extinction curve and assuming an inclination of $\cos{\theta}$=0.75.  Two thin AD models are listed for each sources corresponding to maximum
    ($a_{\star}=0.998$) and minimum ($a_{\star}$=-1) spin; the later correspond to the second row. }\label{table.bfSED}
\vspace*{0.3cm}
\hspace*{-0.5cm}
\resizebox{1.05\linewidth}{!}{
\begin{tabular}{l cr cc ccc ccrc cccccccccc}\toprule
 & \multicolumn{2}{c}{RM results} & \multicolumn{2}{c}{Host Galaxy} & \multicolumn{3}{c}{Power-Law model} & \multicolumn{12}{c}{thin AD model} \\ \cmidrule(r{.75em}l){2-3}\cmidrule(r{.75em}l){4-5}\cmidrule(r{.75em}l){6-8}\cmidrule(r{.75em}l){9-20}
    Object  &  $\log M_{BH}$  &  $\log \dot{\mathcal{M}}$  &  template  &  f$^{\dagger}$  &  E(B-V)  &  ${\beta_{UV}}^{\ddagger}$  &  $\chi^{2}_{\nu}$  &  E(B-V)  &  $\log M_{BH}$  &  $\log \dot{\mathcal{M}}$  &  $\chi^{2}_{\nu}$  &  $\log L_{\text{AGN}}$  &  $L_{\text{Lyman}}/L_{\text{AGN}}$  &  $\kappa_{5100\mAA{}}$  &  $\kappa_{3000\mAA{}}$  &  $\kappa_{1400\mAA{}}$  &  $\langle h\nu\rangle$  &  $\log L_{5100\mAA{}}$  &  $\log L_{2500\mAA{}}$  &  $\log L_{2keV}$  &  $\log L_{5\mu m}$  \\
    & [\msun] & & & & & & & & [\msun] & & & [erg/s] & & & & & Ryd & erg/s & erg/s & erg/s & erg/s
\\ \midrule
      Mrk142    & $6.47\pm^{0.38}_{0.38}$& $2.10\pm^{0.59}_{0.59}$ &  ssp\_11Gyr\_z05& $0.30$& $0.00$& $2.07\pm0.19$& $0.998$& $0.09\pm^{0.02}_{0.02}$& $6.78$& $2.10$& $0.174$& $46.60$& $0.99$& $907.54$& $259.02$& $63.26$& $14.30$& $43.64$& $44.31$&  & 43.61 \\
      &&&&&&&&  $0.09\pm^{0.01}_{0.02}$& $6.80$& $2.16$& $0.167$& $45.74$& $0.93$& $126.18$& $36.88$& $9.42$& $4.24$& $43.64$& $44.30$&  & 43.61 \\
      Mrk335    & $6.93\pm^{0.10}_{0.11}$& $1.39\pm^{0.18}_{0.17}$ &  ssp\_11Gyr\_z05& $0.24$& $0.00$& $1.58\pm0.34$& $0.050$& $0.07\pm^{0.07}_{0.03}$& $7.42$& $0.52$& $0.054$& $45.71$& $0.93$& $64.26$& $30.54$& $12.30$& $5.34$& $43.90$& $44.29$& $43.31$& 43.96 \\
      &&&&&&&&  $0.06\pm^{0.07}_{0.03}$& $7.42$& $0.70$& $0.056$& $44.90$& $0.62$& $10.12$& $4.94$& $2.20$& $1.85$& $43.89$& $44.26$& $43.31$& 43.96 \\
      Mrk382    & $6.50\pm^{0.19}_{0.29}$& $1.20\pm^{0.69}_{0.53}$ &  ssp\_11Gyr\_z02& $0.52$& $0.00$& $2.43\pm0.20$& $0.979$& $0.01\pm^{0.04}_{0.01}$& $6.80$& $0.84$& $0.794$& $45.41$& $0.97$& $174.46$& $66.43$& $23.83$& $8.04$& $43.17$& $43.65$& $42.84$& 43.22 \\
      &&&&&&&&  $0.01\pm^{0.04}_{0.01}$& $6.80$& $1.00$& $0.795$& $44.58$& $0.80$& $26.47$& $10.19$& $3.78$& $2.40$& $43.16$& $43.63$& $42.84$& 43.22 \\
      Mrk486    & $7.24\pm^{0.12}_{0.06}$& $0.67\pm^{0.20}_{0.32}$ &  ssp\_11Gyr\_z05& $0.31$& $0.00$& $1.84\pm0.42$& $0.246$& $0.07\pm^{0.16}_{0.05}$& $7.74$& $-0.28$& $0.245$& $45.23$& $0.85$& $25.86$& $13.55$& $5.62$& $3.08$& $43.82$& $44.16$&  & 43.84 \\
      &&&&&&&&  $0.04\pm^{0.16}_{0.04}$& $7.74$& $-0.10$& $0.251$& $44.42$& $0.32$& $4.28$& $2.44$& $1.36$& $1.45$& $43.79$& $47.07$&  & 43.84 \\
      Mrk493    & $6.14\pm^{0.04}_{0.11}$& $2.06\pm^{0.33}_{0.21}$ &  ssp\_11Gyr\_z05& $0.38$& $0.00$& $1.96\pm0.12$& $1.946$& $0.10\pm^{0.01}_{0.01}$& $6.64$& $2.06$& $0.191$& $46.41$& $0.99$& $24.49$& $287.00$& $67.10$& $14.64$& $45.02$& $44.09$&  & 43.61 \\
      &&&&&&&&  $0.08\pm^{0.01}_{0.01}$& $6.64$& $2.06$& $0.871$& $45.48$& $0.93$& $137.15$& $39.58$& $9.81$& $4.25$& $43.34$& $44.01$&  & 43.61 \\
      Mrk1044   & $6.45\pm^{0.12}_{0.13}$& $1.37\pm^{0.40}_{0.41}$ &  ssp\_11Gyr\_z05& $0.43$& $0.00$& $2.14\pm0.22$& $0.982$& $0.08\pm^{0.02}_{0.06}$& $6.60$& $1.70$& $0.109$& $46.04$& $0.99$& $676.15$& $200.50$& $52.41$& $13.04$& $43.21$& $43.86$&  & 43.34 \\
      &&&&&&&&  $0.07\pm^{0.02}_{0.05}$& $6.64$& $1.66$& $0.118$& $45.08$& $0.90$& $76.80$& $24.01$& $6.87$& $3.35$& $43.19$& $43.81$&  & 43.34 \\
      IRAS04416 & $6.78\pm^{0.31}_{0.06}$& $2.76\pm^{0.16}_{0.67}$ &  ssp\_11Gyr\_z02& $0.38$& $0.05$& $2.98\pm0.09$& $1.946$& $0.07\pm^{0.01}_{0.02}$& $7.28$& $2.54$& $1.331$& $47.55$& $0.99$& $872.42$& $254.02$& $66.27$& $14.39$& $44.61$& $45.27$&  & 44.74 \\
      &&&&&&&&  $0.05\pm^{0.01}_{0.01}$& $7.28$& $2.54$& $1.721$& $46.60$& $0.92$& $111.48$& $33.44$& $9.16$& $4.33$& $44.55$& $45.20$&  & 44.74 \\
      IRASF12397& $6.79\pm^{0.27}_{0.45}$& $2.94\pm^{0.98}_{0.62}$ &  ssp\_11Gyr\_z05& $0.52$& $0.27$& $2.54\pm0.35$& $1.020$& $0.35\pm^{0.05}_{0.03}$& $6.84$& $2.38$& $0.139$& $46.92$& $0.99$& $211.56$& $333.89$& $75.88$& $15.37$& $44.59$& $44.51$& $43.23$& 43.82 \\
      &&&&&&&&  $0.35\pm^{0.06}_{0.03}$& $6.86$& $2.40$& $0.135$& $46.04$& $0.94$& $165.90$& $46.98$& $11.29$& $4.69$& $43.82$& $44.48$& $43.23$& 43.82 \\
      J075101   & $7.16\pm^{0.17}_{0.09}$& $1.57\pm^{0.25}_{0.41}$ &  ssp\_11Gyr\_z02& $0.35$& $0.08$& $2.39\pm0.19$& $1.990$& $0.12\pm^{0.01}_{0.04}$& $7.24$& $1.92$& $0.118$& $46.92$& $0.98$& $399.02$& $130.52$& $41.19$& $11.50$& $44.32$& $44.87$&  & 44.29 \\
      &&&&&&&&  $0.12\pm^{0.01}_{0.03}$& $7.28$& $1.92$& $0.134$& $45.98$& $0.86$& $46.42$& $15.98$& $5.36$& $3.17$& $44.31$& $44.51$&  & 44.29 \\
      J080101   & $6.78\pm^{0.34}_{0.17}$& $2.51\pm^{0.39}_{0.72}$ &  ssp\_11Gyr\_z05& $0.27$& $0.00$& $2.74\pm0.43$& $0.131$& $0.10\pm^{0.08}_{0.09}$& $7.26$& $2.06$& $0.152$& $47.08$& $0.98$& $468.74$& $149.35$& $45.45$& $12.11$& $44.40$& $45.01$&  & 44.51 \\
      &&&&&&&&  $0.11\pm^{0.06}_{0.09}$& $7.24$& $2.34$& $0.152$& $46.36$& $0.91$& $87.91$& $27.22$& $7.86$& $3.99$& $44.41$& $45.04$&  & 44.51 \\
      J081441   & $6.97\pm^{0.23}_{0.27}$& $1.66\pm^{0.63}_{0.57}$ &  ssp\_11Gyr\_z02& $0.56$& $0.00$& $2.28\pm0.24$& $1.331$& $0.05\pm^{0.02}_{0.05}$& $7.00$& $2.10$& $0.159$& $46.84$& $0.99$& $688.69$& $205.30$& $55.06$& $13.42$& $44.00$& $44.65$&  & 44.02 \\
      &&&&&&&&  $0.05\pm^{0.02}_{0.05}$& $7.04$& $2.08$& $0.171$& $45.90$& $0.90$& $79.94$& $24.99$& $7.25$& $3.74$& $44.00$& $44.62$&  & 44.02 \\
      J081456   & $7.44\pm^{0.12}_{0.49}$& $0.85\pm^{1.03}_{0.30}$ &  ssp\_11Gyr\_z02& $0.49$& $0.07$& $1.74\pm0.26$& $1.977$& $0.14\pm^{0.02}_{0.03}$& $7.28$& $1.50$& $0.196$& $46.55$& $0.97$& $220.18$& $80.63$& $28.91$& $9.34$& $44.20$& $44.73$&  & 44.06 \\
      &&&&&&&&  $0.13\pm^{0.01}_{0.03}$& $7.36$& $1.42$& $0.248$& $45.56$& $0.77$& $22.51$& $9.01$& $3.45$& $2.29$& $44.21$& $44.68$&  & 44.06 \\
      J093922   & $6.53\pm^{0.07}_{0.33}$& $2.65\pm^{0.71}_{0.20}$ &  ssp\_11Gyr\_z02& $0.50$& $0.00$& $2.12\pm0.20$& $0.414$& $0.06\pm^{0.01}_{0.01}$& $6.80$& $2.76$& $1.843$& $47.19$& $0.99$& $67.99$& $537.70$& $106.64$& $17.04$& $45.36$& $44.63$& $43.28$& 44.47 \\
      &&&&&&&&  $0.05\pm^{0.01}_{0.00}$& $6.80$& $2.76$& $2.259$& $46.34$& $0.96$& $319.34$& $84.76$& $17.58$& $5.75$& $43.83$& $44.58$& $43.28$& 44.47 \\
      PG0844    & $7.66\pm^{0.15}_{0.23}$& $0.82\pm^{0.57}_{0.42}$ &  ssp\_11Gyr\_z02& $0.22$& $0.00$& $1.45\pm0.36$& $0.494$& $0.17\pm^{0.15}_{0.08}$& $8.00$& $0.28$& $0.362$& $46.05$& $0.88$& $32.48$& $16.90$& $7.09$& $3.85$& $44.54$& $44.93$& $42.56$& 44.17 \\
      &&&&&&&&  $0.16\pm^{0.14}_{0.07}$& $7.98$& $0.50$& $0.365$& $45.26$& $0.42$& $5.49$& $3.01$& $1.54$& $1.55$& $44.52$& $44.88$& $42.56$& 44.17 \\
      PG2130    & $7.05\pm^{0.08}_{0.10}$& $1.75\pm^{0.23}_{0.20}$ &  ssp\_11Gyr\_z02& $0.54$& $0.00$& $2.22\pm0.26$& $0.715$& $0.03\pm^{0.04}_{0.03}$& $7.32$& $1.48$& $0.237$& $46.57$& $0.97$& $204.56$& $76.13$& $27.75$& $9.11$& $44.26$& $44.75$& $43.23$& 44.62 \\
      &&&&&&&&  $0.04\pm^{0.03}_{0.04}$& $7.32$& $1.68$& $0.237$& $45.78$& $0.82$& $32.33$& $11.96$& $4.31$& $2.71$& $44.27$& $44.76$& $43.23$& 44.62 \\
      Mrk110    & $7.10\pm^{0.13}_{0.14}$& $0.85\pm^{0.26}_{0.25}$ &  ssp\_11Gyr\_z05& $0.50$& $0.00$& $1.82\pm0.25$& $0.048$& $0.04\pm^{0.05}_{0.03}$& $7.60$& $0.26$& $0.039$& $45.63$& $0.91$& $43.75$& $22.07$& $9.03$& $4.36$& $43.99$& $44.35$& $43.63$& 43.76 \\
      &&&&&&&&  $0.03\pm^{0.04}_{0.03}$& $7.58$& $0.48$& $0.038$& $44.84$& $0.53$& $7.33$& $3.83$& $1.81$& $1.69$& $43.97$& $44.31$& $43.63$& 43.76 \\
\bottomrule
\addlinespace
\multicolumn{20}{l}{\large $^{\dagger}$ Host galaxy contribution to the observed flux at 5100\AA{}.}\\
\multicolumn{20}{l}{\large $^{\ddagger}$ Optical to UV spectral slope defined such that $L_{\lambda}\propto \lambda^{-\beta_{UV}}$, $\beta_{UV}=2-\alpha_{UV}$
and $L_{\nu}\propto \nu^{-\alpha_{UV}}$.}\\
\end{tabular}
}}
\end{table}
\end{landscape}

\setcounter{table}{0}
\begin{landscape}
\begin{table}
{\renewcommand{\arraystretch}{1.55}
\caption{{\bf sub-Eddington group:} Measured and deduced physical parameters modelling the intrinsic reddening \ebmv{}
    with a classical SMC extinction curve and assuming an inclination of $\cos{\theta}$=0.75.  Two thin AD models are listed for each sources corresponding to maximum ($a_{\star}=0.998$)
    and minimum ($a_{\star}$=-1) spin; the later correspond to the second row. }
\vspace*{0.3cm}
\hspace*{-0.5cm}
\resizebox{1.05\linewidth}{!}{
\begin{tabular}{l cr cc ccc ccrc cccccccccc}\toprule
 & \multicolumn{2}{c}{RM results} & \multicolumn{2}{c}{Host Galaxy} & \multicolumn{3}{c}{Power-Law model} & \multicolumn{12}{c}{thin AD model} \\ \cmidrule(r{.75em}l){2-3}\cmidrule(r{.75em}l){4-5}\cmidrule(r{.75em}l){6-8}\cmidrule(r{.75em}l){9-20}
    Object  &  $\log M_{BH}$  &  $\log \dot{\mathcal{M}}$  &  template  &  f$^{\dagger}$  &  E(B-V)  &  ${\beta_{UV}}^{\ddagger}$  &  $\chi^{2}_{\nu}$  &  E(B-V)  &  $\log M_{BH}$  &  $\log \dot{\mathcal{M}}$  &  $\chi^{2}_{\nu}$  &  $\log L_{\text{AGN}}$  &  $L_{\text{Lyman}}/L_{\text{AGN}}$  &  $\kappa_{5100\mAA{}}$  &  $\kappa_{3000\mAA{}}$  &  $\kappa_{1400\mAA{}}$  &  $\langle h\nu\rangle$  &  $\log L_{5100\mAA{}}$  &  $\log L_{2500\mAA{}}$  &  $\log L_{2keV}$  &  $\log L_{5\mu m}$  \\
    & [\msun] &  & & & & & & & [\msun] &  & & [erg/s] & & & & & Ryd & erg/s & erg/s & erg/s & erg/s
\\ \midrule
      Mrk79     & $7.84\pm^{0.12}_{0.16}$& $-0.60\pm^{0.25}_{0.21}$ &  ssp\_11Gyr\_z05& $0.24$& $0.00$& $1.56\pm0.28$& $0.507$& $0.04\pm^{0.03}_{0.02}$& $7.88$& $-1.06$& $0.549$& $44.59$& $0.72$& $13.42$& $7.20$& $3.15$& $2.20$& $43.46$& $43.79$& $42.39$& 43.86 \\
      &&&&&&&&  $0.00\pm^{0.03}_{0.00}$& $7.80$& $-0.74$& $0.612$& $43.84$& $0.14$& $2.75$& $1.70$& $1.26$& $1.29$& $43.40$& $43.62$& $42.39$& 43.86 \\
      Mrk279    & $7.97\pm^{0.09}_{0.12}$& $-0.86\pm^{0.33}_{0.30}$ &  ssp\_11Gyr\_z02& $0.33$& $0.00$& $1.84\pm0.37$& $0.324$& $0.02\pm^{0.03}_{0.02}$& $7.92$& $-0.62$& $0.268$& $45.07$& $0.79$& $17.77$& $9.43$& $3.99$& $2.46$& $43.82$& $44.15$& $43.40$& 43.69 \\
      &&&&&&&&  $0.00\pm^{0.01}_{0.00}$& $7.80$& $-0.20$& $0.303$& $44.38$& $0.28$& $3.85$& $2.23$& $1.31$& $1.41$& $43.79$& $44.06$& $43.40$& 43.69 \\
      Mrk290    & $7.55\pm^{0.07}_{0.07}$& $-0.82\pm^{0.23}_{0.23}$ &  ssp\_11Gyr\_z02& $0.28$& $0.00$& $1.94\pm0.20$& $0.189$& $0.02\pm^{0.02}_{0.02}$& $8.00$& $-1.14$& $0.079$& $44.63$& $0.69$& $11.68$& $6.31$& $2.81$& $2.08$& $43.56$& $43.89$& $42.85$& 43.58 \\
      &&&&&&&&  $0.00\pm^{0.01}_{0.00}$& $7.80$& $-0.56$& $0.146$& $44.02$& $0.18$& $3.07$& $1.85$& $1.26$& $1.33$& $43.53$& $43.79$& $42.85$& 43.58 \\
      Mrk509    & $8.15\pm^{0.03}_{0.03}$& $-0.50\pm^{0.13}_{0.14}$ &  ssp\_11Gyr\_z02& $0.35$& $0.00$& $2.12\pm0.26$& $0.350$& $0.01\pm^{0.02}_{0.01}$& $7.96$& $0.20$& $0.200$& $45.93$& $0.88$& $31.49$& $16.39$& $6.86$& $3.74$& $44.43$& $44.77$&  & 44.36 \\
      &&&&&&&&  $0.00\pm^{0.01}_{0.00}$& $8.24$& $-0.08$& $0.383$& $44.94$& $0.19$& $3.14$& $1.89$& $1.26$& $1.33$& $44.44$& $44.69$&  & 44.36 \\
      Mrk590    & $7.55\pm^{0.07}_{0.08}$& $-0.28\pm^{0.36}_{0.36}$ &  ssp\_11Gyr\_z05& $0.62$& $0.00$& $2.02\pm4.65$& $1.452$& $0.07\pm^{0.01}_{0.01}$& $7.80$& $-1.12$& $4.553$& $44.45$& $0.73$& $13.61$& $7.30$& $3.19$& $2.22$& $43.32$& $43.59$& $42.65$& 43.39 \\
      &&&&&&&&  $0.00\pm^{0.00}_{0.00}$& $7.80$& $-1.12$& $2.445$& $43.46$& $0.07$& $2.22$& $1.46$& $1.37$& $1.23$& $43.11$& $43.27$& $42.65$& 43.39 \\
      Mrk817    & $7.99\pm^{0.14}_{0.14}$& $-0.88\pm^{0.22}_{0.22}$ &  ssp\_11Gyr\_z05& $0.41$& $0.00$& $2.16\pm0.22$& $0.757$& $0.00\pm^{0.02}_{0.00}$& $8.10$& $-0.98$& $0.452$& $44.89$& $0.70$& $12.17$& $6.55$& $2.90$& $2.11$& $43.80$& $44.12$& $43.19$& 44.14 \\
      &&&&&&&&  $0.00\pm^{0.00}_{0.00}$& $7.84$& $-0.22$& $2.192$& $44.40$& $0.26$& $3.70$& $2.16$& $1.30$& $1.39$& $43.83$& $44.12$& $43.19$& 44.14 \\
      Mrk1511   & $7.29\pm^{0.07}_{0.07}$& $-0.35\pm^{0.24}_{0.24}$ &  ssp\_11Gyr\_z05& $0.44$& $0.00$& $2.48\pm0.19$& $1.832$& $0.00\pm^{0.01}_{0.00}$& $7.00$& $0.36$& $1.544$& $45.13$& $0.94$& $82.35$& $37.08$& $14.51$& $5.72$& $43.21$& $43.64$& $42.84$& 43.48 \\
      &&&&&&&&  $0.00\pm^{0.00}_{0.00}$& $7.10$& $0.34$& $1.855$& $44.22$& $0.61$& $9.77$& $4.80$& $2.15$& $1.83$& $43.23$& $43.64$& $42.84$& 43.48 \\
      NGC5548   & $8.10\pm^{0.16}_{0.16}$& $-1.66\pm^{0.33}_{0.33}$ &  ssp\_11Gyr\_z05& $0.34$& $0.00$& $1.57\pm0.47$& $0.057$& $0.08\pm^{0.04}_{0.02}$& $7.80$& $-0.80$& $0.030$& $44.77$& $0.78$& $17.03$& $9.05$& $3.84$& $2.41$& $43.54$& $43.88$& $42.53$& 43.55 \\
      &&&&&&&&  $0.01\pm^{0.07}_{0.01}$& $7.92$& $-0.90$& $0.014$& $43.80$& $0.09$& $2.34$& $1.52$& $1.32$& $1.25$& $43.43$& $43.62$& $42.53$& 43.55 \\
      NGC7469   & $6.92\pm^{0.84}_{0.84}$& $0.84\pm^{1.31}_{1.35}$ &  ssp\_11Gyr\_z02& $0.70$& $0.00$& $1.90\pm0.28$& $0.167$& $0.05\pm^{0.11}_{0.04}$& $7.30$& $0.36$& $0.025$& $45.43$& $0.93$& $61.84$& $29.54$& $11.84$& $5.11$& $43.64$& $44.02$& $40.64$& 43.86 \\
      &&&&&&&&  $0.04\pm^{0.10}_{0.04}$& $7.30$& $0.54$& $0.024$& $44.62$& $0.61$& $9.77$& $4.80$& $2.15$& $1.83$& $43.63$& $43.99$& $40.64$& 43.86 \\
      PG1229    & $8.03\pm^{0.24}_{0.23}$& $-1.04\pm^{0.52}_{0.55}$ &  ssp\_11Gyr\_z02& $0.40$& $0.00$& $2.60\pm0.16$& $0.710$& $0.00\pm^{0.00}_{0.00}$& $7.80$& $-0.38$& $1.815$& $45.19$& $0.83$& $23.01$& $12.11$& $5.05$& $2.86$& $43.83$& $44.23$& $43.43$& 44.09 \\
      &&&&&&&&  $0.00\pm^{0.00}_{0.00}$& $8.00$& $-0.50$& $5.231$& $44.28$& $0.15$& $2.82$& $1.73$& $1.26$& $1.30$& $43.83$& $44.23$& $43.43$& 44.09 \\
      PG1617    & $8.79\pm^{0.15}_{0.28}$& $-1.51\pm^{0.58}_{0.33}$ &  ssp\_11Gyr\_z02& $0.33$& $0.00$& $2.41\pm0.17$& $1.236$& $0.00\pm^{0.00}_{0.00}$& $8.40$& $-0.62$& $1.588$& $45.55$& $0.71$& $12.68$& $6.81$& $2.98$& $2.12$& $44.45$& $44.77$&  & 44.63 \\
      &&&&&&&&  $0.00\pm^{0.00}_{0.00}$& $8.70$& $-0.92$& $17.081$& $44.56$& $0.01$& $1.58$& $1.26$& $2.28$& $1.15$& $44.36$& $44.42$&  & 44.63 \\
      Fairall9  & $8.09\pm^{0.07}_{0.12}$& $-0.70\pm^{0.31}_{0.21}$ &  ssp\_11Gyr\_z02& $0.32$& $0.00$& $2.14\pm0.43$& $0.020$& $0.01\pm^{0.05}_{0.01}$& $7.98$& $-0.10$& $0.006$& $45.65$& $0.84$& $24.79$& $13.03$& $5.46$& $3.16$& $44.25$& $44.60$& $43.63$& 44.57 \\
      &&&&&&&&  $0.00\pm^{0.02}_{0.00}$& $8.24$& $-0.32$& $0.161$& $44.70$& $0.14$& $2.72$& $1.69$& $1.26$& $1.29$& $44.26$& $44.56$& $43.63$& 44.57 \\
      MCG+06    & $6.92\pm^{0.14}_{0.12}$& $-0.35\pm^{0.37}_{0.45}$ &  ssp\_11Gyr\_z05& $0.51$& $0.00$& $2.54\pm0.24$& $1.676$& $0.00\pm^{0.01}_{0.00}$& $6.60$& $0.48$& $0.899$& $44.85$& $0.96$& $144.10$& $57.10$& $20.89$& $7.13$& $42.69$& $43.13$&  & 43.12 \\
      &&&&&&&&  $0.00\pm^{0.01}_{0.00}$& $6.74$& $0.34$& $1.306$& $43.86$& $0.69$& $13.74$& $6.23$& $2.63$& $2.02$& $42.72$& $43.13$&  & 43.12 \\
\bottomrule
\addlinespace
\multicolumn{20}{l}{\large $^{\dagger}$ Host galaxy contribution to the observed flux at 5100\AA{}.}\\
\multicolumn{20}{l}{\large $^{\ddagger}$ Optical to UV spectral slope defined such that $L_{\lambda}\propto \lambda^{-\beta_{UV}}$,  $\beta_{UV}=2-\alpha_{UV}$
and $L_{\nu}\propto \nu^{-\alpha_{UV}}$.}\\
\end{tabular}
}}
\end{table}
\end{landscape}

\setcounter{figure}{0}
\begin{figure*}
    \centering
\includegraphics[width=0.33\linewidth]{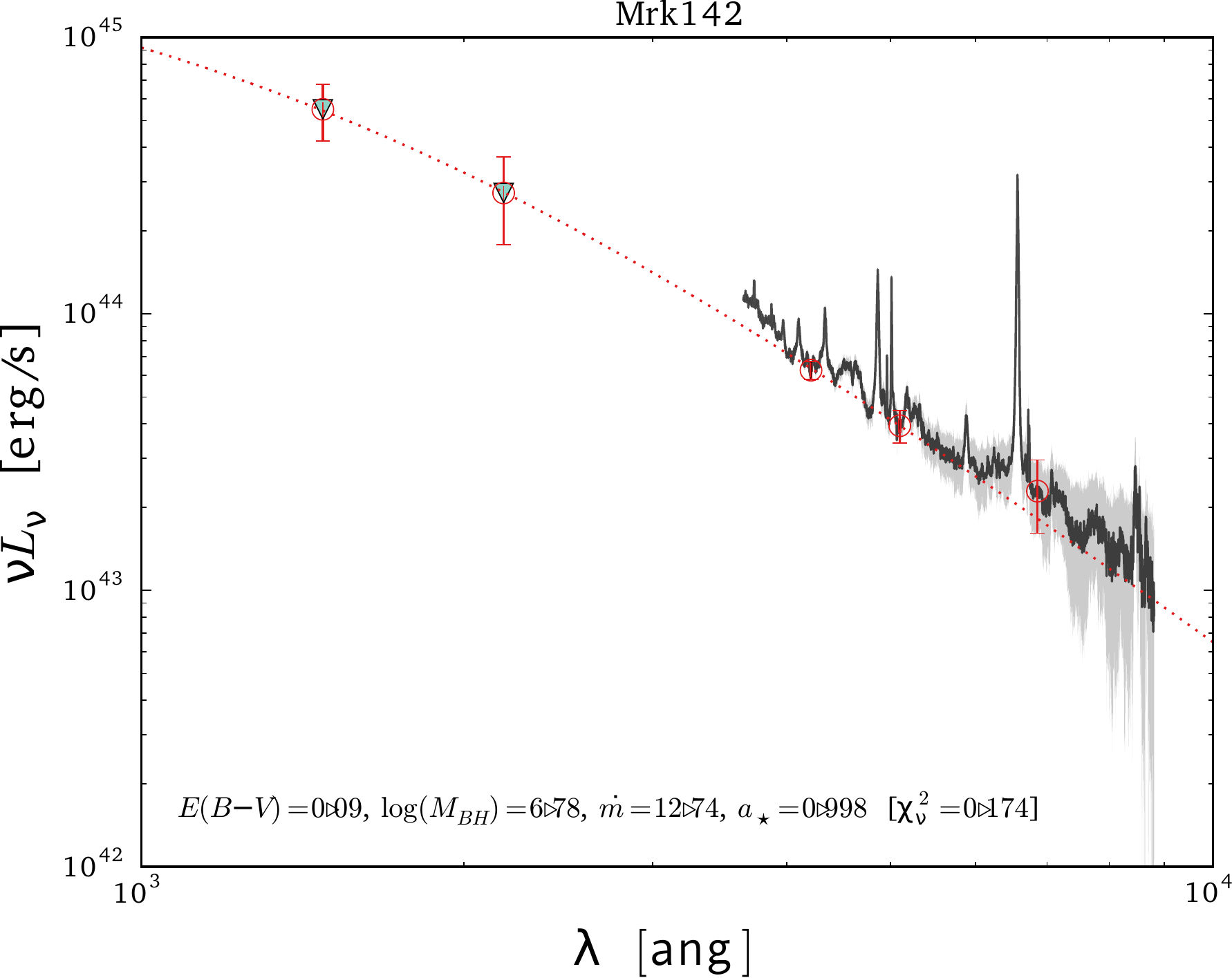}
\includegraphics[width=0.33\linewidth]{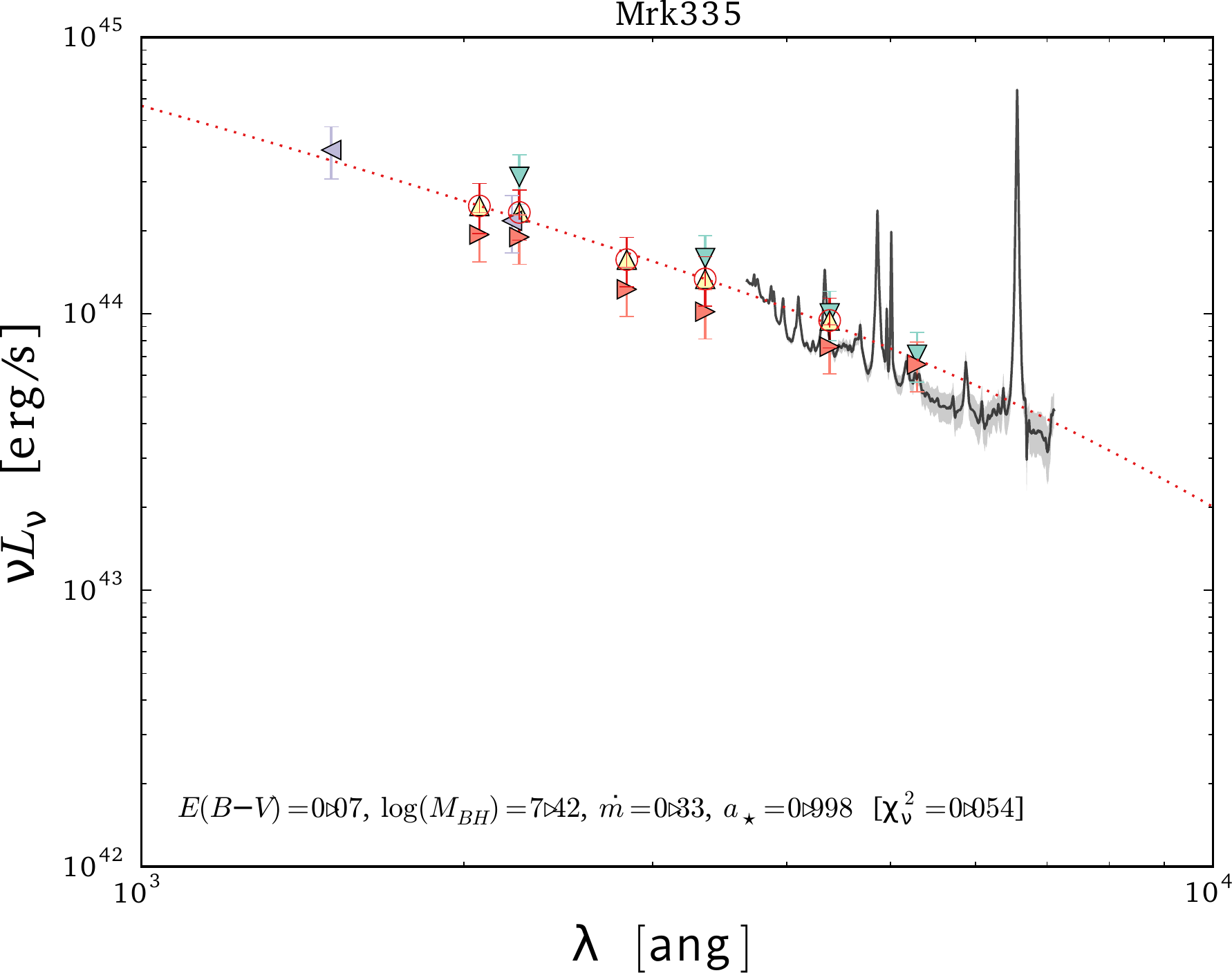}
\includegraphics[width=0.33\linewidth]{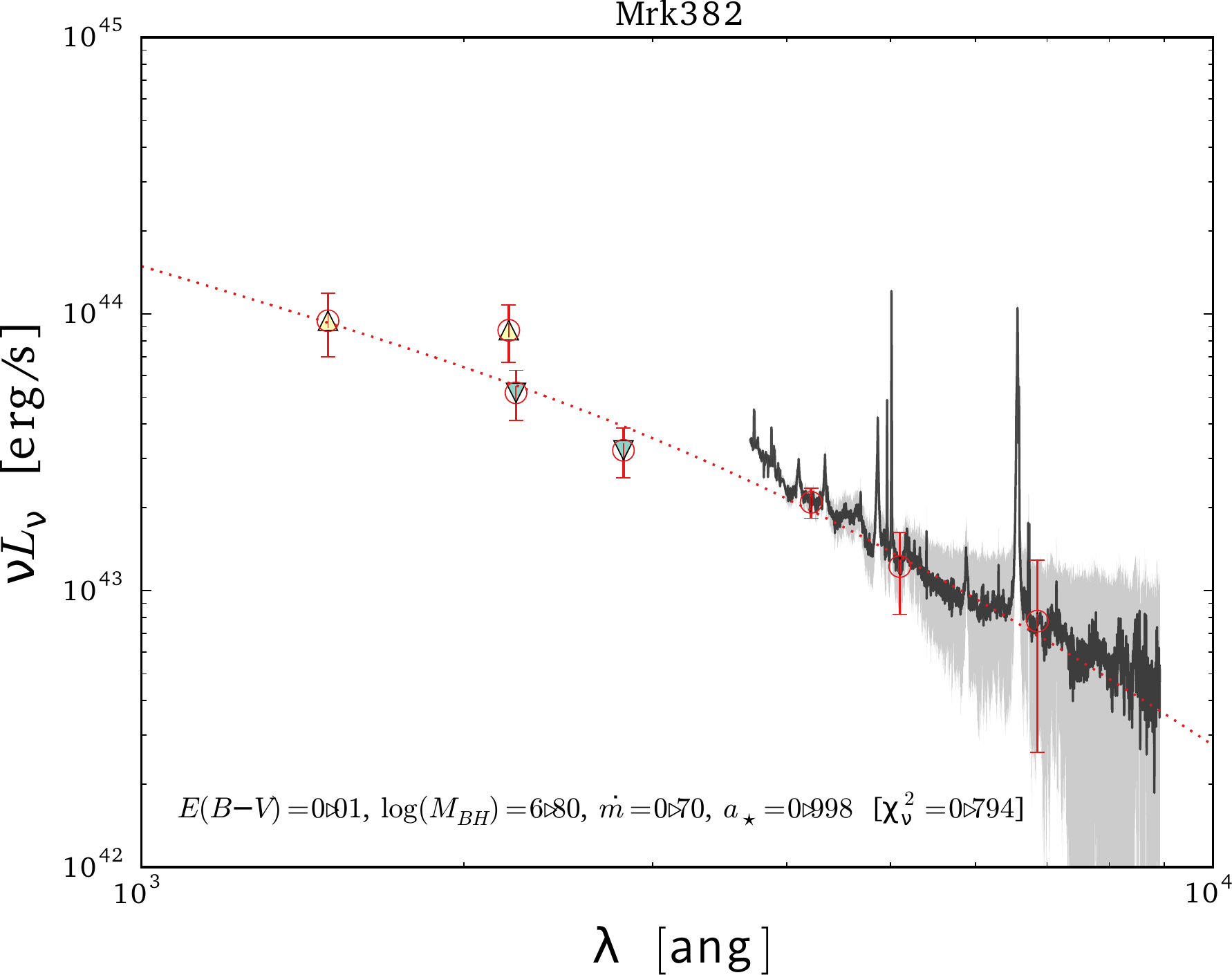}\\
\includegraphics[width=0.33\linewidth]{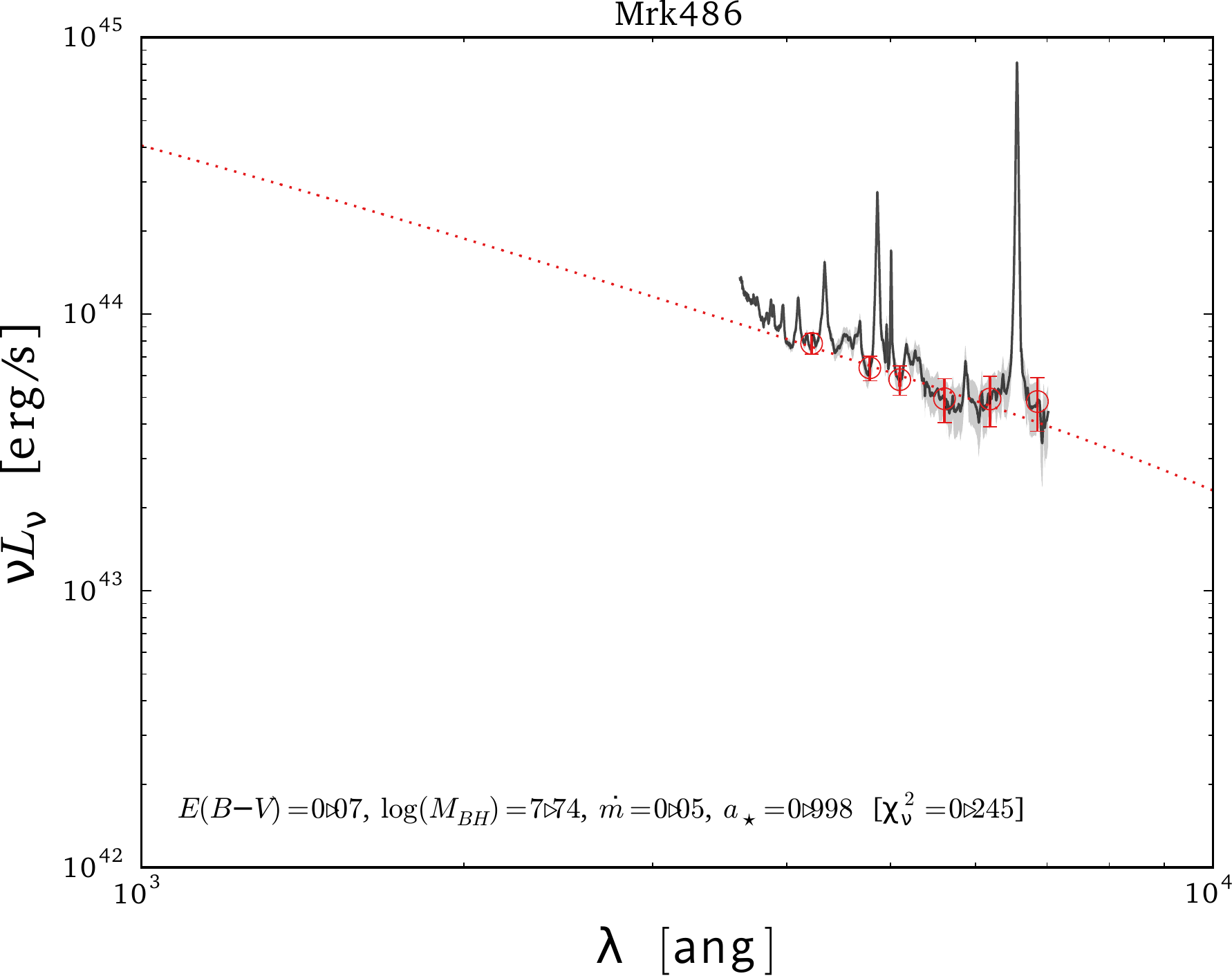}
\includegraphics[width=0.33\linewidth]{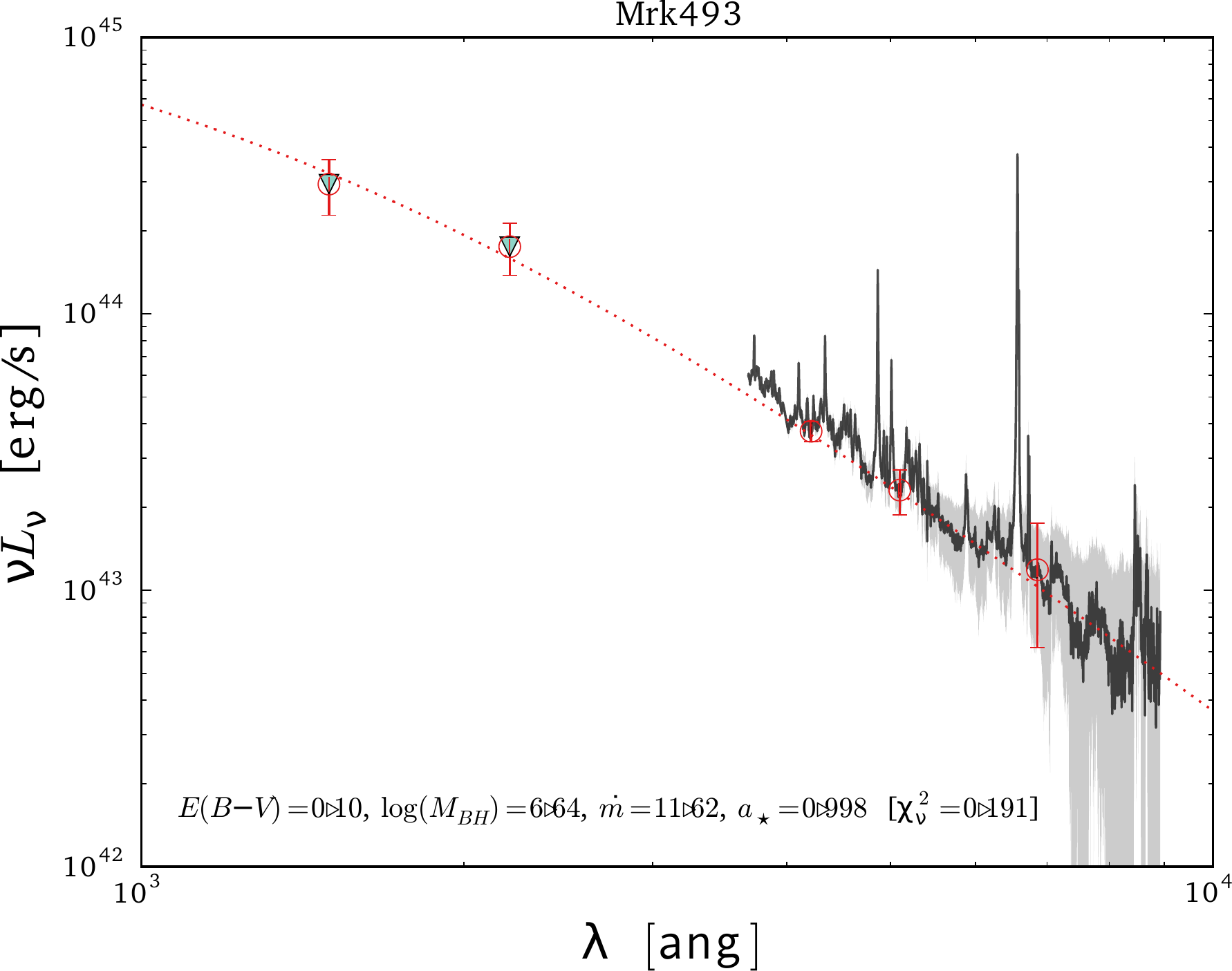}
\includegraphics[width=0.33\linewidth]{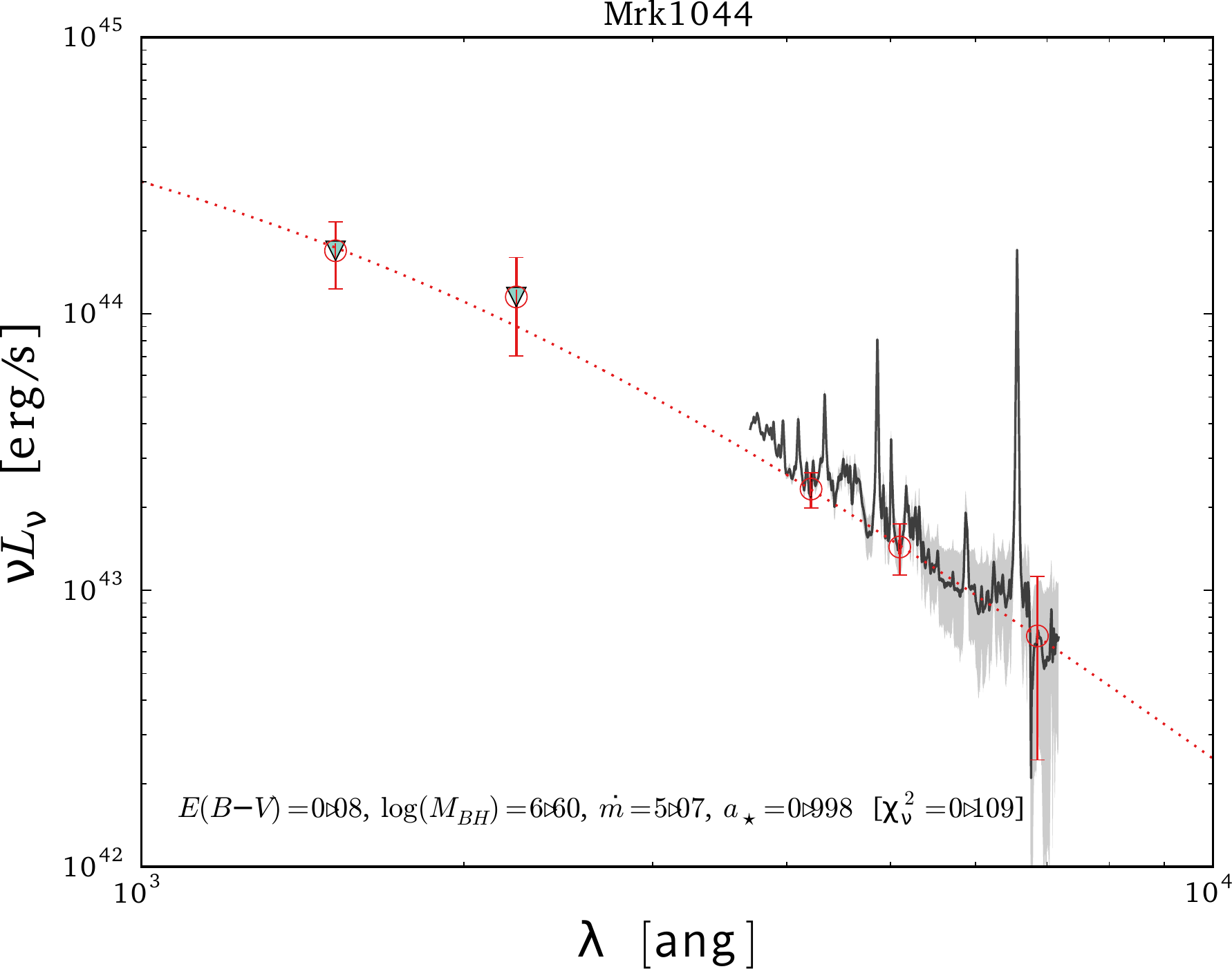}\\
\includegraphics[width=0.33\linewidth]{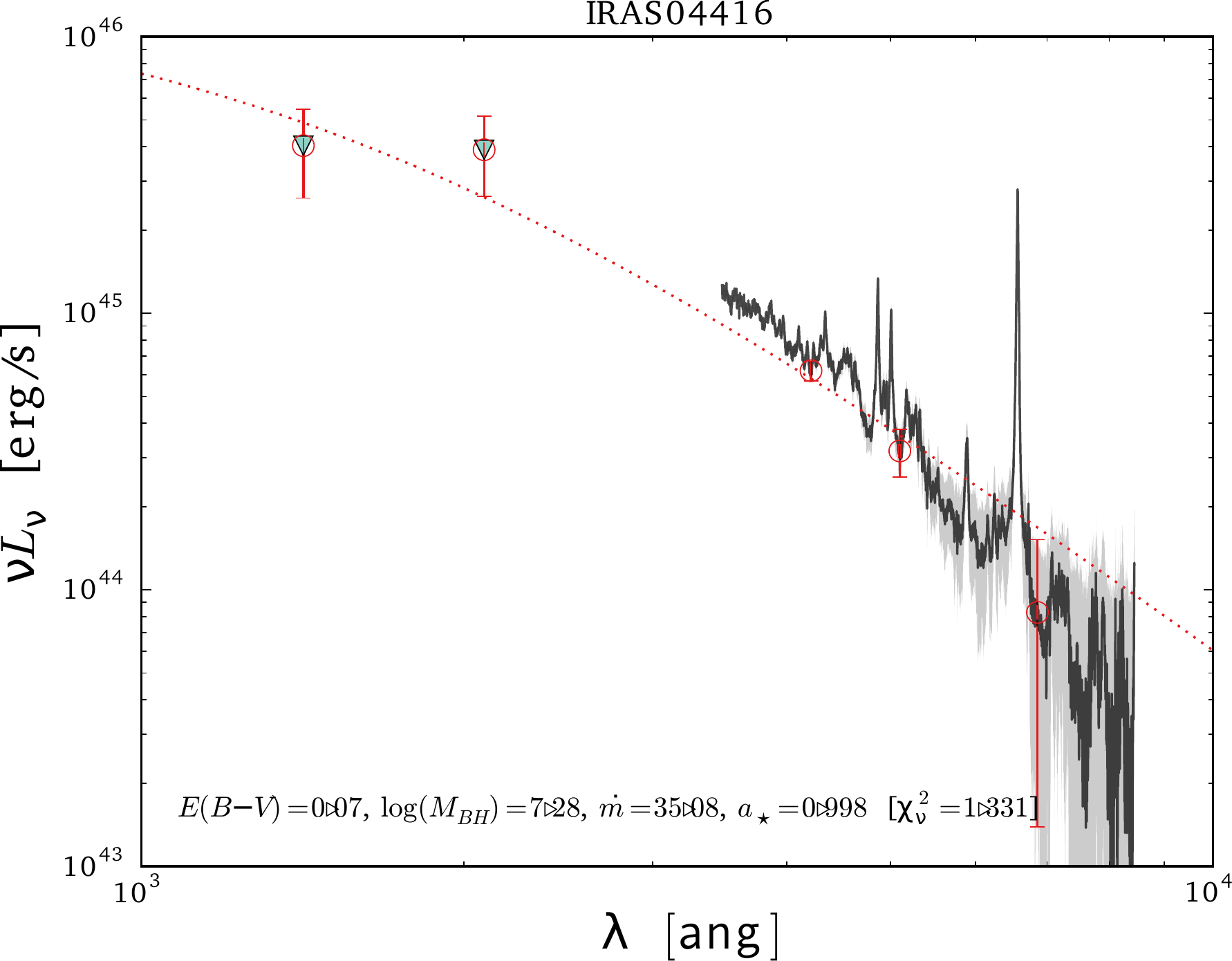}
\includegraphics[width=0.33\linewidth]{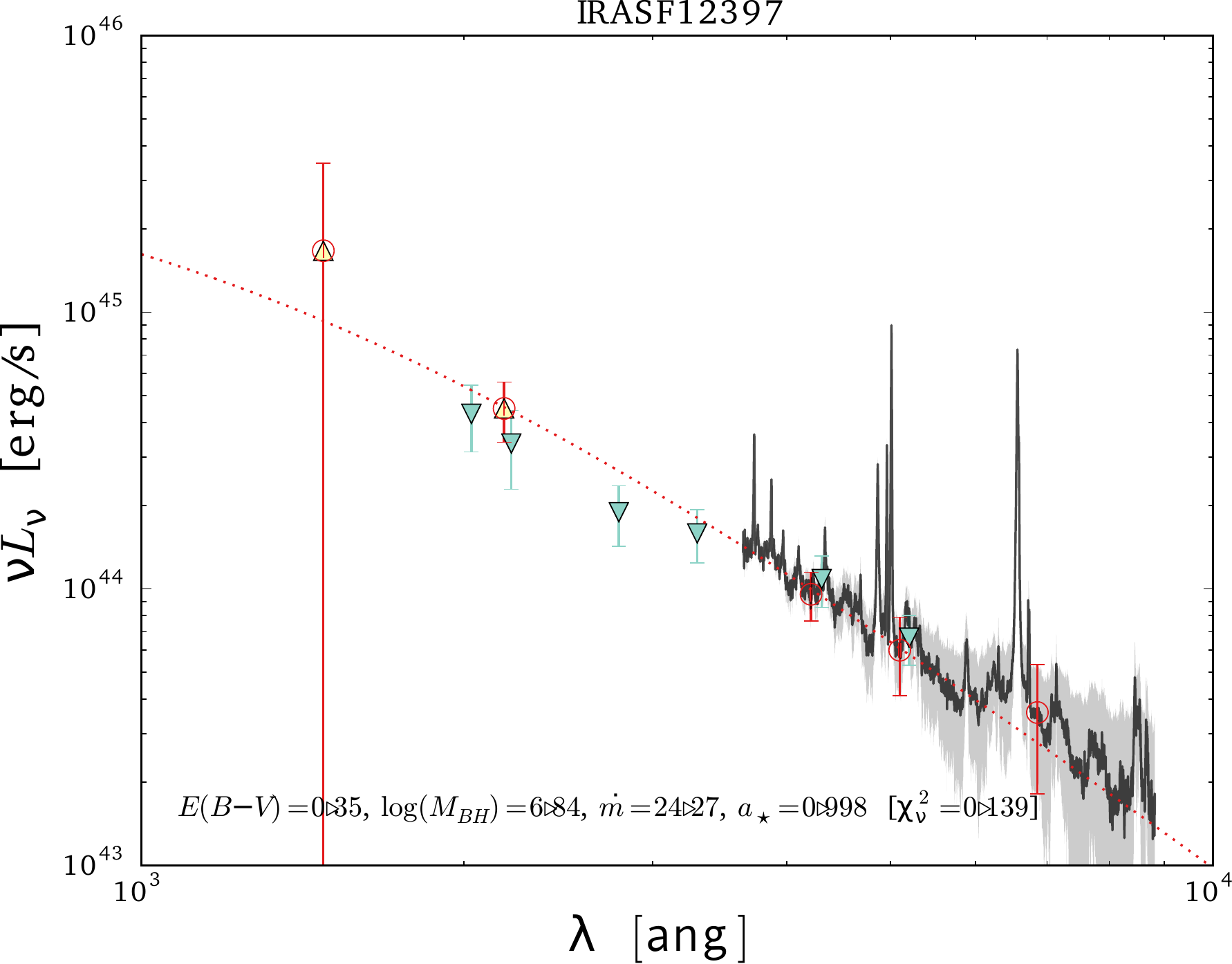}
\includegraphics[width=0.33\linewidth]{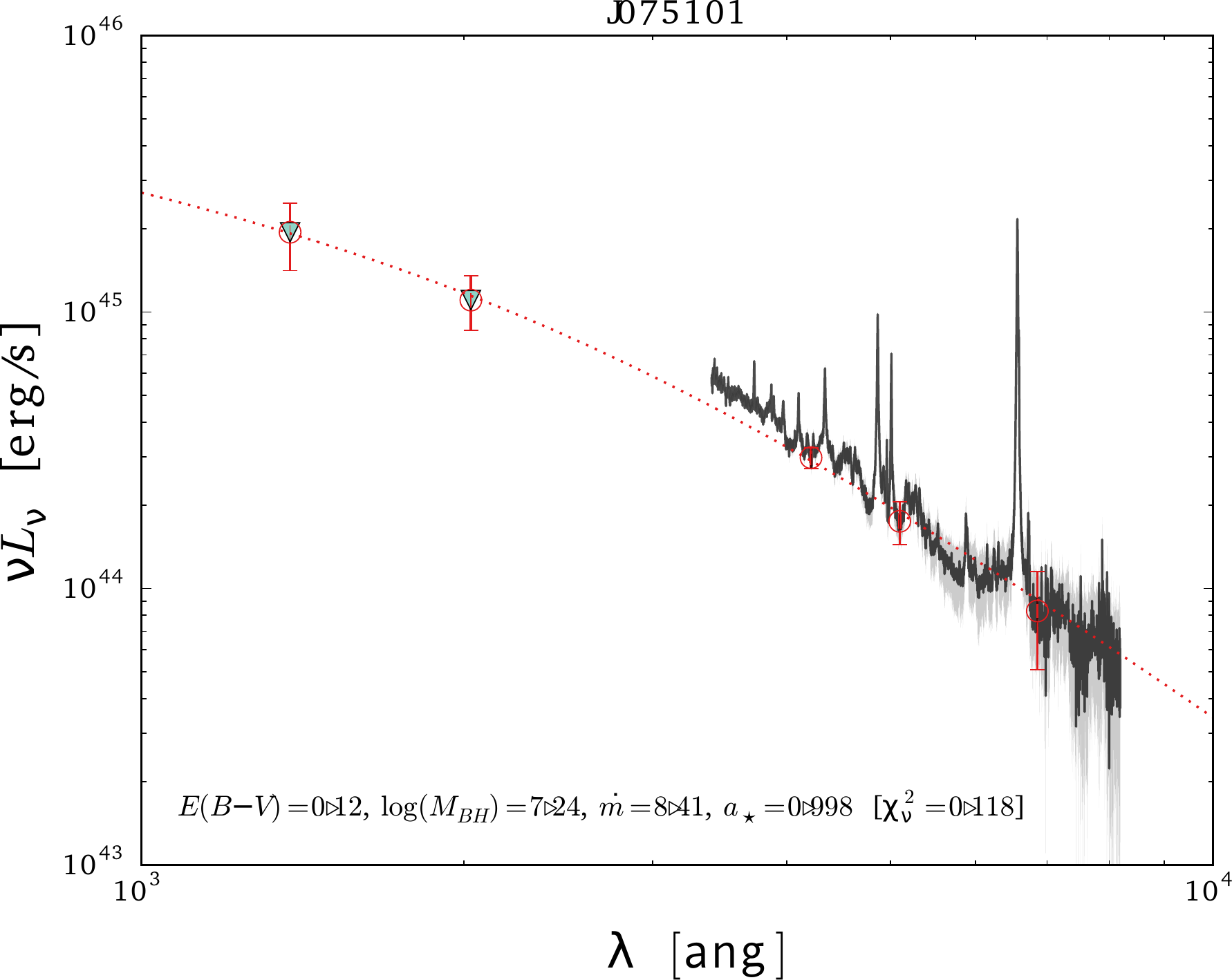}\\
\includegraphics[width=0.33\linewidth]{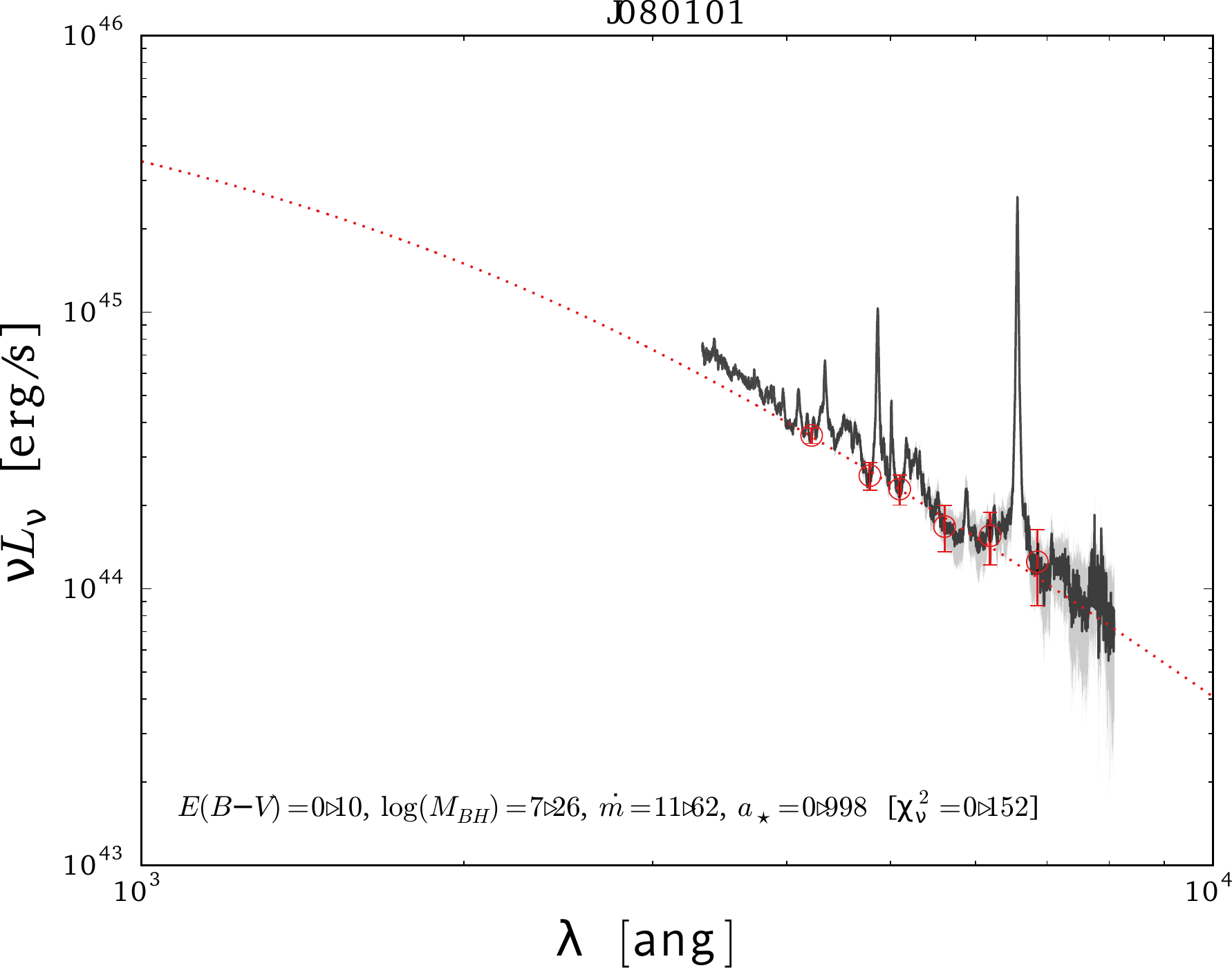}
\includegraphics[width=0.33\linewidth]{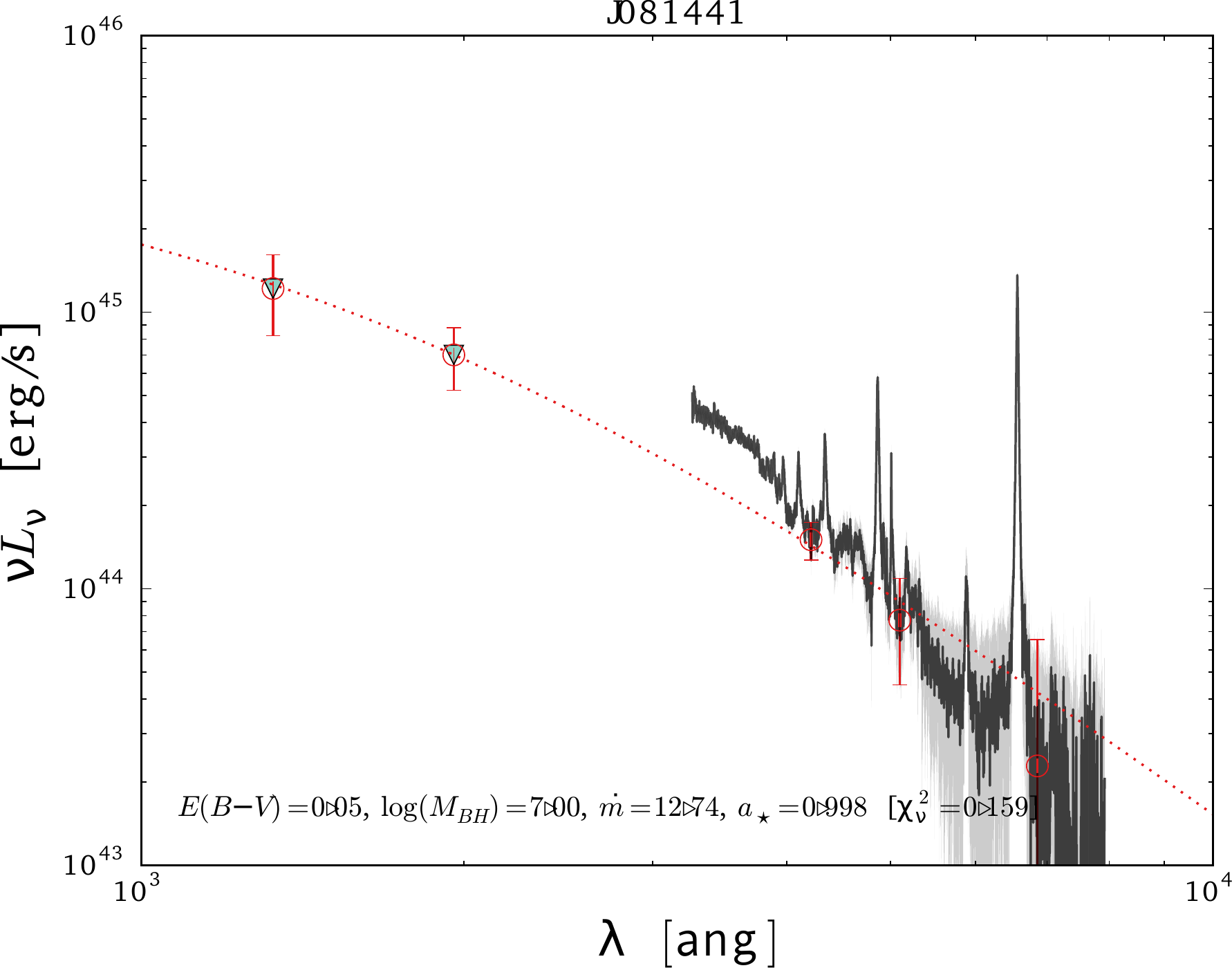}
\includegraphics[width=0.33\linewidth]{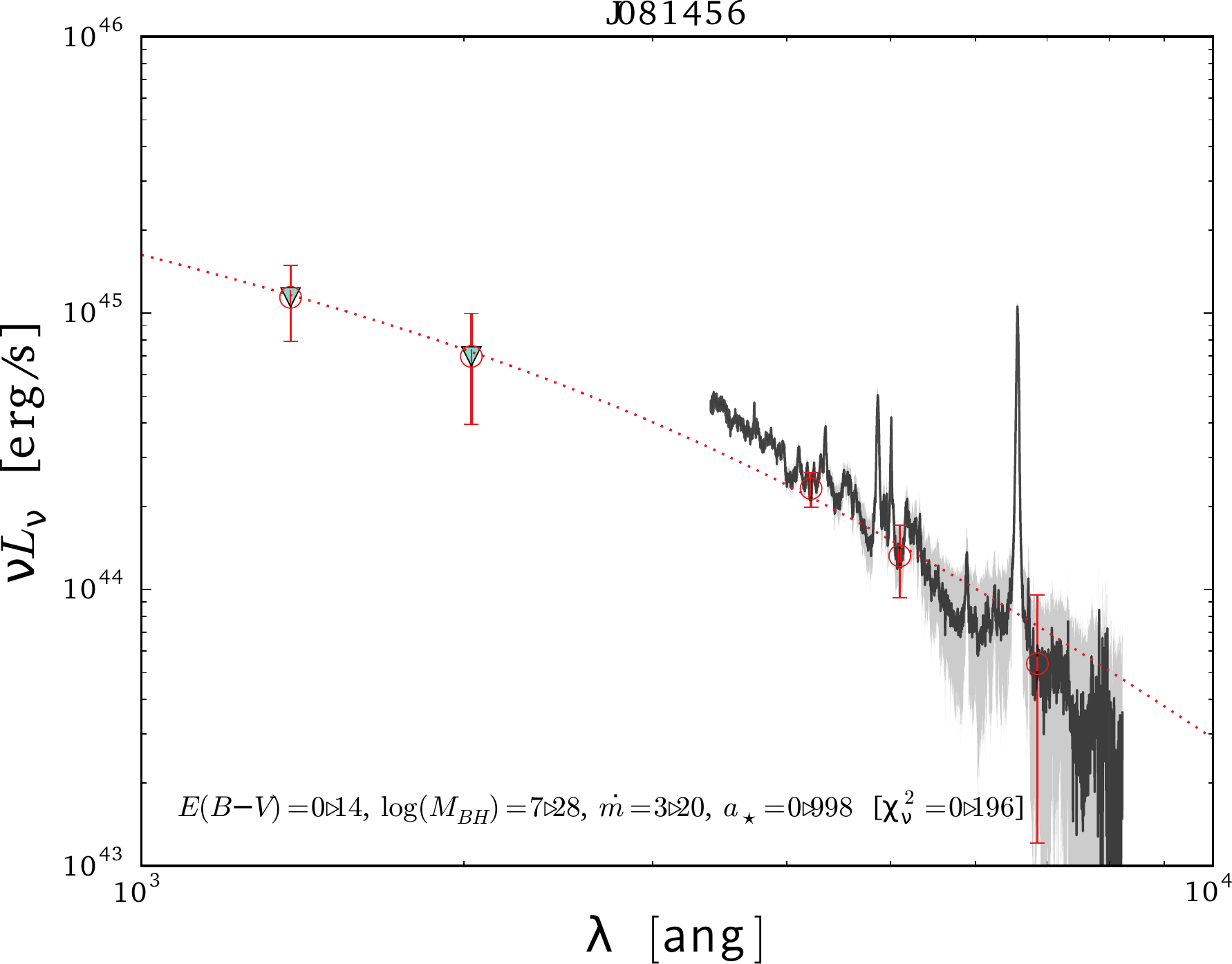}\\
\caption{Intrinsic optical-to-UV SED with the best-fit thin AD model over-plotted assuming
$a_{\star}=0.998$ (red-dotted curves) and $\cos{\theta}$=0.75. The SED is corrected for Galactic
extinction, host-galaxy emission and corrected for intrinsic reddening which
was modelled with a classical SMC extinction curve. The order of the objects is the same as in
Table~\ref{table.bfSED}. Red empty circles represent the continuum regions used for fitting
the models to the spectra, and the full points show the available photometric data points:
GALEX FUV 1551\AA{}, OM UVW2 2120\AA{}, GALEX NUV 2306\AA{}, OM UVM2 2310\AA{}, OM UVW1 2910\AA{},
OM U 3440\AA{}, OM B 4500\AA{},  OM V 5430\AA{}. }\label{figure.bfSED}
\end{figure*}

\begin{figure*}
\includegraphics[width=0.33\linewidth]{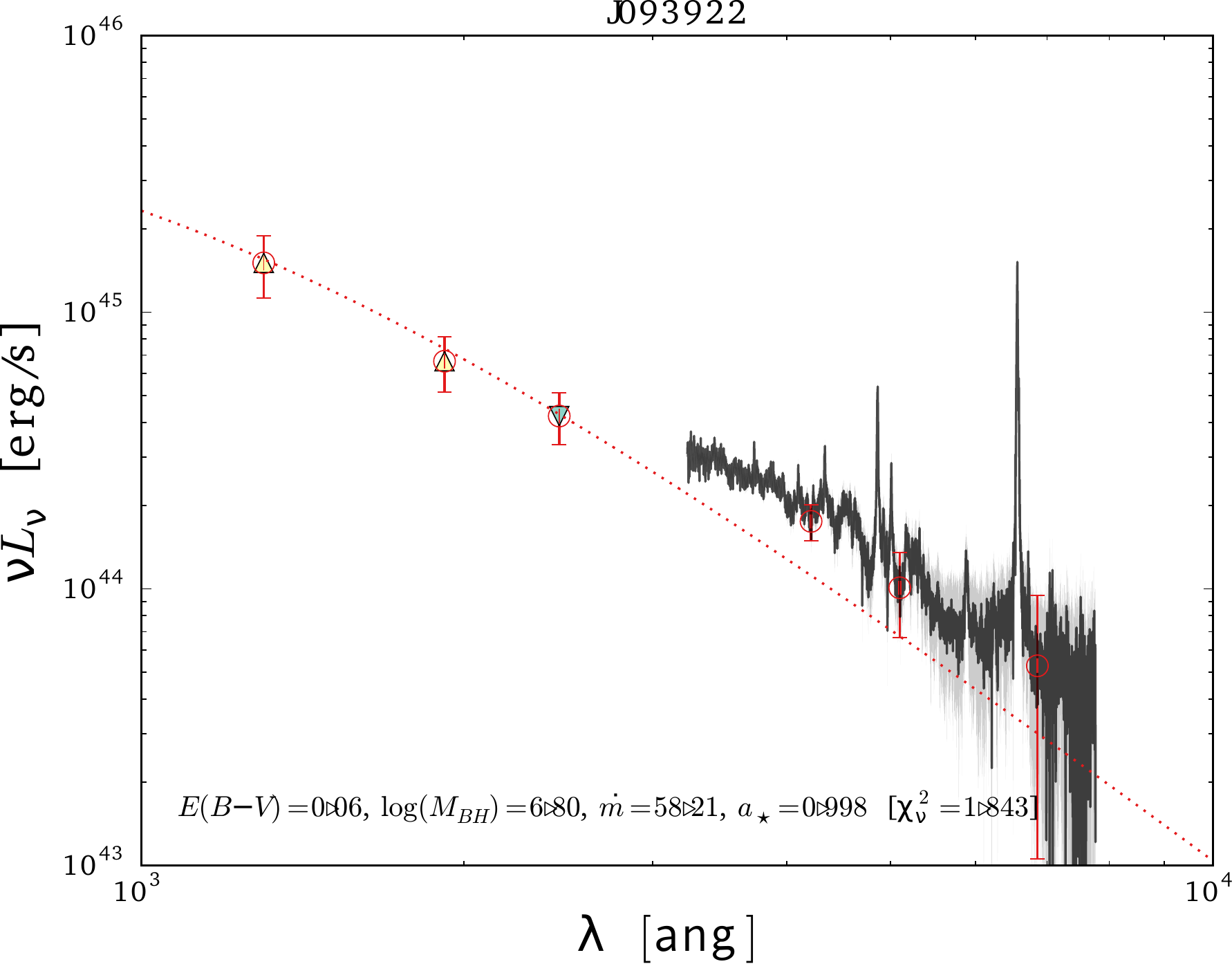}
\includegraphics[width=0.33\linewidth]{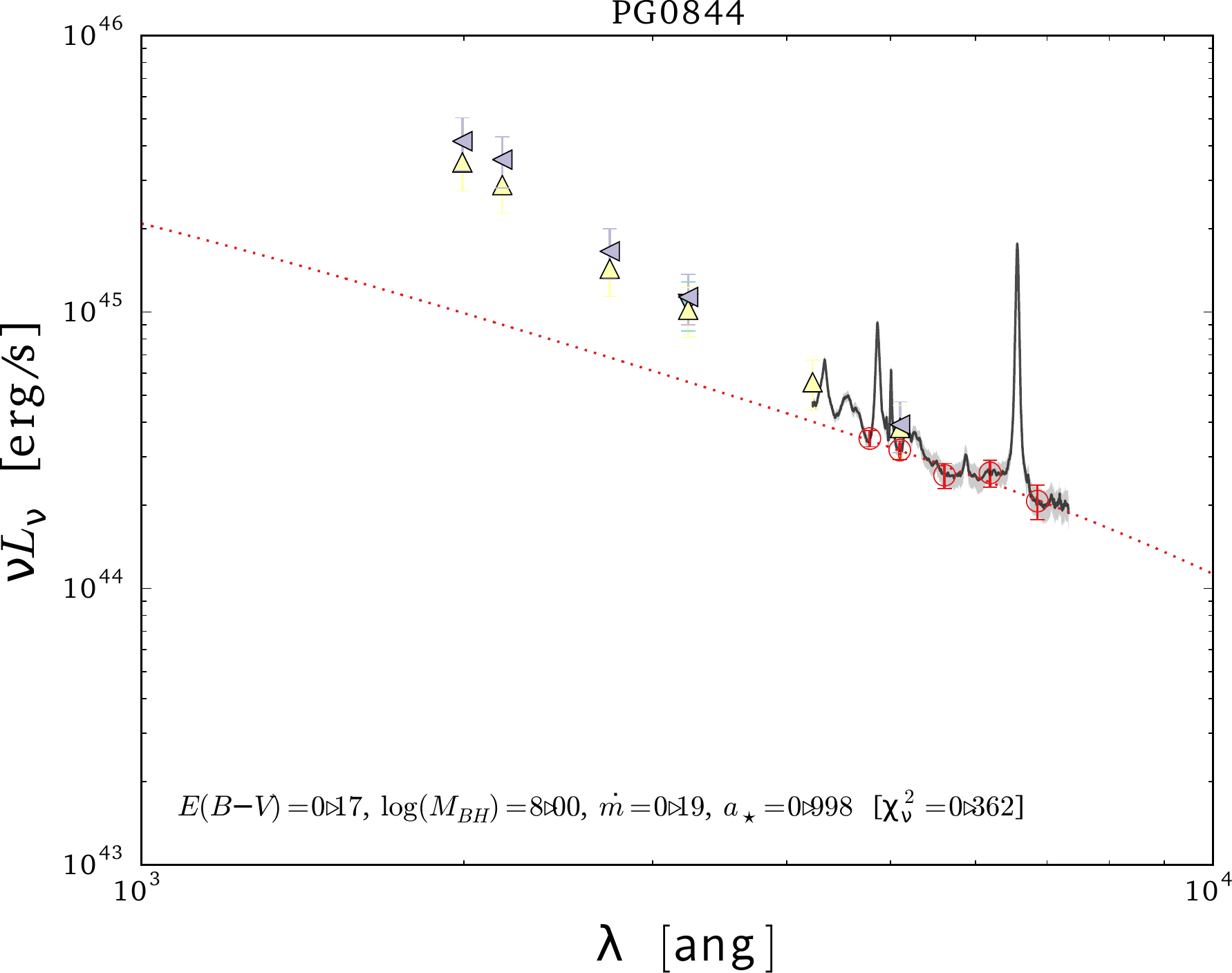}
\includegraphics[width=0.33\linewidth]{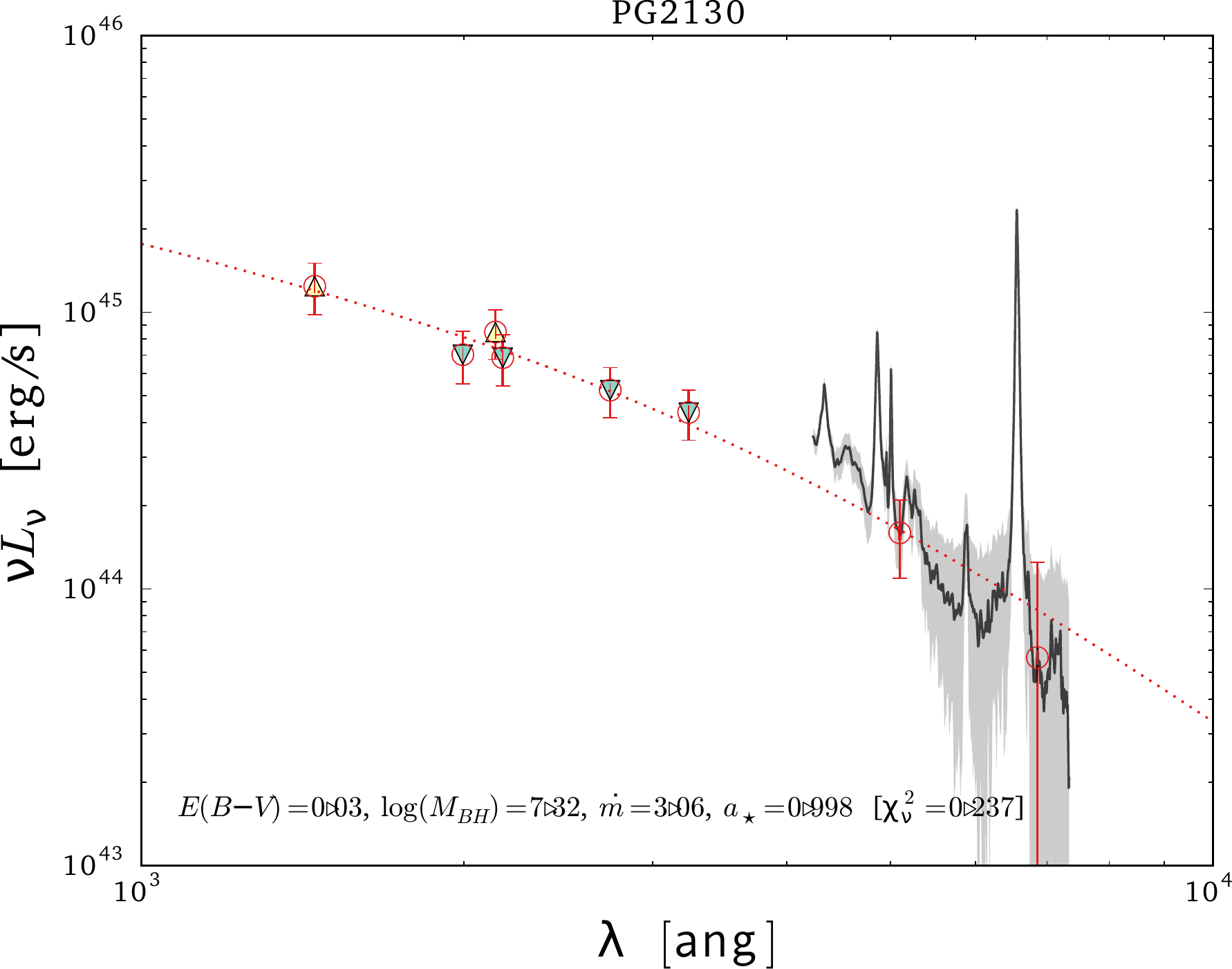}\\
\includegraphics[width=0.33\linewidth]{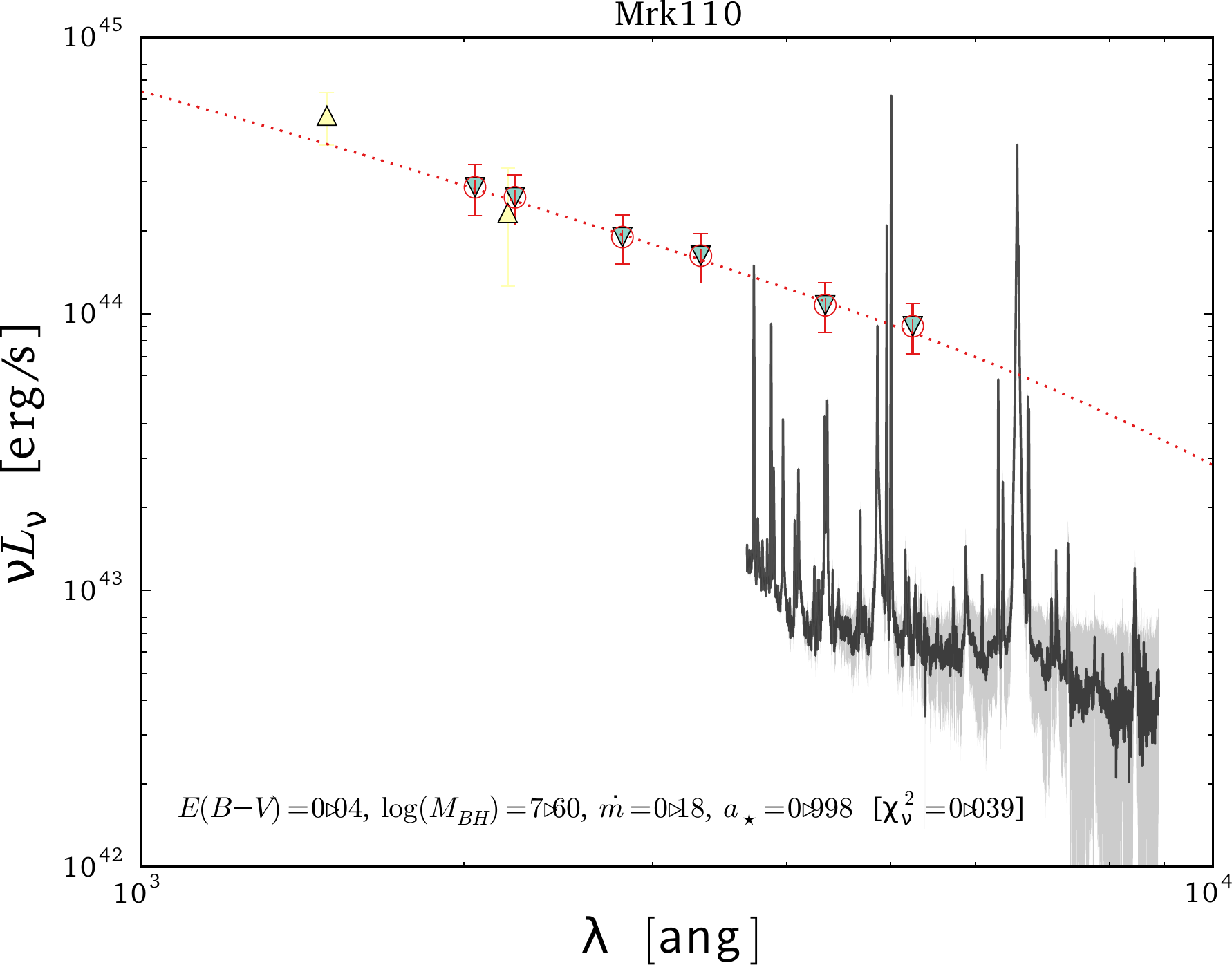}
\includegraphics[width=0.33\linewidth]{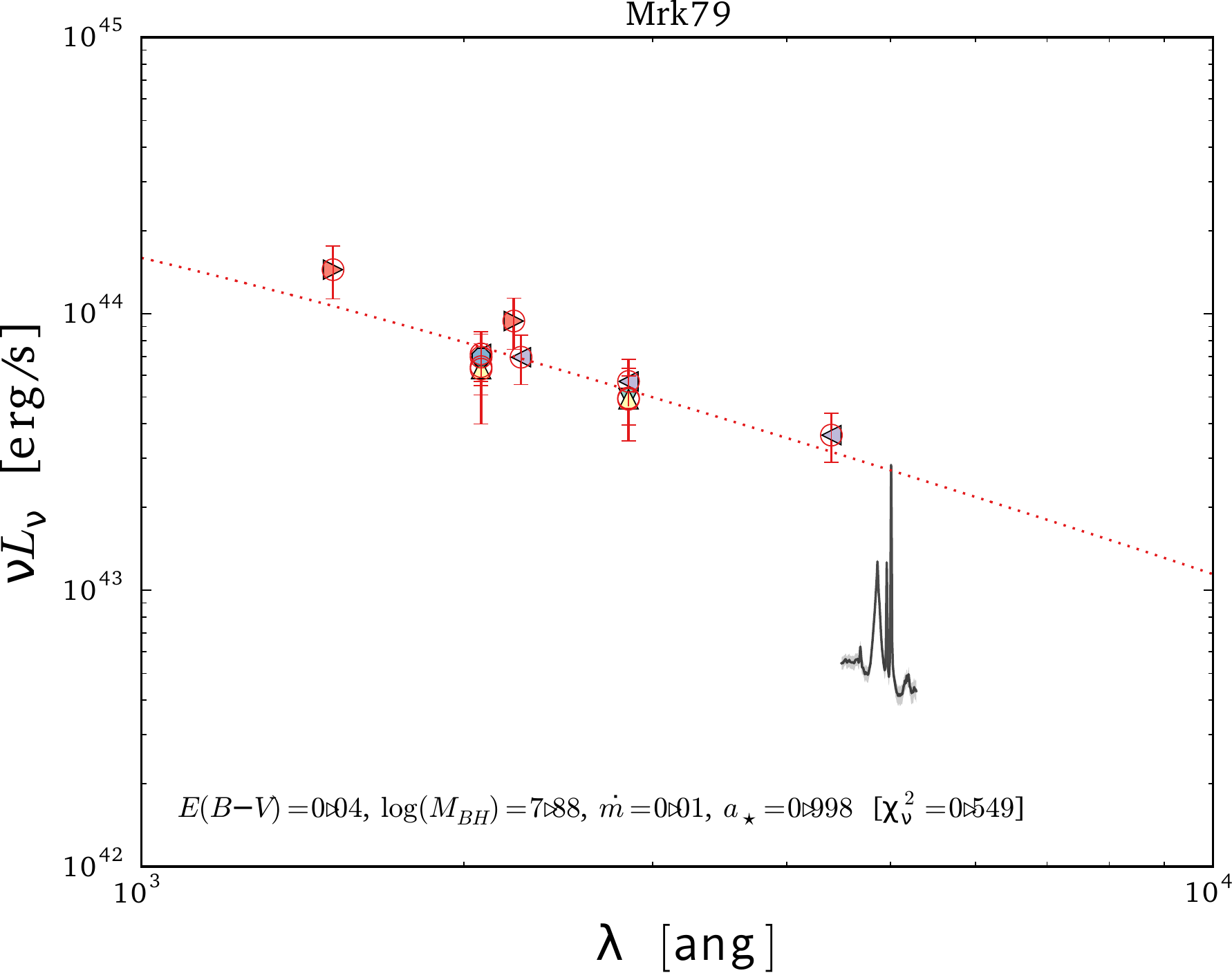}
\includegraphics[width=0.33\linewidth]{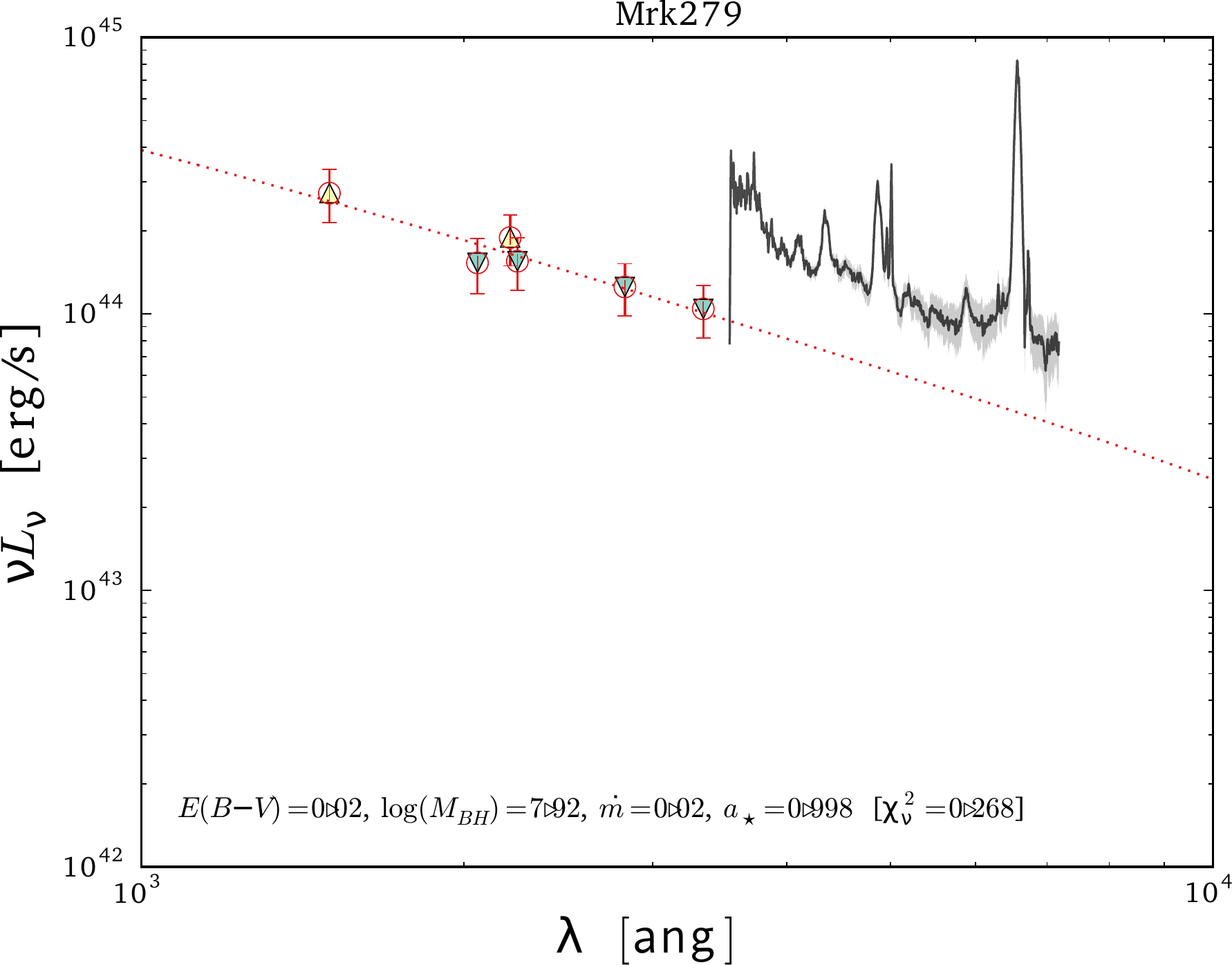}\\
\includegraphics[width=0.33\linewidth]{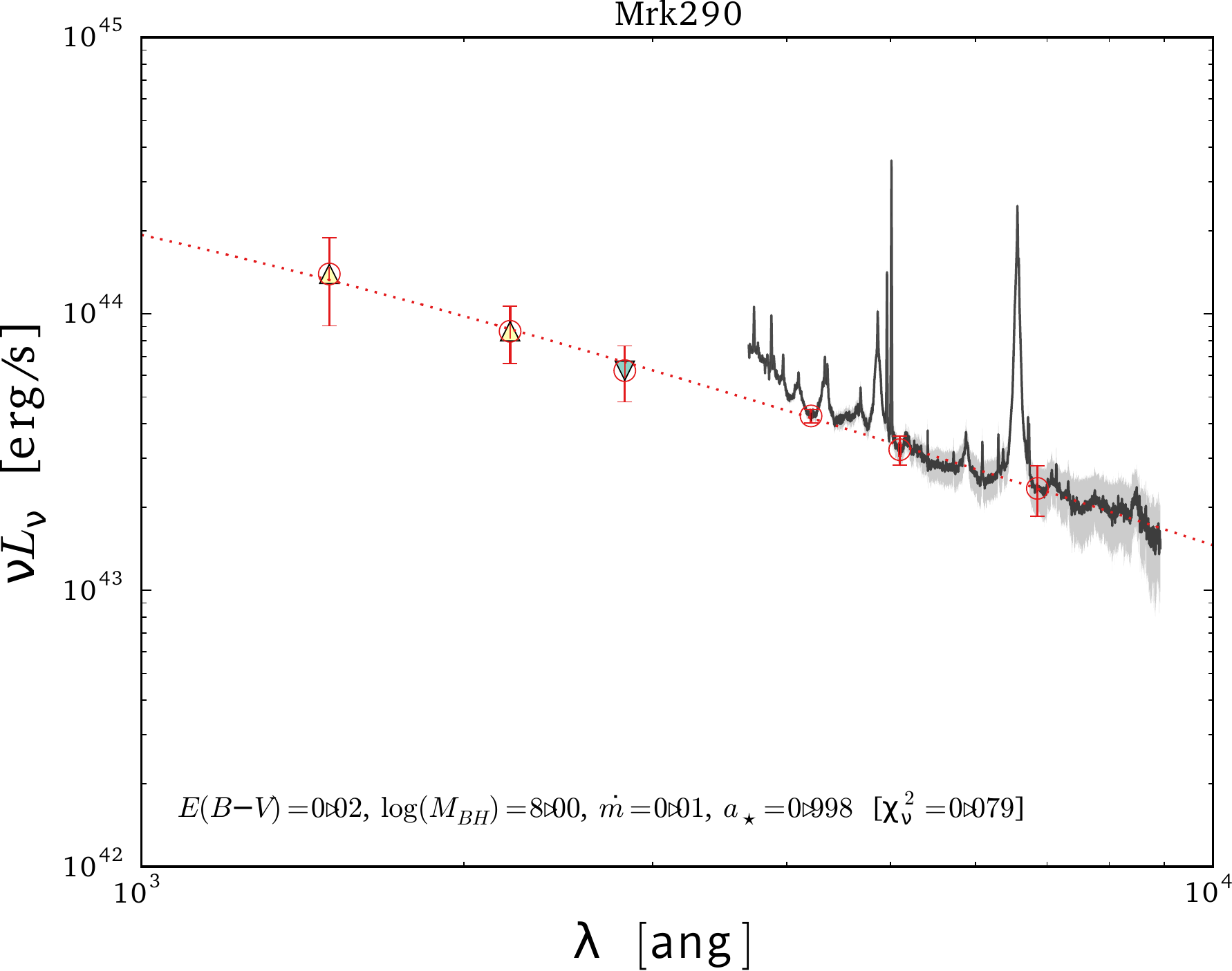}
\includegraphics[width=0.33\linewidth]{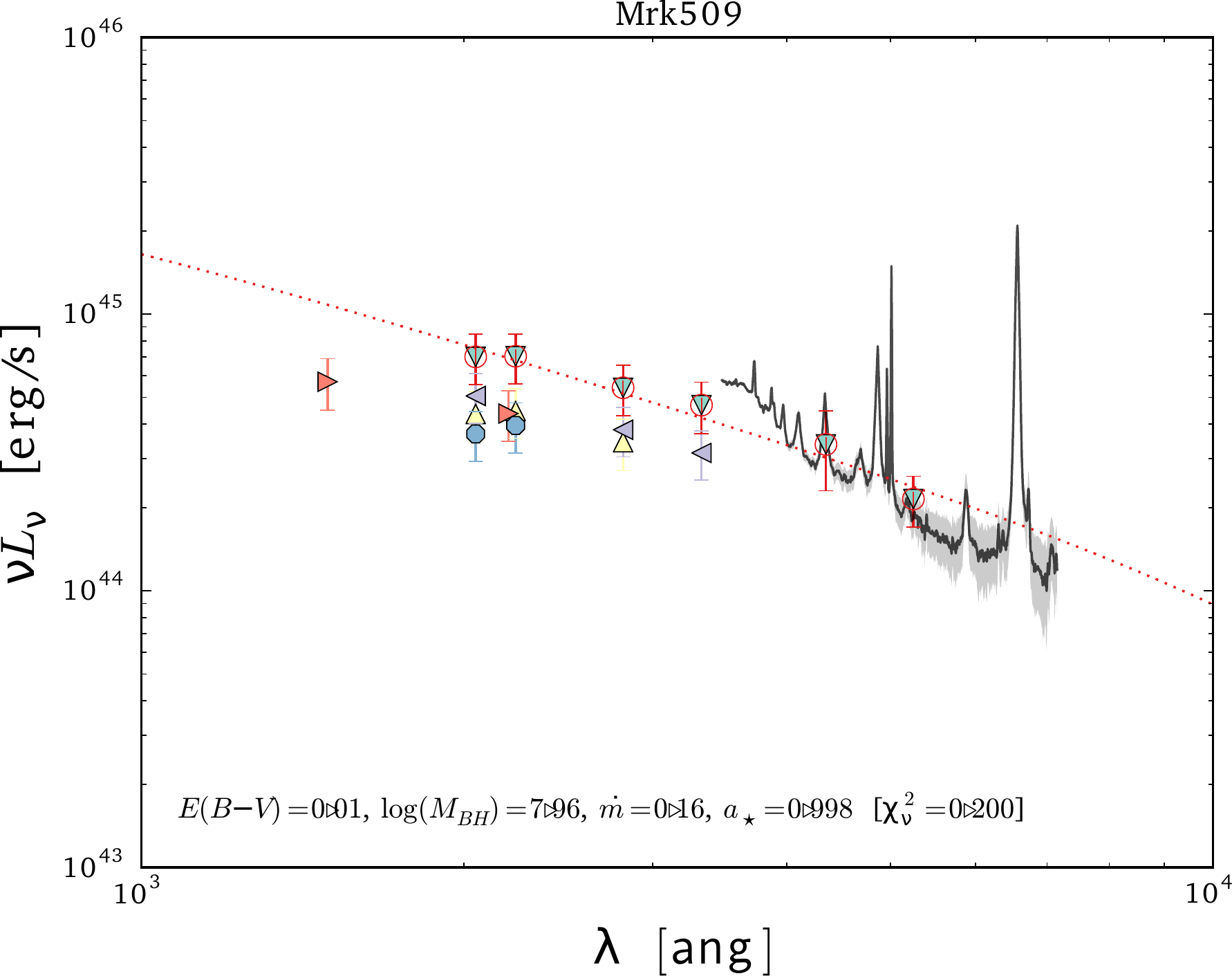}
\includegraphics[width=0.33\linewidth]{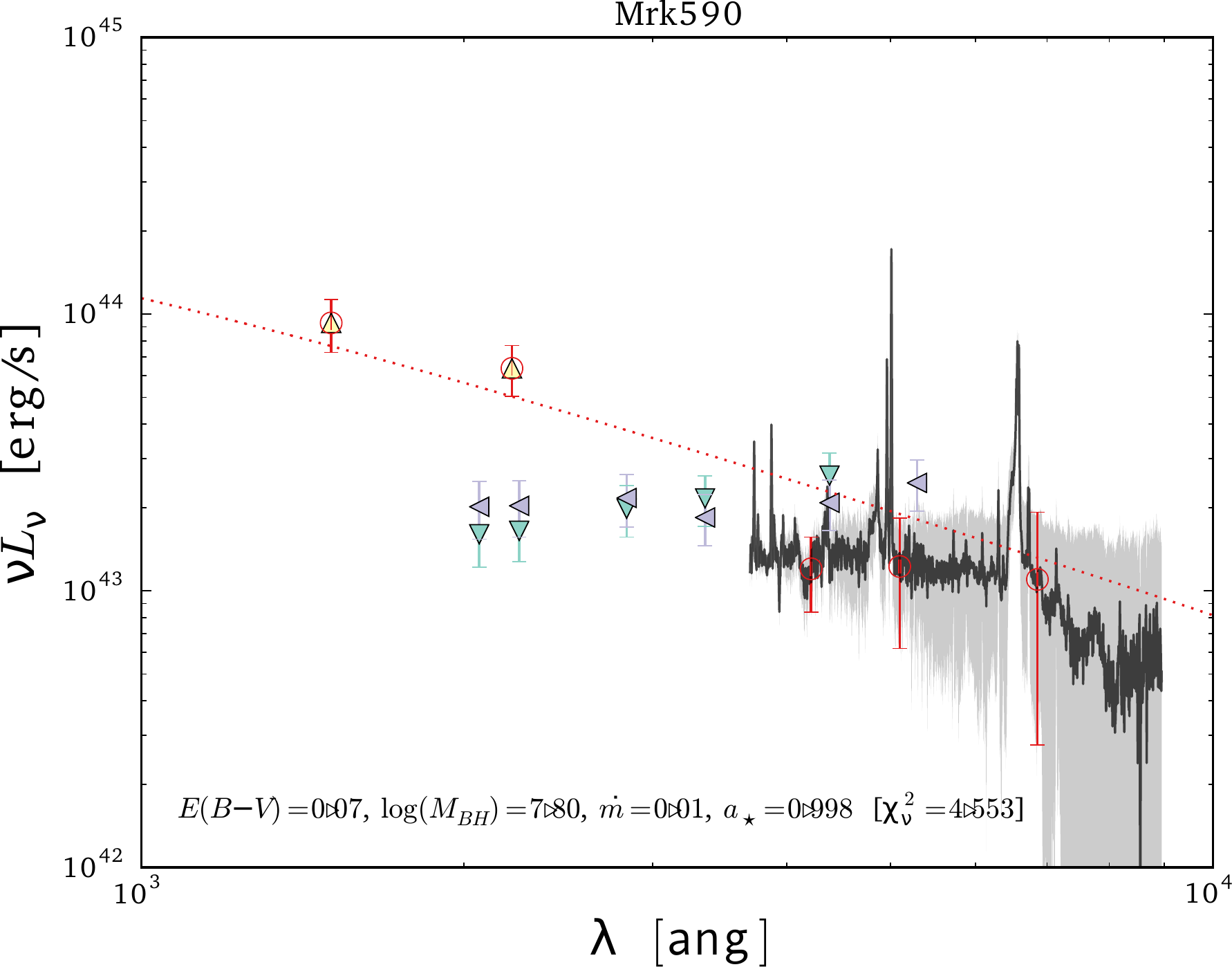}\\
\includegraphics[width=0.33\linewidth]{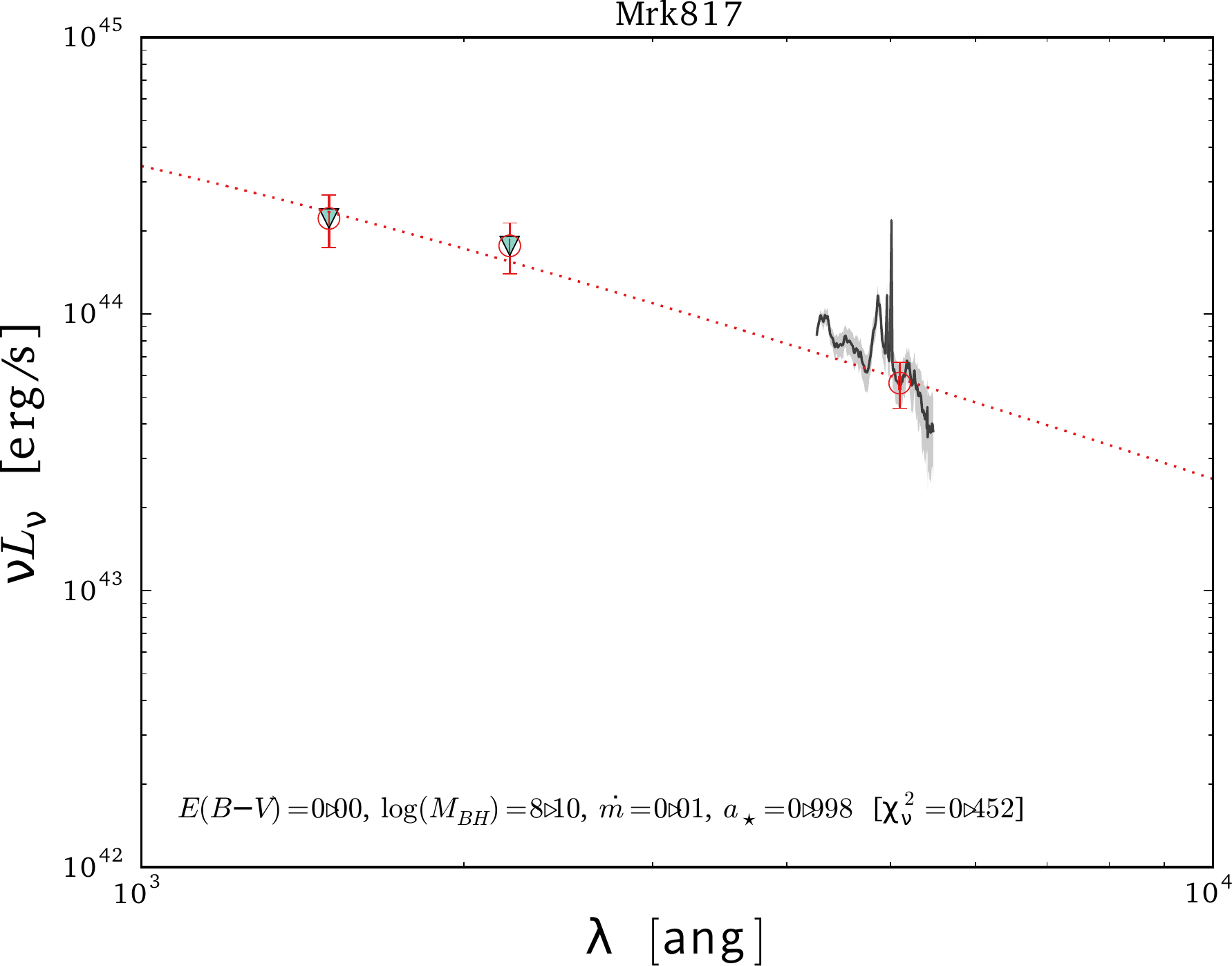}
\includegraphics[width=0.33\linewidth]{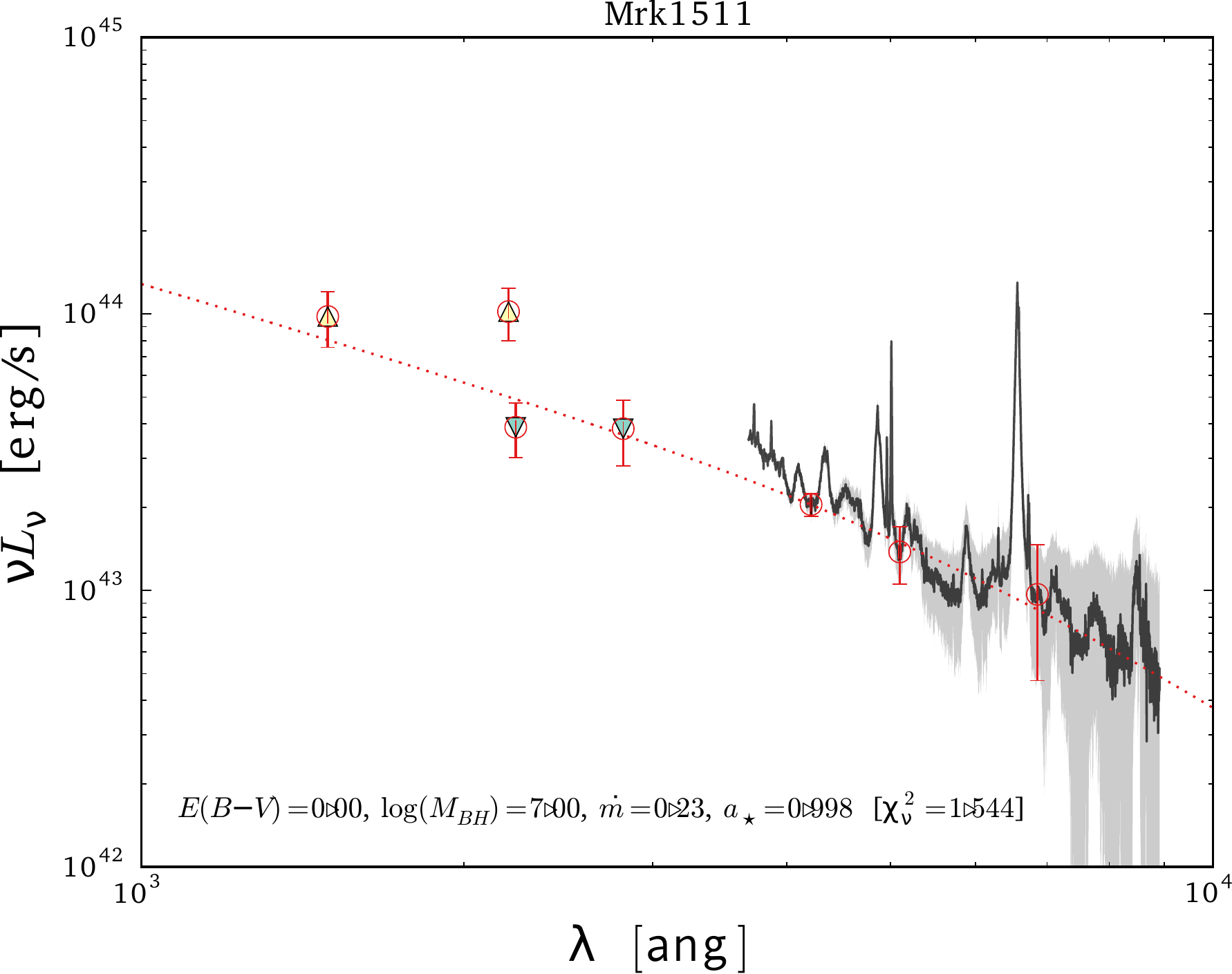}
\includegraphics[width=0.33\linewidth]{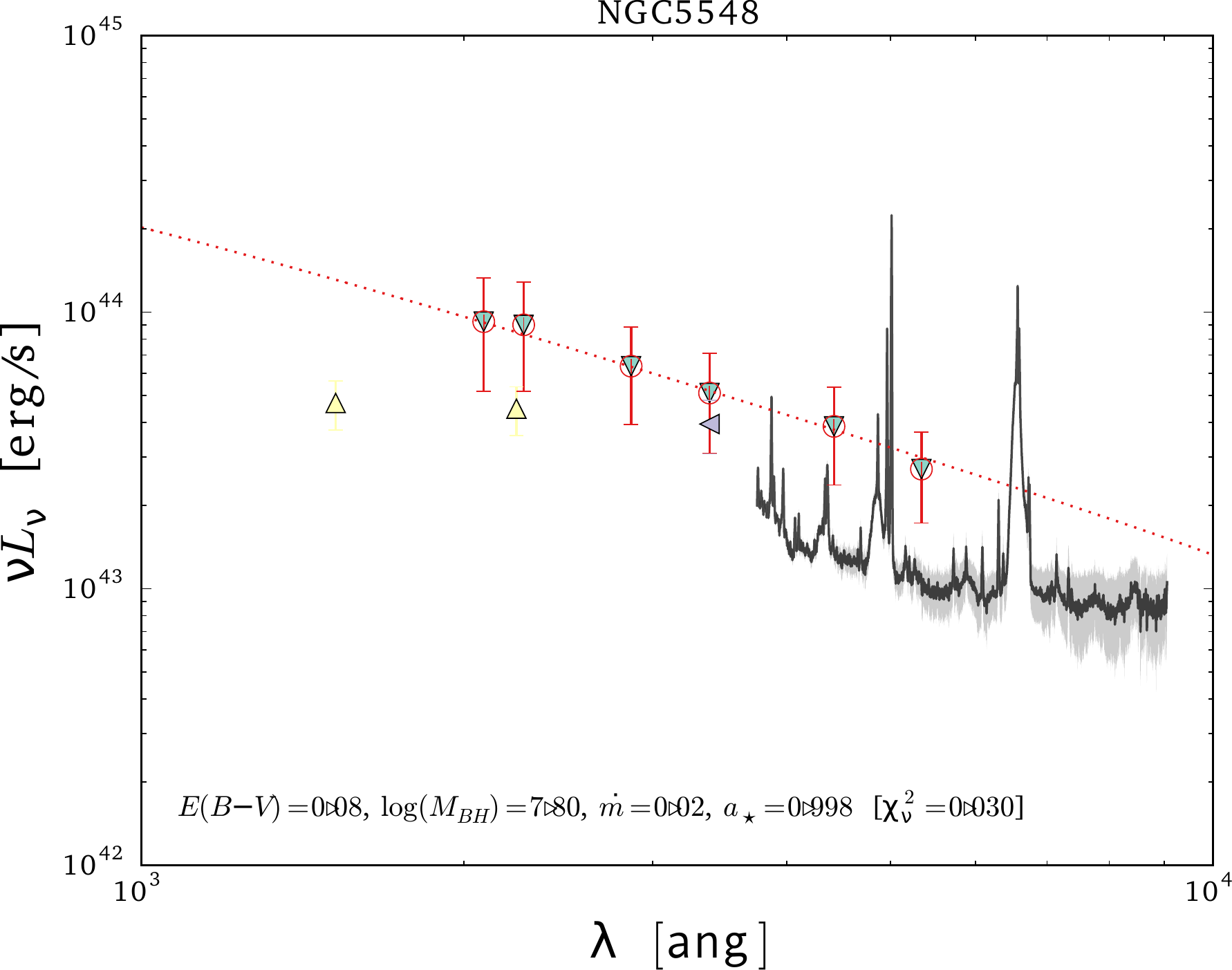}\\
\includegraphics[width=0.33\linewidth]{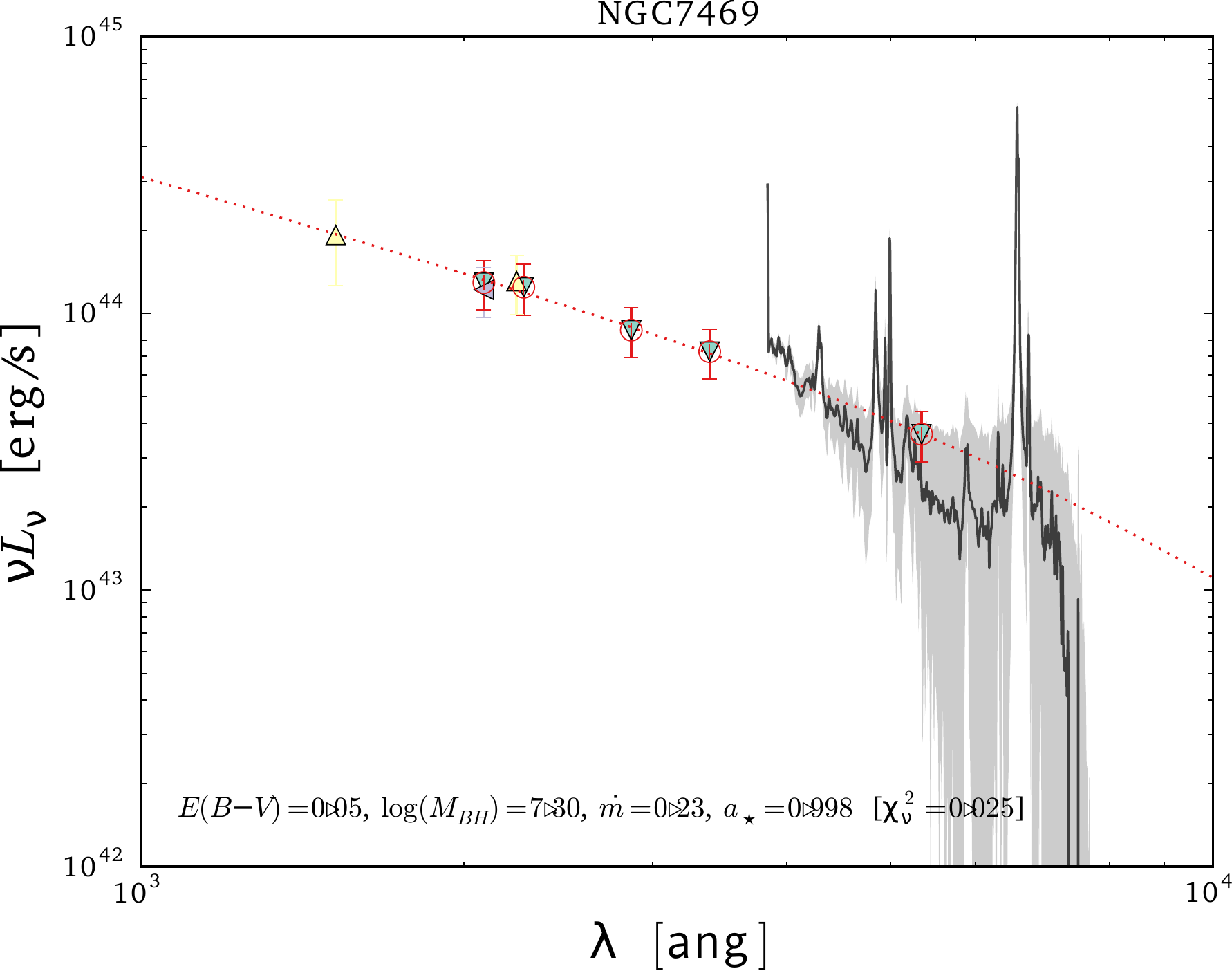}
\includegraphics[width=0.33\linewidth]{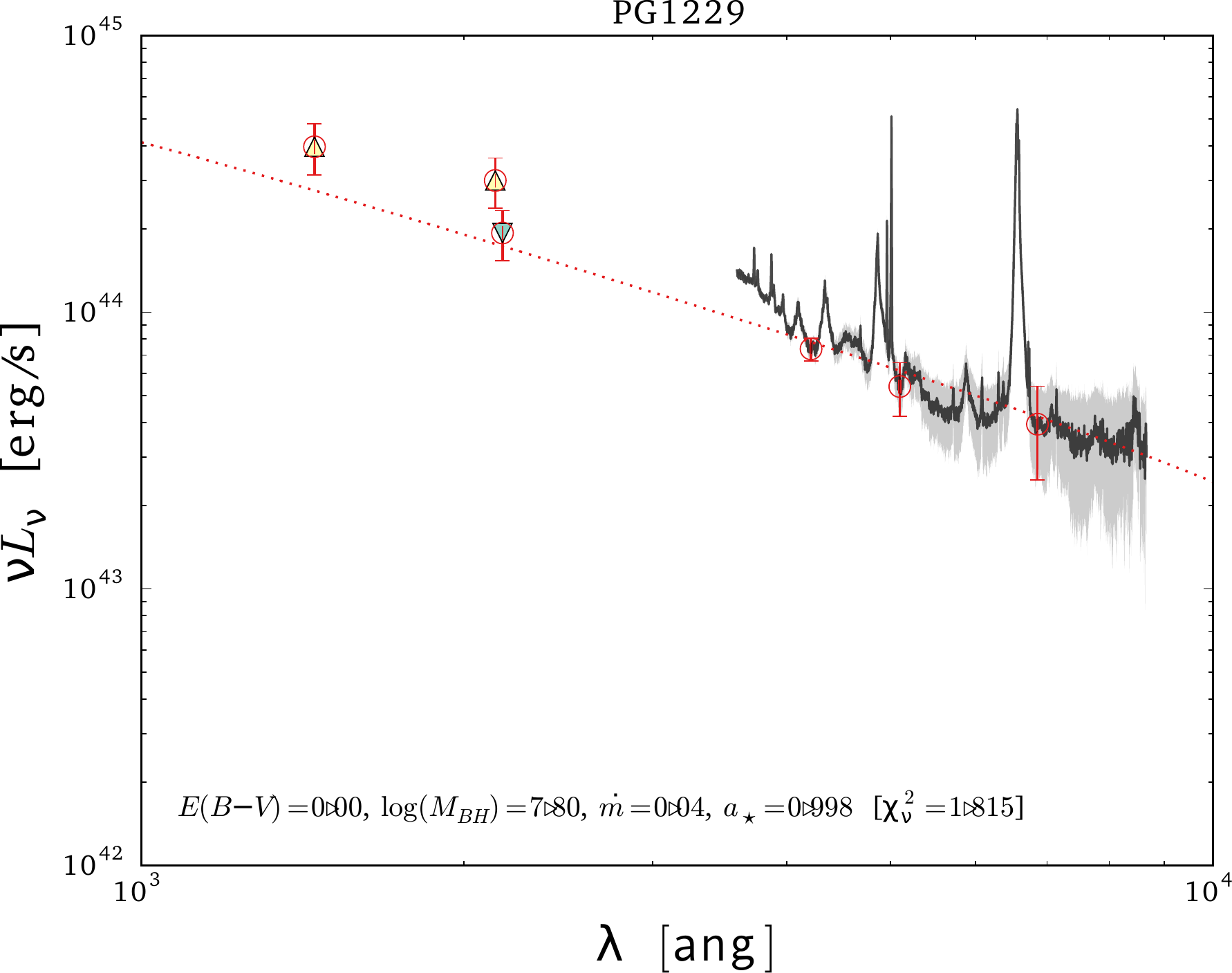}
\includegraphics[width=0.33\linewidth]{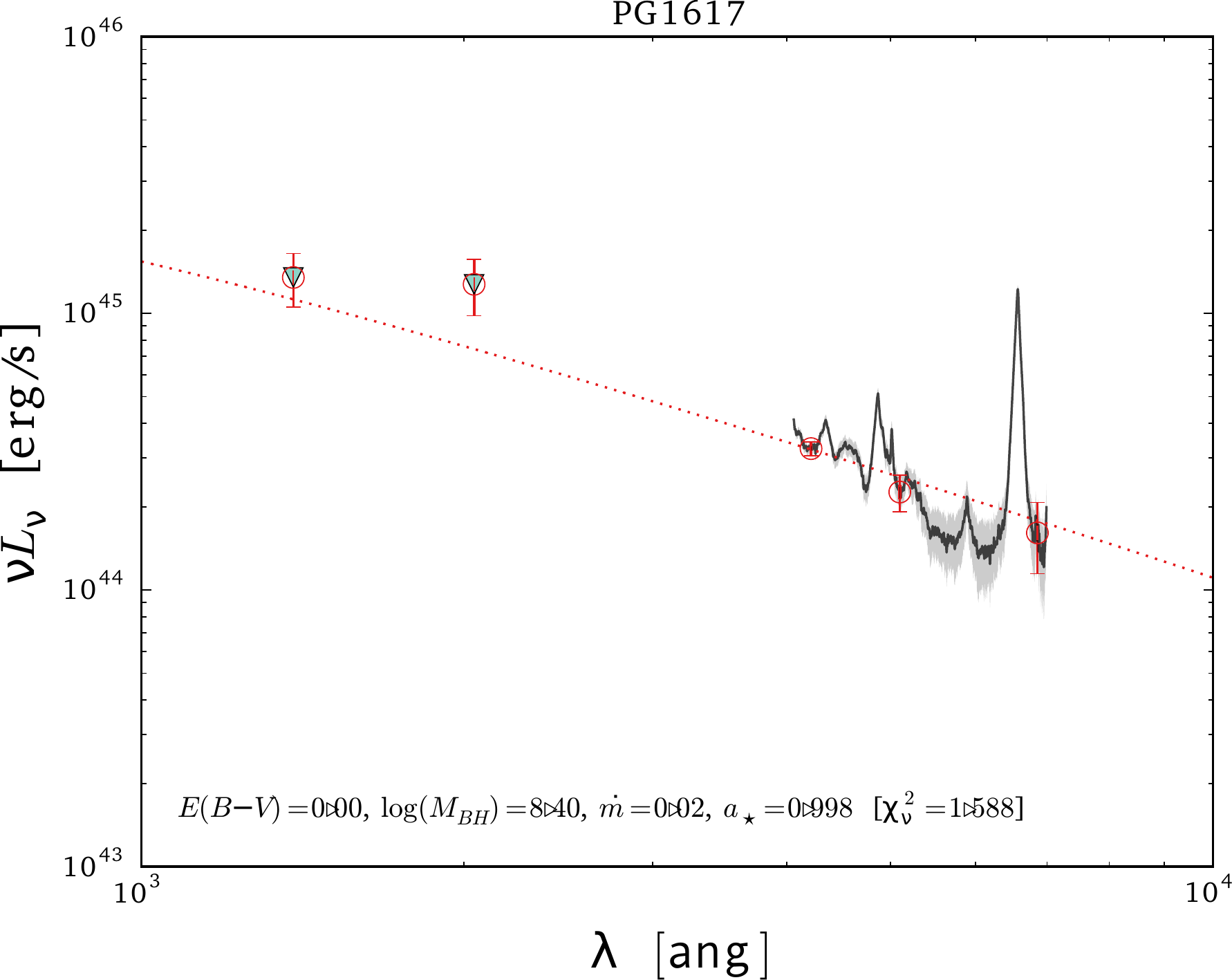}\\
\caption*{continue \dots}
\end{figure*}

\begin{figure*}
\includegraphics[width=0.33\linewidth]{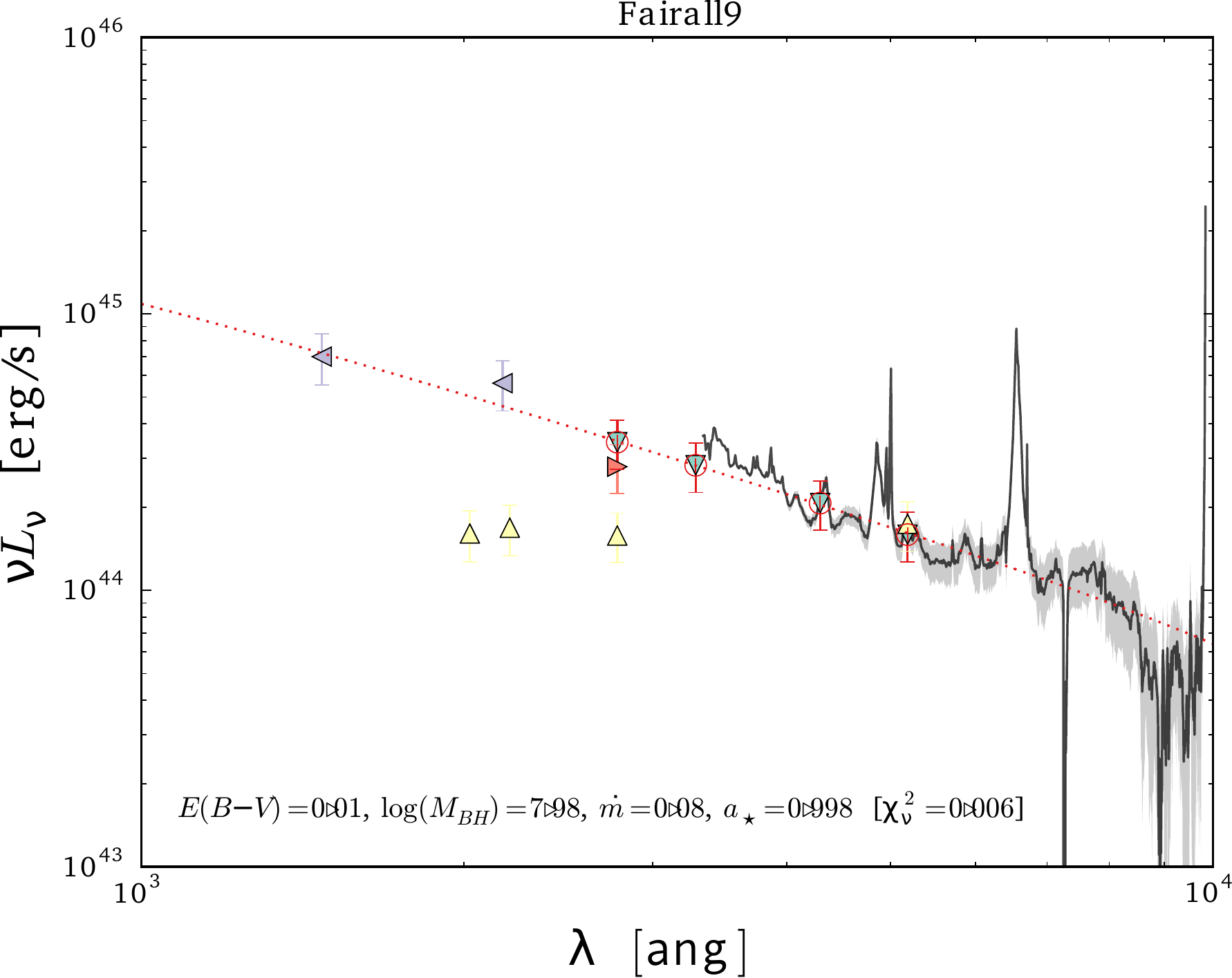}
\includegraphics[width=0.33\linewidth]{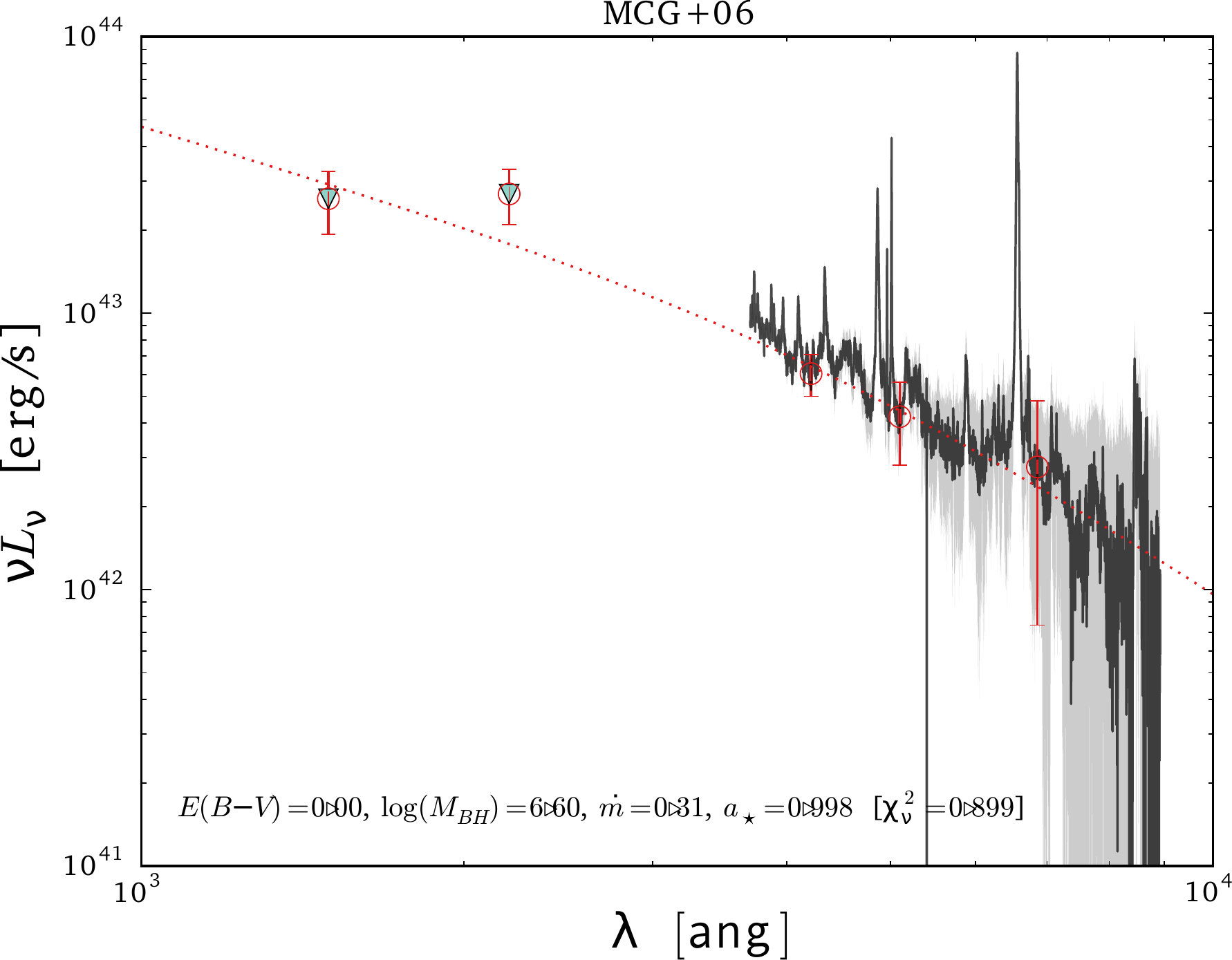}
\caption*{continue}
\end{figure*}

\section{Notes on individual objects}
\label{appendix.Notes}
Additional information regarding available data. Table~\ref{table.bfSEDcomparison} show the best-fit thin
AD model for those sources with more than one possible dataset.

\begin{table*}
\caption{Measured and deduced physical parameters}\label{table.bfSEDcomparison}
\label{table.bfSED2}
\resizebox{1.0\textwidth}{!}{
\begin{tabular}{llccccccccccl}\toprule
& \multicolumn{2}{c}{Host Galaxy} && \multicolumn{9}{c}{\sc thin AD model} \\ \cmidrule(r{.75em}l){2-3}\cmidrule(r{.75em}l){5-12}
Object & template & f && $A_{V}$ & $\log{M_{BH}}$ & $\log{\dot{M}}$ & spin & $\eta$ & $L_{\text{AGN}}$ &  $\dot{m}$ & $\chi^{2}_{\nu}$ & Dataset \\
    &    &   &&     &$[M_{\odot}]$& [$M_{\odot}$/yr] & &   & $10^{44}$erg/s &\\ \midrule
    Mrk335  & ssp\_11Gyr\_z05 & 0.24 &  & $0.15^{+0.07}_{-0.02}$ & 7.30 & -0.85 & -1.000 & 0.038 & 3.04  & 0.27 & 0.556 & GALEX+spec \\
            & ssp\_11Gyr\_z05 & 0.24 &  & $0.22^{+0.45}_{-0.22}$ & 7.40 & -0.80 &  0.600 & 0.091 & 8.17  & 0.24 & 0.044 & OM (2000)  \\
            & ssp\_11Gyr\_z05 & 0.24 &  & $0.34^{+0.41}_{-0.20}$ & 7.40 & -0.65 & -0.200 & 0.052 & 6.59  & 0.34 & 0.014 & OM (2007)  \\
            & ssp\_11Gyr\_z05 & 0.24 &  & $0.37^{+0.28}_{-0.14}$ & 7.40 & -0.75 & -0.200 & 0.052 & 5.24  & 0.27 & 0.121 & OM (2009)  \\
 IRASF12397 & ssp\_11Gyr\_z05 & 0.52 &  & $1.63^{+0.26}_{-0.13}$ & 7.11 &  0.04 & -0.200 & 0.052 & 32.37 & 3.25 & 0.189 & GALEX+spec \\
            & ssp\_11Gyr\_z05 & 0.52 &  & $1.89^{+0.00}_{-0.12}$ & 7.28 &  0.04 &  0.000 & 0.057 & 35.49 & 2.22 & 1.698 & OM (2005)  \\
    Mrk279  & ssp\_11Gyr\_z02 & 0.33 &  & $0.06^{+0.17}_{-0.06}$ & 7.90 & -1.60 &  0.998 & 0.321 & 4.57  & 0.01 & 0.337 & GALEX+OM   \\
            & ssp\_11Gyr\_z02 & 0.33 &  & $0.00^{+0.33}_{-0.00}$ & 8.40 & -1.65 &  0.998 & 0.321 & 4.07  & 0.00 & 0.277 & spec       \\
    Mrk509  & ssp\_11Gyr\_z02 & 0.35 &  & $0.03^{+0.06}_{-0.03}$ & 7.70 & -0.85 &  0.998 & 0.321 & 25.69 & 0.11 & 0.123 & GALEX+spec \\
            & ssp\_11Gyr\_z02 & 0.35 &  & $0.21^{+0.11}_{-0.21}$ & 7.90 & -0.45 &  0.998 & 0.321 & 64.52 & 0.17 & 0.035 & OM (2009)  \\
    NGC5548 & ssp\_11Gyr\_z05 & 0.34 &  & $0.25^{+0.12}_{-0.25}$ & 8.00 & -2.70 &  0.998 & 0.321 & 0.36  & 0.00 & 1.324 & GALEX+spec \\
            & ssp\_11Gyr\_z05 & 0.34 &  & $0.44^{+0.34}_{-0.44}$ & 8.50 & -2.20 &  0.998 & 0.321 & 1.15  & 0.00 & 0.070 & OM (2013)  \\
    NGC7469 & ssp\_11Gyr\_z02 & 0.70 &  & $0.14^{+0.43}_{-0.14}$ & 7.10 & -1.05 &  0.600 & 0.091 & 4.59  & 0.27 & 0.389 & GALEX+spec \\
            & ssp\_11Gyr\_z02 & 0.70 &  & $0.34^{+0.28}_{-0.25}$ & 7.70 & -1.30 &  0.600 & 0.091 & 2.58  & 0.04 & 0.092 & OM (2009)  \\
\bottomrule
\end{tabular}
}
\end{table*}

\label{IndividualNotes}
\subsubsection*{Mrk~335}
Mrk~335 was observed three times by the OM aboard of the {\it XMM-Newton} which allow us to fit three independent and simultaneous datasets
and compare it with the non-simultaneous SED form by the mean optical spectrum and the GALEX photometry. The simultaneous datasets show
variability at a level of 20\% over a 9 year period. The best-fit thin AD model results achieved with the non-simultaneous SED are firmly
consistent with the shapes of the simultaneous SED and both contemporaneous and non-contemporaneous SED shapes claim for
a considerable amount of intrinsic extinction in order to ensure satisfactory fit ($A_V=0.37-0.15$).
The contemporaneous SED datasets show larger $A_V$ values.

\subsubsection*{IRAS~F12397+3333}
IRAS~F12397+3333 is an IR luminous source and as such shows \emph{redder} optical SED compared with most AGN, including those in the sample.
Both simultaneous and non-simultaneous SED fit results show that a satisfactory fit will not be reached unless a significant dust attenuation is assumed.
The hypothesis of notable intrinsic reddening is supported by polarization measurements, as well as by a steep Balmer Decrement \citep{Grupe1998,DuPu2015}.

\subsubsection*{J093922}
The good agreement between the near-UV GALEX flux with the OM photometric point (temporally separated by $\sim$1 year) allow us to use both in
conjunction with the optical spectrum and assume that it is the most convenient approach to model the optical/UV underlying continuum.

\subsubsection*{PG~2130}
The good agreement between the near-UV GALEX flux with the two OM photometric points (UVW2, UVM2) and the low statistics of the simultaneous dataset
lead us to believe that a non-simultaneous dataset is the best procedure to study the optical/UV emission. The simultaneous SED shape, which consist of only four
photometric points, is consistent to that found by fitting this non-contemporaneous dataset.

\subsubsection*{PG~0844}
PG~0844 is the only source where the optical/UV broadband SED could not be modelled by thin AD spectrum.
The inconsistent on the time-lag measurements from line-to-line \citep[see][]{Peterson1999} as well as the residuals on the
Balmer narrow-emission lines in this object turns the reverberation mapping results to be of rather low quality. Thus, the
\hbeta{} BLR size measurement is considered unreliable and does not enable us to find a satisfactory fit to the optical/UV
OM photometric points. We believe that the unreliable time-lag measurements enable to find an accurate black hole mass measurements.

\subsubsection*{Mrk~110}
The optical spectrum was taken in the lowest flux state, whilst both GALEX and XMM-Newton observed Mrk~110
in higher flux states. The optical observations show variability of a factor $\sim$10 over
a 3 year period. Therefore the simultaneous SED composed by six OM photometric points is the only possible dataset.
Although the GALEX photometry was not fitted they are in good agreement with the contemporaneous
optical/UV SED shape.

\subsubsection*{Mrk~79}
The optical spectrum was taken in the lowest flux state, whilst both GALEX and XMM-Newton observed Mrk~79
in a higher flux states. The narrower spectral window of Mrk~79 and the large observed variability between the mean optical
spectrum and the photometry rules out the spectrum.
Taking advantage of the good agreement between the near-UV GALEX flux and the OM photometry, a non-simultaneous dataset
is probably the best SED to study the accretion disc emission. Indeed, the low number of simultaneous photometric points
does not allow to construct a contemporaneous dataset.

\subsubsection*{Mrk~279}
The OM photometric flux at the effective wavelength 3444\AA{} seems to be in a lower flux state when it is compared with the
optical spectrum. This 20\% of variability between both spectrum and OM photometry prevent to join both datasets. Two non-simultaneous
SED were constructed: on the one hand, GALEX in conjunction with the simultaneous OM photometric band; on the other, a fit over the
single optical spectrum adding another two line-free points. The fitting results of both datasets are in good agreement and no intrinsic
reddening was necessary to get a satisfactory thin AD fit.

\subsubsection*{Mrk~509}
There is a variability of a factor $\sim$2 between simultaneous OM observations over a 9 year period. From the OM observations, we can see
how both optical spectrum and GALEX photometry seem to follow the lowest flux state rather than a higher one. In addition to the
contemporaneous data set given by the OM observation in the 2009, we also fitted a non-simultaneous dataset which consists of the
GALEX photometric data in conjunction with the optical spectrum. The fitting results show that for the simultaneous SED fit the intrinsic
reddening is higher than that claimed for the non-simultaneous fit.

\subsubsection*{NGC~5548}
There is a variability of a factor $\sim$4 between the OM photometric point at the effective wavelength 5430\AA{} and the optical spectrum.
The non-simultaneous dataset (GALEX plus the optical spectrum) was also fitted in order to be compared with the contemporaneous SED given
by the OM dataset. Both constructed SED are in good agreement and the best fit in both demand for some dust attenuation.

\subsubsection*{Fairall 9}
As has been seen in previous works, the OM observations manifest variability of a factor $\sim$2.5 at the effective wavelength 2910\AA{}.
We decline to the simultaneous SED due to the poor statistic of the contemporaneous OM dataset. The good agreement between the
highest OM flux state with the GALEX data makes the non-simultaneous SED the best data to perform our study.

\subsubsection*{NGC~7469}
The contemporaneous photometric points are in good agreement with both the GALEX data and the optical spectrum. The simultaneous SED fit is
in good agreement with the non-simultaneous dataset and also require a similar dust extinction correction, being both fits highly
consistent.

\subsubsection*{Mrk~382, J093922, Mrk~290, Mrk~1511, PG~1229}
The good agreement between the near-UV GALEX flux with the OM photometric point allow us to use both in
conjunction with the optical spectrum as the most convenient approach to model the optical/UV underlying continuum.

\subsubsection*{Mrk~590}
Note that there is a real disagreement between the OM photometry and the optical spectrum. We decide
not to use the OM photometry.

\bsp    
\label{lastpage}
\end{document}